\documentclass[aps,prd,preprint,preprintnumbers,groupedaddress,eqsecnum,showpacs]{revtex4-1}
\pdfoutput=1
\usepackage[T1]{fontenc}
\usepackage{latexsym}
\usepackage{amsmath}
\usepackage{graphicx, color}
\input{definitions.command}

\usepackage{ulem}

\begin{document}

\preprint{
OU-HET 788,   
KEK-CP 288,  
KIAS-PREPRINT-P13052
}

\flushbottom

\title{Polyakov loops and the Hosotani mechanism on the lattice}
\author{Guido Cossu}
\email[]{cossu@post.kek.jp}
\affiliation{Theory Center, IPNS, High Energy Accelerator Research Organization (KEK),\\
 Tsukuba, Ibaraki 305-0810, Japan}

\author{Hisaki Hatanaka}
\email[]{hatanaka@kias.re.kr}
\affiliation{School of Physics, KIAS,\\ Seoul 130-722, Republic of Korea}

\author{Yutaka Hosotani}
\email[]{hosotani@phys.sci.osaka-u.ac.jp}
\affiliation{Department of Physics, Osaka University,\\ Toyonaka, Osaka 560-0043, Japan} 

\author{Jun-Ichi Noaki}
\email[]{noaki@post.kek.jp}
\affiliation{Theory Center, IPNS, High Energy Accelerator Research Organization (KEK),\\
 Tsukuba, Ibaraki 305-0810, Japan}

\begin{abstract}
We explore the phase structure and symmetry breaking in 
four-dimensional SU(3) gauge theory with one spatial compact dimension 
on the lattice ($16^3 \times 4$ lattice) in the presence of fermions 
in the adjoint representation with periodic boundary conditions.  
We estimate numerically the density plots of the Polyakov loop eigenvalues phases, 
which reflect the location of minima of the effective potential in the Hosotani mechanism. 
We find strong indication that the four phases found on the lattice
correspond to $SU(3)$-confined, $SU(3)$-deconfined, $SU(2) \times U(1)$, and 
$U(1)\times U(1)$ phases predicted by the one-loop perturbative calculation.
The case with fermions in the fundamental representation with general boundary conditions,
equivalent to the case of imaginary chemical potentials,
is also found to support the $Z_3$ symmetry breaking in the effective potential analysis.\\

\end{abstract}

\maketitle

\section{Introduction}

Symmetry breaking mechanisms play a central role in the 
unification of gauge forces. The gauge symmetry of a unified theory must
be partially and spontaneously broken at low energies to describe the nature.
In the standard model (SM) of electroweak interactions, the Higgs scalar field
induces the symmetry breaking.  There are other mechanisms of gauge symmetry breaking.
In technicolor theories, strong technicolor gauge forces induce
condensates of fermion-antifermion
pairs in the same manner as in QCD, which in turn breaks the gauge symmetry.

In addition to these mechanism there is another intriguing scenario of
dynamical gauge symmetry breaking by adding compact extra dimensions.  
Let us start with a gauge theory defined in space-time with extra
spatial dimensions.
In brief, when the extra dimensional space is not simply-connected, the
non-vanishing phases $\theta_H$ of the Wilson line integral of gauge
fields along a non-contractible loop in these extra dimensions 
can break the symmetry of the vacuum at one loop level~\cite{YH1,Davies1,Davies2,YH2}.
These phases $\theta_H$ are the 
Aharonov-Bohm (AB) phases in the extra dimensional space, which, despite its
vanishing field strengths, affect physics leading to gauge symmetry breaking.
This is the so called the Hosotani mechanism.
The values of $\theta_H$ are determined dynamically.

These AB phases $\theta_H$ play the role of the Higgs field in the SM. 
Indeed, the 4D Higgs boson 
appears as 4D fluctuations of $\theta_H$, or the zero-mode of the extra-dimensional component
of gauge potentials $A_M$.  This leads to a scenario of gauge-Higgs 
unification~\cite{Hatanaka1998}.
The 4D Higgs boson is a part of gauge fields in higher dimensions.  
Its mass is generated radiatively at the quantum level and turns out to
be finite, free from divergences.  
Recently, the Hosotani mechanism has been applied to the electroweak interactions 
\cite{Nomura1,Csaki2,ACP,Csaki3,MSW, HS2,Lim2007_1,Lim2007_2,Lim2007_3,HOOS,HKT,unitarity,HNU}.
The gauge-Higgs unification scenario gives several predictions to be
tested at LHC/ILC 
\cite{HTU2,Adachi2012,Funatsu,Maru2013a,Maru2013b}.
It should be pointed out, however, that the Hosotani mechanism as a
mechanism of gauge symmetry
breaking has been so far established only in perturbation theory. 
It is based on the evaluation of
the effective potential $V_\eff (\theta_H)$ at the one-loop level.  
It is still not clear whether the mechanism operates at the non-perturbative level. 
This paper is a first investigation on the non-perturbative realization
of the Hosotani mechanism using lattice calculations.

Lattice QCD has been accepted as a successful non-perturbative scheme
describing strong interactions.
It provides a reliable method for investigating strong gauge interaction
dynamics from first principles, establishing the color confinement and
the chiral symmetry breaking in QCD, for example.
In a recent work, Cossu and D'Elia~\cite{Cossu} (inspired by the
semi-classical study~\cite{Unsal}) considered the case of SU(3) lattice
gauge theory with fermions in the adjoint representation.   
They showed in a four dimensional lattice with one compact dimension
({\it i.e.} much smaller than the
others in a finite volume), that the presence of periodic fermions in the adjoint representation
leads to new phases in the space of the gauge coupling and 
fermion mass parameters. They found four different phases
by measuring Polyakov loops average values.
Besides the usual confined and deconfined phases, found using anti-periodic
boundary condition (finite temperature),
they identified two new phases, called the split and reconfined phases.  

In this paper, we would like to point out the connection between the
phases identified by Cossu and D'Elia and the Hosotani mechanism~\cite{HosotaniGUT2012}
by showing that the deconfined phase, the split phase, and 
the reconfined phase introduced in ref.~\cite{Cossu} correspond to the SU(3) phase, 
the SU(2)$ \times $U(1) phase, and the U(1)$\times $U(1) phase in the
language of the Hosotani mechanism.  
Results of measurements of Polyakov loops in numerical simulations on 
the lattice,
with fermions in the adjoint and fundamental representation,  are
interpreted in terms of the effective potential of the AB phases.  
A clear connection to the location of minima of the effective potential
$V_\eff (\theta_H)$ can be identified by the density plots of eigenvalue 
phases of Polyakov loops.
We refine the connection by generalizing the boundary conditions for 
fermions in the fundamental representation, 
which corresponds to introducing an imaginary chemical potential.
The analysis of the present paper paves the way for establishing the Hosotani mechanism on the lattice.
Once established, it can be applied to the electroweak unification and
the grand unification of electroweak and strong
interactions to achieve a paradigm of gauge unification without recourse to elementary scalar fields.
Compared to the SM employing the Higgs mechanism, the scenario with 
the Hosotani mechanism has the advantage that interactions of the Higgs boson,
which is a part of  gauge fields,  
are dictated by the gauge principle and that its finite mass is generated radiatively, free from 
divergence,  thus solving the gauge hierarchy problem.
Phase structure and the Hosotani mechanism in SU(3) gauge theory, including chiral symmetry 
breaking, have been discussed recently~\cite{Misumi1}.

The definition of a lattice gauge theory in more than 4 dimensions is afflicted with
a subtle problem of finding the corresponding continuum theory.  There have been many investigations in this
direction~\cite{IK1,IK2,Forcrand,Debbio,IKY1,IKY2,Knechtli:2011gq}.
Lattice gauge theory on orbifolds has been under intensive study for applications to 
electroweak interactions in mind.
In this paper we take advantage of the fact that the Hosotani mechanism works
in any dimensions such as $R^n \times S^1$, so we focus on the
four-dimensional case ($R^3 \times S^1$) in which the lattice gauge theory has been firmly established.

The paper is organized as follows. 
In Section~\ref{sec:ABphases}, after introducing the AB phases
$\theta_H$ in our setup, we explain the relationship between $\theta_H$
and the Polyakov loops. We also briefly describe the lattice formulation of the theory.
Section~\ref{sec:sym-breaking} contains discussion on the gauge symmetry
breaking by the Hosotani mechanism and classification of $\theta_H$'s
according to the pattern of the symmetry breaking.
A perturbative prediction of $\theta_H$ is given in
Sec.~\ref{sec:PertResults} from the analysis of the effective
potential $V_\eff (\theta_H)$ at the one loop level. 
The results to be compared with the lattice calculation are derived here in the 
specific case of $R^3 \times S^1$ and in the presence of massless and massive fermions 
in the adjoint and fundamental representations with general boundary conditions.
Our lattice simulations are presented in Sec.~\ref{sec:LatticeResults}.
In Secs.~\ref{AdjointSection} and \ref{FundamentalSection}, the simulations with adjoint fermions and
fundamental fermions are discussed.
We obtain the phase structures for both cases and discuss their connection 
to the perturbative prediction.
Section~\ref{Discussion} is devoted to discussion.

\section{Aharonov-Bohm phases in SU($N$) theory \label{sec:ABphases}}

Let us begin the analysis by presenting the relation between the Aharonov-Bohm phases in the extra-dimensions
and a relevant quantity measured in lattice gauge theory calculations: the Polyakov loop. We define these basic
observables and show how we can obtain information on the Hosotani mechanism  from lattice measurements.

\subsection{Continuum gauge theory on $R^{d-1} \times S^1$}
\label{sec:ABphasesInContinuum}

As the simplest realization of the Hosotani mechanism, we consider SU(3) gauge theory coupled with 
fermions in the fundamental representation ($\psi_\fund$)
and/or in the adjoint representation ($\psi_\ad$) in $d$-dimensional flat space-time with one spatial 
dimension compactified on $S^1$ \cite{Hatanaka1999,Hosotani2005}.   
The circle $S^1$ has coordinate $y$ with a radius $R$ so that
$y \sim y + 2\pi R$.  In terms of these quantities the Lagrangian density is given by:
\beeq
\cL = - \frac{1}{2} \Tr F_{MN} F^{MN} + \bar \psi_\fund (\cD_\fund - m_\fund) \psi_\fund
+ \Tr \bar \psi_\ad (\cD_\ad - m_\ad) \psi_\ad 
\label{Lagrangian1}
\eneq
where $\cD_\fund$ and $\cD_\ad$ denote covariant Dirac operators.
The gauge potentials $A_M = (A_\mu, A_y)$ ($\mu = 1, \cdots, d-1$) and
fermions $\psi_\fund, \psi_\ad$ satisfy the following boundary conditions:
\begin{align}
\begin{split}
A_M ( x, y + 2\pi R) &= V A_M ( x, y)  V^{-1},\\
\psi_\fund  ( x, y + 2\pi R) &= e^{i \alpha_\fund} \, V \, \psi_\fund  ( x, y), \\
\psi_\ad ( x, y + 2\pi R)  &= e^{i \alpha_\ad} \, V \, \psi_\ad  ( x, y)  V^{-1}, 
\end{split}
\label{eq:BoundCond_1}
\end{align}
where $V \in SU(3)$.  
With these boundary conditions the Lagrangian density
is single-valued on $S^1$, namely $\cL (x, y + 2\pi R) = \cL (x, y )$, so that physics is 
well-defined on the manifold $R^{d-1} \times S^1$. It has been proven (see \cite{YH2}) 
that physics is independent of $V$ at the quantum level. We adopt $V=I$ 
in the most part of the arguments below. 
Setting $\alpha_\fund=\alpha_\ad=0$ corresponds to periodic fermions, whereas 
$\alpha_\fund=\alpha_\ad=\pi$ to anti-periodic fermions.
In the Matsubara formalism of finite temperature field theory the imaginary time corresponds
to $S^1$ with boundary conditions $\alpha_\fund=\alpha_\ad=\pi$.
When $S^1$ represents a spatial dimension, $\alpha_\fund$ and
$\alpha_\ad$ can take arbitrary values and become important 
in the calculation of the effective potential.

There is a residual gauge invariance, given the boundary conditions (\ref{eq:BoundCond_1}).
Under a gauge transformation $A_M' = \Omega A_M \Omega^{-1} + (i/g) \Omega \dd_M \Omega^{-1}$,
$\psi_\fund' = \Omega \psi_\fund$, and $\psi_\ad' = \Omega \psi_\ad \Omega^{-1}$, 
the boundary condition (\ref{eq:BoundCond_1}) with $V=I$ is maintained, provided
\beeq
\Omega(x, y+ 2\pi R) = \Omega(x, y) ~.
\label{BC2}
\eneq
The zero mode, or a constant configuration, of $A_y$ satisfies (\ref{eq:BoundCond_1}), but it cannot be
gauged away in general.  To see this, consider a Wilson line integral along $S^1$
\beeq
W (x)= P \exp \left( i g \int_0^{2\pi R} dy \, A_y(x,y) \right) ~, 
\label{eq:Wilson1}
\eneq
which covariantly transforms under residual gauge transformations as
\beeq
W'(x) = \Omega(x, 0) W(x) \Omega(x, 2\pi R)^{-1} = \Omega(x, 0) W(x) \Omega(x, 0)^{-1} ~.
\label{Wilson2}
\eneq
Consequently the eigenvalues of $W$ are gauge invariant.  They are denoted by
\beeq
\big\{ e^{i \theta_1},  e^{i \theta_2},   e^{i \theta_3} \big\}  \quad
\hbox{where ~}
\sum_{j=1}^3 \theta_j = 0 ~~ (\bmod ~ 2\pi)~.
\label{ABphase1}
\eneq
Constant configurations of $A_y \not= 0$ with $A_\mu=0$ yield vanishing
field strengths $\langle F_{MN}\rangle=0$, but in general give 
$W \neq I$, or nontrivial $\theta_H$.
This class of configurations is not gauge equivalent to $A_M =0$ if the boundary conditions
(\ref{eq:BoundCond_1}) are maintained. 
The $\theta_j $'s are the elements of AB phase $\theta_H$ in the 
extra dimension. These are the dynamical degrees of freedom 
of the gauge fields affecting physical quantities
as in the Aharonov-Bohm effect in quantum mechanics.
The constant modes of $A_y$ factorize as 
\begin{equation}
 A_y^{\rm const} 
= \frac{1}{2\pi g R }\cdot K \begin{pmatrix} \theta_1 & &\\ & \theta_2 & \\ & & \theta_3 \end{pmatrix} K^{-1}
~~,~~ K \in SU(3) ~.
\label{AB2}
\end{equation}
Since the gauge transformation
\beqn
&&\hskip -1.cm
\Omega(y) = K \, 
\begin{pmatrix} e^{i n_1y/R} &&\cr & e^{i n_2y/R} &\cr && e^{i n_3y/R} \end{pmatrix}  \, K^{-1} ~, \cr
\noalign{\kern 10pt}
&&\hskip 0.cm
n_j : \hbox{an integer} ~~,~~ \sum_j n_j = 0 ~,
\label{AB3}
\eeqn
satisfying eq.~(\ref{BC2}) transforms $\theta_j$ to $\theta_j' = \theta_j + 2\pi n_j$, 
the periodicity of $\theta_j$ with the period $2\pi$ is encoded by the gauge invariance.

We write $W_3$ and $W_8$ to denote the Wilson line (in eq.~(\ref{eq:Wilson1})) for 
$A_y^{\rm const}$ and its counterpart in the adjoint representation, respectively.
Accordingly, by taking trace for relevant indices, the spatial average
of the Polyakov loops $P_3$ and $P_8$ are defined as 
\begin{align}
\label{eq:PolyakovFund}
P_3 &= \frac{1}{3} \Tr W_3 
= \frac{1}{3} \Big\{ e^{i \theta_1} + e^{i \theta_2} + e^{i \theta_3} \Big\},\\
\label{eq:PolyakovAdj}
P_8 &= \frac{1}{8} \Tr W_8 
= \frac{1}{4}  \Big\{ 1 + \cos (\theta_1 - \theta_2) +  \cos (\theta_2 - \theta_3) 
+ \cos (\theta_3 - \theta_1) \Big\}.
\end{align}
As discussed in the next section, it is possible to read off the information on the
non-perturbative behavior of $\theta_H$ from the Polyakov loop  calculated
on the lattice.

\subsection{Gauge theory on the $N_x^3 \times N_y$ lattice}
\label{LatticeGeneral}

We carry out a non-perturbative study of SU(3) gauge theory coupled with 
fermions 
in ($R^3\times S^1$) by numerical simulations of a lattice gauge theory.
On the $N_x^3\times N_y$ Euclidean lattice with an isotropic lattice
spacing $a$, we compactify the extra dimension
of size $N_y$ by imposing the appropriate boundary conditions as in
(\ref{eq:BoundCond_1}), where $R=N_ya/2\pi$.
However in a lattice simulation each of the space-like directions
 $N_x$ is always finite and periodic. 
In order for the space-time boundary conditions not to cause any finite size artifact,
$N_x$ is set to be sufficiently larger than $N_y$. 
A ratio of $N_x/N_y=4$ is used throughout this article.

We now describe some basic facts on the formulation of gauge theories on the lattice
for the sake of the reader not familiar with the subject.
A building block of the action on the lattice is the link variable 
$U_{(x,y),M}$, namely the parallel transporter of the gauge field 
connecting $(x,y)$ and 
$(x,y)+a\hat{M}$, where $\hat{M}$ denotes the unit vector in the 
$M$-direction. 
Using the plaquette {\it i.e.} the smallest closed path in the $MN$ plane 
\begin{eqnarray}
 P_{(x,y),MN}= U_{(x,y)M}U_{(x,y)+\hat{M},N}
  U_{(x,y)+\hat{N},M}^\dagger U_{(x,y),N}^\dagger,
\end{eqnarray}
the simplest gauge action (Wilson gauge action) is written as 
\begin{eqnarray}
 S_g[U] = \beta\sum_{x,y,M<N}\left(1-\frac{1}{3}\re\Tr P_{(x,y),MN}\right),
\label{eq:LatticeGauge}
\end{eqnarray}
where $\beta$ and the bare coupling constant $g_0$ are related by
$\beta=6/g_0^2$. The parameter $\beta$ determines the lattice spacing
through the $\beta$-function~\cite{QCDonTheLattice}. 
Equation (\ref{eq:LatticeGauge}) reduces to the continuum action as $a \rightarrow 0$
 {\it i.e.} $\beta \rightarrow +\infty$ by asymptotic freedom.
The Dirac operator $D_{\rm R}(U;m_{\rm R}a)$ for representation 
${\rm R}\in \{\fund,\ad\}$, is given as a function of $U$ and the
bare mass $m_{\rm R}a$.
The lattice fermion action with $2 N_{\rm R}$ degenerate flavors is
\begin{eqnarray}
 S_f[U] = N_{\rm R} \ln \det
  \left[D_{\rm R}(U;m_{\rm R}a)^\dagger D_{\rm R}(U;m_{\rm R}a)\right]
\label{eq:FermionAction}
\end{eqnarray}
after integrating out the fermion fields and exponentiating 
the resulting determinant.
Using the lattice action $S = S_g+S_f$,
we apply the Hybrid Monte Carlo (HMC) algorithm~\cite{Duane1987,Gottlieb1987}
 for the numerical simulation to generate an ensemble of statistically
 independent gauge configurations distributed with the Boltzmann weight $e^{-S}$.

We study the phase diagram with adjoint fermions in the plane 
$(\beta, m_{\ad}a)$ with periodic boundary conditions in the $y$-direction, {\it i.e.} $\alpha_\ad=0$.
In  the fundamental fermions case we generated configurations 
with several values of the couple $(\beta,\alpha_{\fund})$,
fixing the bare mass to $m_\fund a=0.10$, where $\alpha_{\rm fd}$ 
is introduced through the boundary conditions
\begin{eqnarray}
 e^{-i\alpha_\fund}U_{(x,y+N_y),4}=U_{(x,y),4}.\label{alpha_lattice}
\end{eqnarray}

Among the several quantities that can be measured on the lattice,
we are mainly interested in the Polyakov loop in both representations. 
The discretized versions of (\ref{eq:PolyakovFund}) and (\ref{eq:PolyakovAdj}) are 
given by
\begin{align}
 P_3 &= \frac{1}{3N_x^3}\sum_{x}\Tr W_3^{\rm latt}(x) 
 =\frac{1}{3N_x^3}\sum_{x}\Tr\prod_{y=1}^{N_y}
 U_{(x,y),4},\label{eq:LatticePLFund}\\ 
 P_8 &= \frac{1}{8N_x^3}\sum_{x}\Tr W_8^{\rm latt}(x) =\frac{1}{8N_x^3}\sum_{x}\Tr\prod_{y=1}^{N_y}
  U^{(8)}_{(x,y),4}\label{eq:LatticePLAdj}
\end{align}
where $U_{(x,y),M}^{(8)}$ is the link variable in the adjoint (real)
representation 
\begin{eqnarray}
 U^{(8)\, ab}_{(x,y),M} = (U^{(8)\, ab}_{(x,y),M})^\dagger
  =\frac{1}{2}\Tr[\lambda^a U_{(x,y),M}\lambda^b U^\dagger_{(x,y),M}]
\end{eqnarray}
with $\lambda^a$ the Gell-Mann matrices. Note that 
$P_8$ is a real number while $P_3$ is complex in general.

Generally speaking, there is a potential concern about the connection 
between the lattice theory and the continuum theory.
The continuum theory is achieved in the large $\beta=6/g_0^2$ 
limit. However, to keep physical quantities fixed in reaching the continuum
limit we need larger lattice volumes and smaller bare fermion masses, which is
a computationally demanding task. For this reason as a starting point of our
project, we restrict ourselves 
to the study of the parameter-dependence of the Polyakov loops 
in the fixed lattice volume $N_x^3\times N_y = 16^3\times 4$
with the constant bare mass parameters $m_{\rm ad} a$ and $m_\fund a$
chosen independently from $\beta$. The choice of the parameters is not
intended to keep the physics constant in the $\beta \rightarrow +\infty$ limit.
In other words, in this first investigation we obtain the phase diagram 
in the lattice parameter space and infer the connection to continuum 
theory predictions
without attempting an extrapolation to the continuum limit, left for future 
studies.

\section{Symmetry breaking\label{sec:sym-breaking}}

To see the effect of the AB phases on the spectrum of gauge bosons we expand the fields 
of the SU(3) gauge theory on $R^{d-1} \times S^1$ in Kaluza-Klein modes of the extra-dimension:
\begin{align}
\begin{split}
A_M(x, y) &= \frac{1}{\sqrt{2 \pi R}} \sum_{n=-\infty}^\infty 
 A_M^{(n)}(x) \, e^{i ny/R}, \\
\psi_\fund (x,y) &= \frac{1}{\sqrt{2 \pi R}} \sum_{n=-\infty}^\infty
\psi_\fund^{(n)} (x) \, e^{i (n + \alpha_\fund/2\pi)  y/R}, \\
\psi_\ad (x,y) &= \frac{1}{\sqrt{2 \pi R}} \sum_{n=-\infty}^\infty
\psi_\ad^{(n)} (x) \, e^{i (n + \alpha_\ad/2\pi)  y/R}.
\label{expansion1}
\end{split}
\end{align}
The $A_M^{(0)} (x)$ fields are the zero modes in the $(d-1)$-dimensional space-time.
These lowest modes are massless at the tree level.  Some of them can acquire masses at the quantum
level. The modes $A_y^{(n)}(x)$ ($n\not= 0$) can be gauged away,
but $A_y^{(0)}(x)$ cannot (see discussion in Sec.~\ref{sec:ABphasesInContinuum},
where we have seen that the $A_y^{(0)}$ modes represent
the AB phase $\theta_H$). A different set of the elements 
$(\theta_1,\theta_2,\theta_3)$ leads to a different mass spectrum and different phase in physics.

It has been shown~\cite{YH2} that on $R^{d-1}\times S^1$ one can take $K=I$ in
(\ref{AB2}) without loss of generality.
With this background, each KK mode has the following mass-squared in the $(d-1)$-dimensional 
space-time.
\beqn
A_\mu^{(n)} &:& \left(m_A^{(n)}\right)_{jk}^2 = \frac{1}{R^2} \Big( n + \frac{\theta_j - \theta_k}{2\pi} \Big)^2 ~, \cr
\noalign{\kern 10pt}
\psi_\fund^{(n)} &:& \left(m_\fund^{(n)}\right)_j^2 = \frac{1}{R^2} \Big( n +  \frac{\theta_j + \alpha_\fund}{2\pi} \Big)^2 
   + m_\fund^2 ~, \cr
\noalign{\kern 10pt}
\psi_\ad^{(n)} &:& \left(m_\ad^{(n)}\right)_{jk}^2=\frac{1}{R^2} \Big( n +  \frac{\theta_j  - \theta_k 
     + \alpha_\ad}{2\pi} \Big)^2     + m_\ad^2 ~.
\label{spectrum1}
\eeqn
Note that the diagonal components of $A_\mu^{(0)}$ always remain massless.
$A_y^{(n)}$  has the same mass spectrum as   $A_\mu^{(n)}$ at the tree level.
In particular, massive states of $A_y^{(n)}$ are absorbed as longitudinal components 
of the corresponding massive vector bosons $A_\mu^{(n)}$ in the Stueckelberg field formalism.
Massless modes of $A_y^{(0)}$ remain physical, and acquire finite masses through loop corrections. 
We will come back to this issue in the subsection~\ref{sec:HiggsMass}. 

Although the mass spectra \eqref{spectrum1} seem to be based on a specific gauge, 
the spectra themselves are gauge-invariant.
To see it more concretely,   consider a gauge transformation
which  eliminates the vacuum expectation values (VEVs) of $A_y$ and therefore $\theta_i$; 
\beqn
&&\hskip -1.cm
A'_y = \Omega A_y \Omega^{-1} + (i/g) \Omega \partial_y \Omega^{-1}, \cr
\noalign{\kern 10pt}
&&\hskip -1.cm
\langle A'_y \rangle = 0,
\quad
\Omega(y) =  \exp (- i g y \langle A_y \rangle).
\label{large-gauge-transf}
\end{eqnarray}
In the new gauge the fields are not periodic anymore.  
$A'_M$, $\psi'_\fund$, $\psi'_\ad$ satisfy the boundary conditions (\ref{eq:BoundCond_1})
with
\beeq
V = \Omega(y+2\pi R) \Omega(y)^{-1} = 
\begin{pmatrix} e^{-i \theta_1} \cr & e^{-i \theta_2} \cr  && e^{-i \theta_3} \end{pmatrix} .
\label{newBC}
\eneq
Due to the nontrivial boundary condition $V$, 
the Kaluza-Klein masses from the $y$-direction momenta change, which compensates 
the eliminated mass terms coming from the AB phases. 
The resultant mass spectra \eqref{spectrum1} remain intact under the gauge 
transformation \eqref{large-gauge-transf}.
In general, under any gauge transformation the change of the VEV of $A_y$ is compensated 
by the change of $y$-direction momenta so that the mass spectra remain invariant.
The statement is valid at the quantum level as well, as explained in Section \ref{sec:PertResults}.

From the gauge boson mass of the zero-mode $(m_A^{(0)})^2$,
we can infer the remaining gauge symmetry realization after the compactification.
Because the mass is given by the difference $\theta_j-\theta_k$, it is
classically expected that the mass spectrum becomes SU(3) asymmetric
unless $\theta_1=\theta_2=\theta_3\ ({\rm mod}\ 2\pi ) $.
However, as a dynamical degree of freedom, $\theta_H$ has quantum fluctuations. 
In the confined phase, these fluctuations are large enough for the SU(3)
symmetry to remain intact.
For a moderate gauge coupling and sufficiently small $R$, $\theta_H$ may take 
nontrivial values to break SU(3) symmetry depending on the fermion content.
To determine which value of $\theta_H$ is realized at the quantum level, 
it is convenient to evaluate the effective potential $V_\eff (\theta_H)$,
whose global minimum is given by the VEVs of $\theta_H$.
In Sec.~\ref{sec:PertResults}, we present our study of $V_\eff (\theta_H)$ at the
one-loop level and demonstrate that VEVs of $\theta_H$ are 
located at certain values for given fermion content.
This picture is also supported by lattice simulations presented in 
Sec.~\ref{sec:LatticeResults}.

In the rest of this section, we discuss configurations of
$\theta_H$ which are relevant in the study of $V_\eff (\theta_H)$.
Besides the configuration in the confined phase, there are three classes
as follows.
Note that there is no intrinsic way to distinguish $\theta_1,\theta_2$
and $\theta_3$. All permutations of them within each configuration are equivalent.\\

\noindent
 $\underline{{A: \rm SU(3)} \hbox{ symmetric configurations}}$

As was discussed earlier on eq.~(\ref{spectrum1}), $\theta_1=\theta_2=\theta_3$
leads to SU(3) symmetry of the $(d-1)$-dimensional space-time. We label the
three possibilities as $A_{1,2,3}$, whose properties are
\beqn
A_1 &:& \theta_H = (0,0,0),\hspace{2cm} P_3 = 1,\
\ \ \ \ \ \ P_8 = 1, \nonumber\\
A_2 &:& \theta_H = (\twothird \pi,\twothird \pi,\twothird \pi),
\hspace{1.2cm} P_3 = e^{2\pi i /3},\ P_8 = 1, \\
A_3 &:& \theta_H = (-\twothird \pi,-\twothird \pi,-\twothird \pi),
\ P_3 = e^{-2\pi i /3},\ P_8 = 1.\nonumber
\label{pointA}
\eeqn
Lattice simulations show that these kinds of configurations appear in the
deconfined phase. They are realized in a system with
fermions in either adjoint or fundamental representation.\\

\noindent
 $\underline{B: {\rm SU(2)} \times {\rm U(1)} \hbox{ symmetric configurations}}$

When two elements of $\theta_H$ are the same and the third one is different, 
zero-elements of $(m_A^{(0)})^2_{jk}$ form a $2\times 1$ block structure,
which imply that SU(3) symmetry is broken into SU(2)$\times$U(1)
symmetry. This is realized by configurations $B_{1,2,3}$,
\beqn
B_1 &:& \theta_H = (0,\pi, \pi),
\hspace{1.7cm} P_3 = -\onethird,\ \ \ \ \ \ P_8 =  0,\nonumber\\
B_2 &:& \theta_H = (\twothird \pi,-\onethird \pi,-\onethird \pi),\ \ 
P_3= \onethird e^{-\pi i/3},\  P_8 = 0,\\
B_3 &:& \theta_H = (-\twothird \pi, \onethird \pi, \onethird \pi),
\ \ \ \ P_3 = \onethird e^{\pi i/3},\ \ \ P_8 =  0.\nonumber
\label{pointB}
\eeqn

In terms of $P_3$, this configuration seems to be realized in the ``split'' phase 
observed in ref.~\cite{Cossu}, where a system with periodic fermions in
the adjoint representation on the lattice is studied. Further discussion on 
the correspondence between the $B$ phase and the split phase are presented
in Sec.~\ref{sec:Directmeasure}.\\

\noindent
$\underline{C: {\rm U(1)} \times {\rm U(1)} \hbox{ symmetric configurations}}$

If $\theta_1, \theta_2$ and $\theta_3$ are different from each other,
there are two independent massless fields in the diagonal components in $A_\mu^{(0)}$
yielding the ${\rm U(1)} \times {\rm U(1)}$ gauge symmetry.
This situation is realized by 
\beqn
 \theta_H = (0, \twothird \pi, - \twothird \pi),\
 \ P_3  =  0,\ \  P_8 = - \tfrac{1}{8}.
\label{pointC}
\eeqn
\noindent
The appearance of such a configuration is signaled by $P_3=0$ in a weaker 
gauge coupling region than the confined phase X. This is the ``reconfined'' phase
found in ref.~\cite{Cossu}.

\begin{figure}[t]
  \centering
  \includegraphics[width=0.4\textwidth]{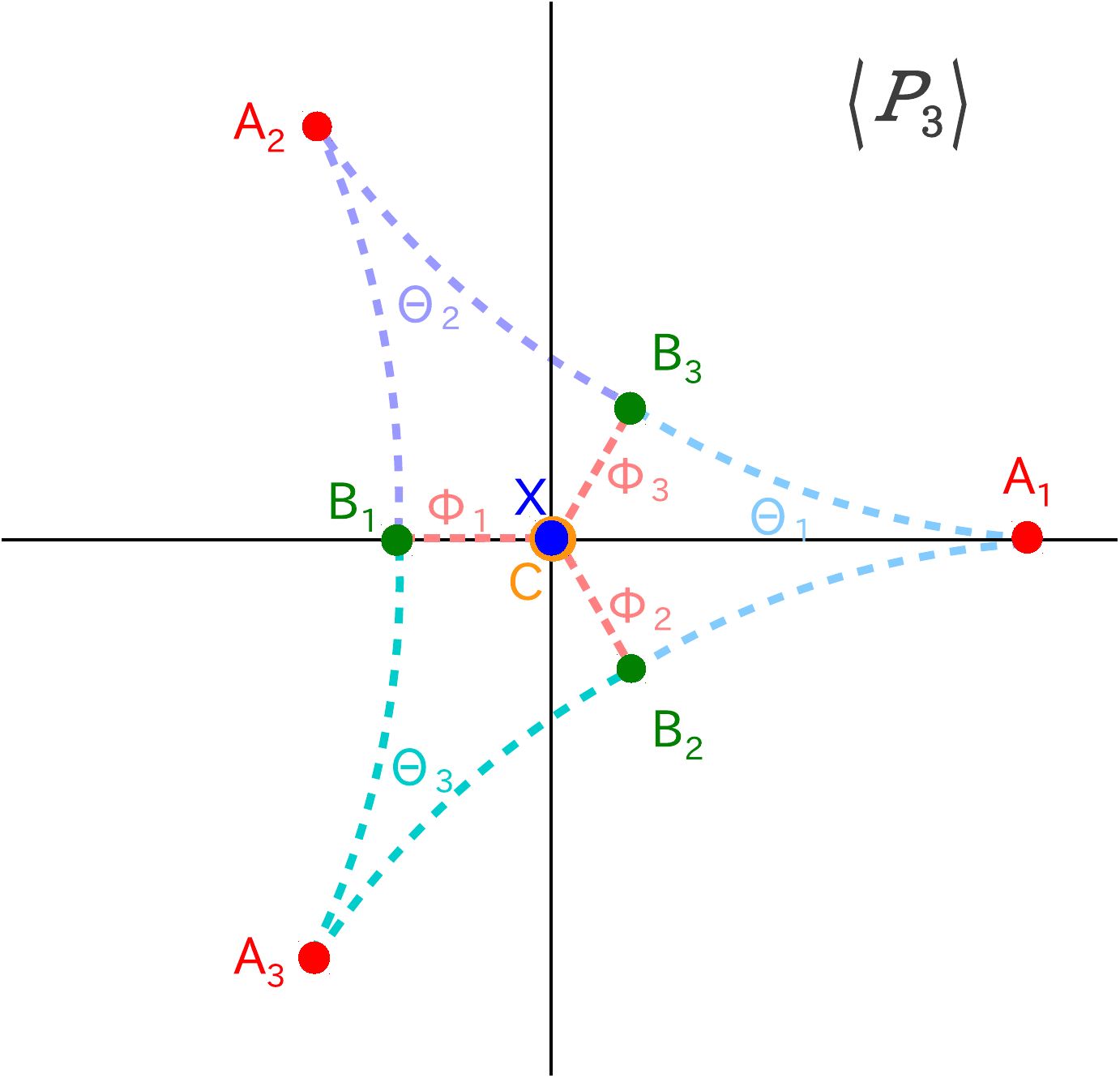}
  \caption{A sketch of the possible values of Polyakov loop in each
    phase. $(A_1,A_2,A_3)$ and $(B_1,B_2,B_3)$ form ${\rm Z}_3$ triplets. 
    $\Theta$ and $\Phi$ are interpolating configurations useful in the
    description of the mixed fermion content case.}
  \label{P3sketch}
\end{figure}

Figure~\ref{P3sketch} shows all relevant values of $P_3$ in the complex plane.
$(A_1,A_2,A_3)$ and $(B_1,B_2,B_3)$ form $Z_3$
triplets.
In lattice simulations one measures both the VEVs and eigenvalue distributions of 
$P_3$ in eq.~(\ref{eq:LatticePLFund}) and $P_8$ in
eq.~(\ref{eq:LatticePLAdj}).  The absolute value of $ P_3 $ is strongly affected by
quantum fluctuations of $\theta_H$ and is reduced at strong gauge couplings.
The phase of $ P_3$, on the other hand, is less affected by quantum fluctuations in the weak 
coupling regime so that transitions from one phase to another can be
seen as changes in the 
phase of $ P_3 $ and density plots of eigenvalue phases of $P_3$. The former has been found in ref.~\cite{Cossu}.
The classifications of $A$, $B$ and $C$
 are summarized in Table~\ref{table-phase}, where
we also include the confined phase, denoted by $X$, in which $\theta_H$ fluctuate and
take all possible values, with equal probability.

The $Z_3$ center symmetry is yet another global symmetry. 
If the action has this symmetry as in the pure gauge theory or
only with adjoint fermions, then its spontaneous breaking is possible.
The magnitude of $P_3$ is the order parameter in this case.
This symmetry is broken in the $A$-phase and the $B$-phase while 
it is unbroken in the $C$-phase.
The phases can be classified by their global SU(3) and $Z_3$ symmetries
as follows,
$X$: (SU(3) symmetric, $Z_3$ symmetric), $A$: (SU(3) symmetric, $Z_3$ broken), 
 $B$: (SU(3) broken, $Z_3$ broken) and $C$: (SU(3) broken, $Z_3$ symmetric).

\begin{table}[bth]
\begin{center}
\renewcommand{\arraystretch}{1.4}
\begin{tabular}{|c|c|c|c|c|}
\hline
 & \myarray{\theta_H=(\theta_1, \theta_2, \theta_3)}{\rm and ~permutations} 
 & $P_3$ & $P_8$ & \myarray{\rm Global\ Symmetry,}{\rm Phase}\\
\hline
$X$ & \myarray{\rm Large ~quantum}{\rm fluctuations} & $0$ &
	     $-\frac{1}{8}$ & \myarray{\rm SU(3),}{\rm confined} \\
\hline
\myarray{A_1}{A_{2,3}} & \myarray{(0,0,0)}{(\pm \twothird \pi, \pm \twothird \pi, \pm \twothird \pi)} 
& \myarray{1}{e^{\pm 2\pi i /3}} & 1 & \myarray{\rm SU(3),}{{\rm deconfined}} \\
\hline
\myarray{B_1}{B_{2,3}} & \myarray{(0,\pi, \pi)}{(\pm \twothird \pi, \mp \onethird \pi, \mp \onethird \pi)} 
& \myarray{- \onethird}{\onethird e^{\mp \pi i/3}} & 0 & \myarray{\rm SU(2) \times U(1),}{{\rm split}} \\
\hline
$C$  & $(0,\twothird \pi , - \twothird \pi)$ & 0
& $- \frac{1}{8}$ & \myarray{\rm U(1)\times U(1),}{{\rm reconfined}} \\
\hline
\end{tabular}
\caption{ Classification of the location of the global minima of 
$V_\eff (\theta_H)$.
In the last column the names of the corresponding phases termed in
 ref.~\cite{Cossu} are also listed for $X, A, B$ and $C$.\label{table-phase}}
\end{center}
\end{table}

\section{Perturbative results\label{sec:PertResults}}

In this section, we present the analysis on the effective potential
$V_{\rm eff}(\theta_H)$ as a function of the AB phase $\theta_H$.
The location of its global minimum defines the VEVs of $\theta_H\ ({\rm mod}\ 2\pi)$.
We first give the formula for $V_\eff (\theta_H)$ at one-loop level, and then
discuss the relationship between $\theta_H$ and the Polyakov loops $P_3$
and $P_8$.

\subsection{One-loop effective potential} 

The first step is to separate the gauge field $A_y$
into its vacuum expectation value $\langle A_y \rangle$ and the quantum 
fluctuation $A_y^q$.
At one-loop level, the effective potential is obtained by the determinant 
of the logarithm of the quadratic action.

In the background gauge, which is defined by the gauge fixing 
\begin{eqnarray}
 {\cal L}_{\rm gf} &=& -\frac{1}{2\alpha}\Tr F[A]^2,\nonumber\\
F[A] &=& \partial_\mu A^\mu + i g [\la A_\mu \ra , A^\mu ] 
\end{eqnarray}
and the gauge parameter $\alpha=1$ (Feynman-'t Hooft gauge),
the one-loop effective potential in $R^{d-1}$, after a Wick rotation, is given by
\begin{eqnarray}
V_\eff^{\rm g+gh}(\theta_H) &=& \frac{d-2}{2 \cdot ({\rm volume})_{d-1}} \,
 \ln \det[ -D_{\rm g+gh}^2] ~,
\nonumber
\\
\noalign{\kern 10pt}
V_\eff^{\rm R}(\theta_H) &=& - \frac{2^{{\lfloor d/2\rfloor}-1}}{({\rm volume})_{d-1}} \, 
\ln \det[-D_{\rm R}^2] ~, 
\end{eqnarray}
for the contributions coming from gauge and ghost fields and from fermions
in the representation ${\rm R}\in \{\fund,\ad\}$.
$({\rm volume})_{d-1}$ denotes the volume of $R^{d-1}$.
$d-2$ counts the number of  physical degrees of freedom of a gauge boson.
$\lfloor d/2 \rfloor$ gives the largest integer which is equals to or smaller than $d/2$ and 
thus $2^{\lfloor d/2 \rfloor}$ counts the number of  degrees of freedom of a Dirac fermion 
in $d$-dimensional  space-time.

In the background configuration
\begin{eqnarray}
\langle A_y \rangle &=& \frac{1}{2\pi R g}  \, \text{diag} \,
(\theta_1, \theta_2, \theta_3),
\quad
\sum_{i=1}^3 \theta_j = 0 ~,
\label{vevAy}
\end{eqnarray}
the expressions for $-D_{\rm g+gh}^2$, $-D_\fund^2$ and $-D_\ad^2$ are given by 
\begin{eqnarray}
[-D_{\rm g+gh}^2]_{jk} &=& -\partial_\mu \partial^\mu - \left[\partial_y 
+ i\left(\frac{\theta_j - \theta_k}{2\pi R}\right)\right]^2,\nonumber\\
 \left[-D_\fund^2\right]_{j} &=&
-\partial_\mu \partial^\mu
- \left[\partial_y 
+ i\left(\frac{\theta_j + \alpha_\fund}{2\pi R}\right)\right]^2
+ m_\fund^2\,,\nonumber
\\
 \left[-D_\ad^2\right]_{jk} &=&
-\partial_\mu \partial^\mu- \left[\partial_y 
   + i\left(\frac{\theta_j -\theta_k + \alpha_{\rm ad}}{2\pi R}\right)\right]^2 
+ m_{\rm ad}^2\,.
\end{eqnarray}
Thus the one-loop effective potential becomes
\begin{eqnarray}
V_\eff(\theta_H) &=& V_\eff^{\rm g+gh}(\theta_H) + N_\fund V_\eff^{\fund}(\theta_H)
+ N_\ad V_\eff^{\ad}(\theta_H) ~, \label{VeffWithFermions}\\
\noalign{\kern 10pt}
V_\eff^{\rm g+gh}(\theta_H)
&\equiv& \frac{d-2}{2} \sum_{j,k=1}^{3} \sum_{n=-\infty}^{\infty}
\int \frac{d^{d-1}p}{(2\pi)^{d-1}}
\ln [p^2 + (m_A^{(n)})^2_{jk}] ~,   \cr
\noalign{\kern 10pt}
V_\eff^{\fund}(\theta_H)
&\equiv& - 2^{{\lfloor d/2\rfloor}-1} \sum_{j=1}^{3} \sum_{n=-\infty}^{\infty}
\int \frac{d^{d-1}p}{(2\pi)^{d-1}}
\ln [p^2 + (m_\fund^{(n)})^2_j]  ~,  \cr
\noalign{\kern 10pt}
V_\eff^{\rm ad}(\theta_H)
&\equiv& - 2^{{\lfloor d/2\rfloor}-1}  \sum_{j,k=1}^{3} \sum_{n=-\infty}^{\infty}
\int\frac{d^{d-1}p}{(2\pi)^{d-1}}
\ln [p^2 + (m_\ad^{(n)})^2_{jk}] ~,
\label{effV2}
\end{eqnarray}
where $(m_A^{(n)})^2$,  $(m_\fund^{(n)})^2$ and $(m_\ad^{(n)})^2$ are given in
(\ref{spectrum1}),  and $N_\fund$ and $N_\ad$ are the numbers of fermions in 
the fundamental and adjoint representations, respectively.
\begin{figure}[t]
  \centering
  \includegraphics[height=4.5cm]{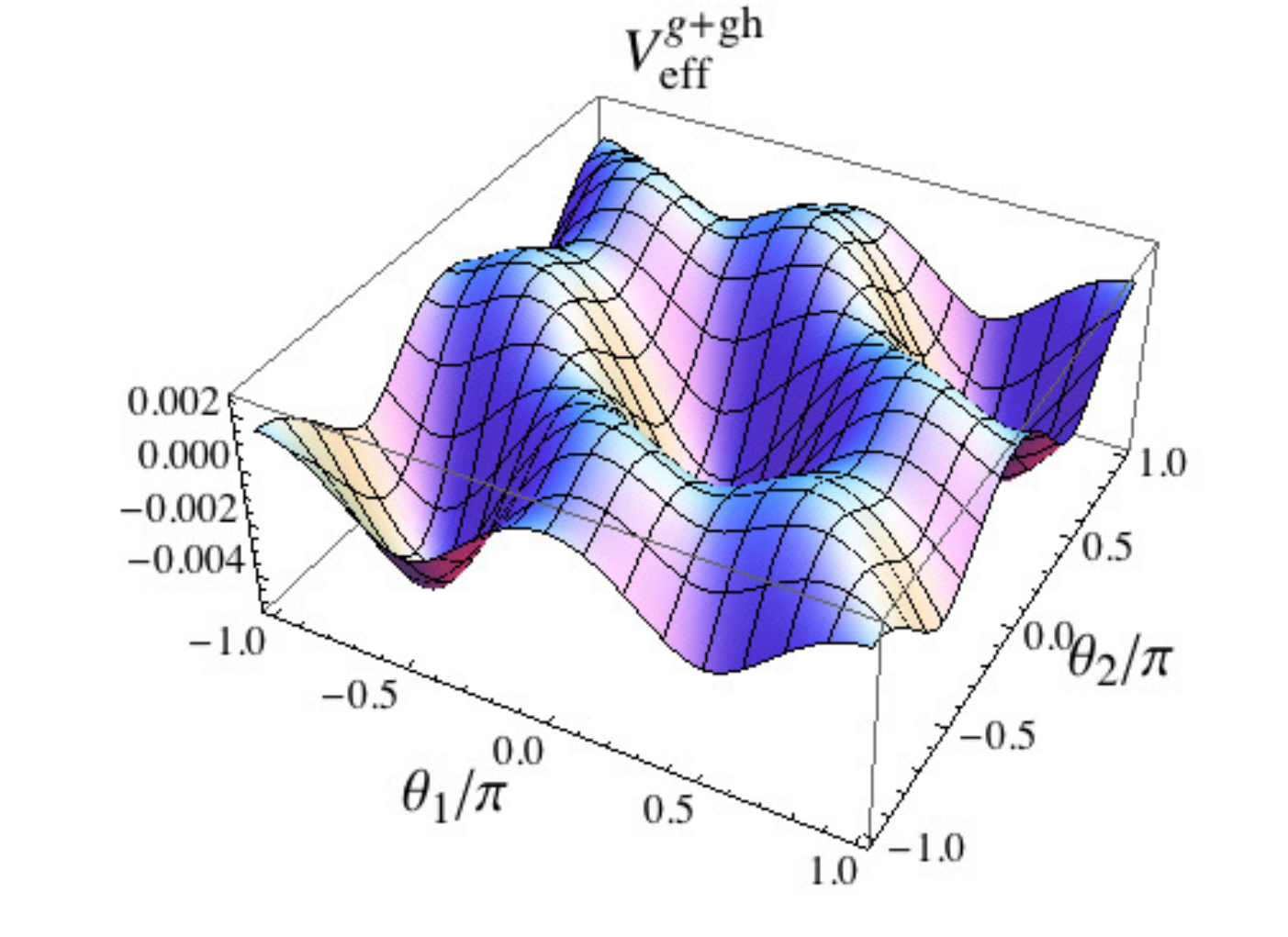}
\hspace{-5mm}
  \includegraphics[height=4.5cm]{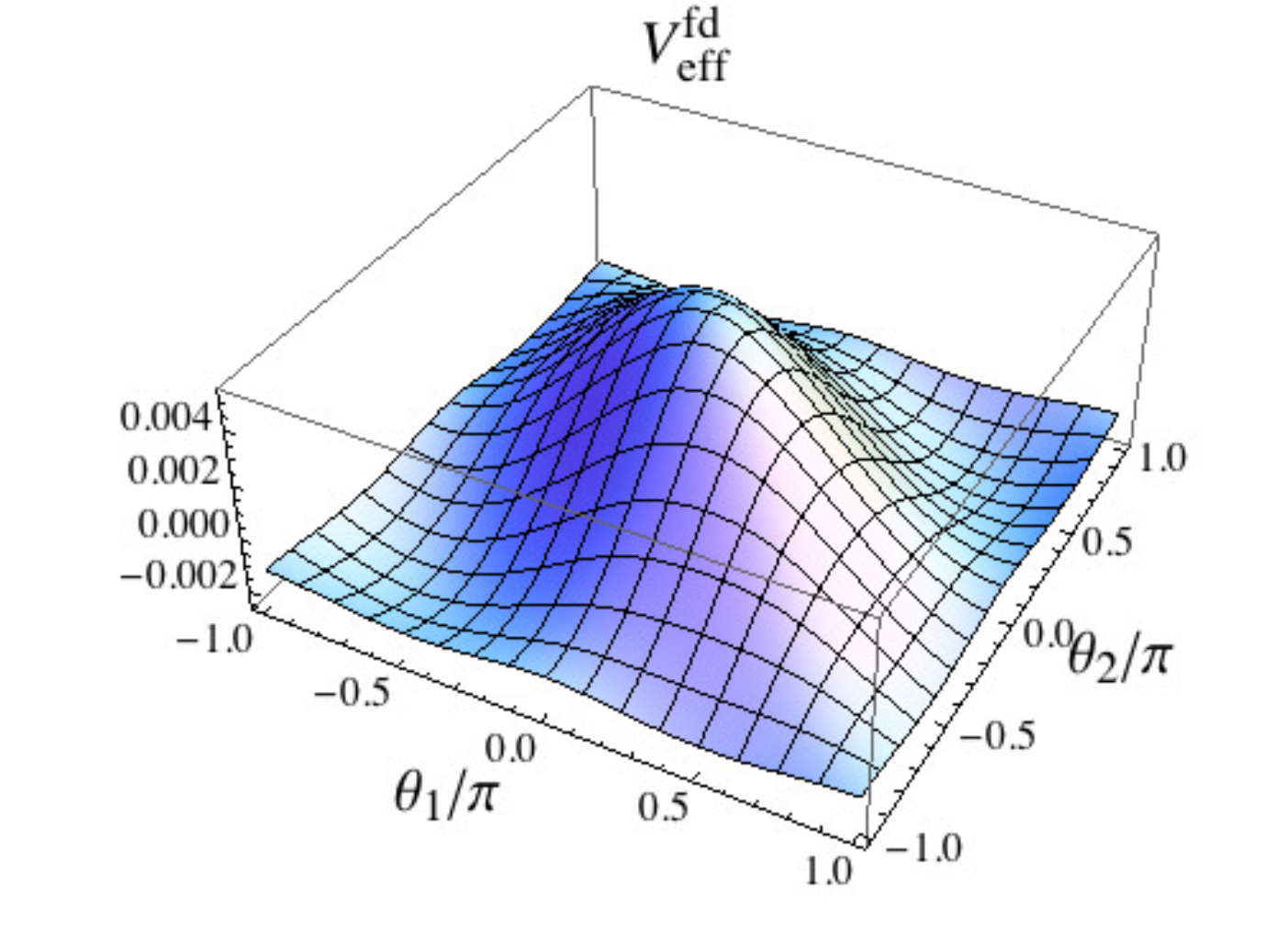}
\hspace{-5mm}
  \includegraphics[height=4.5cm]{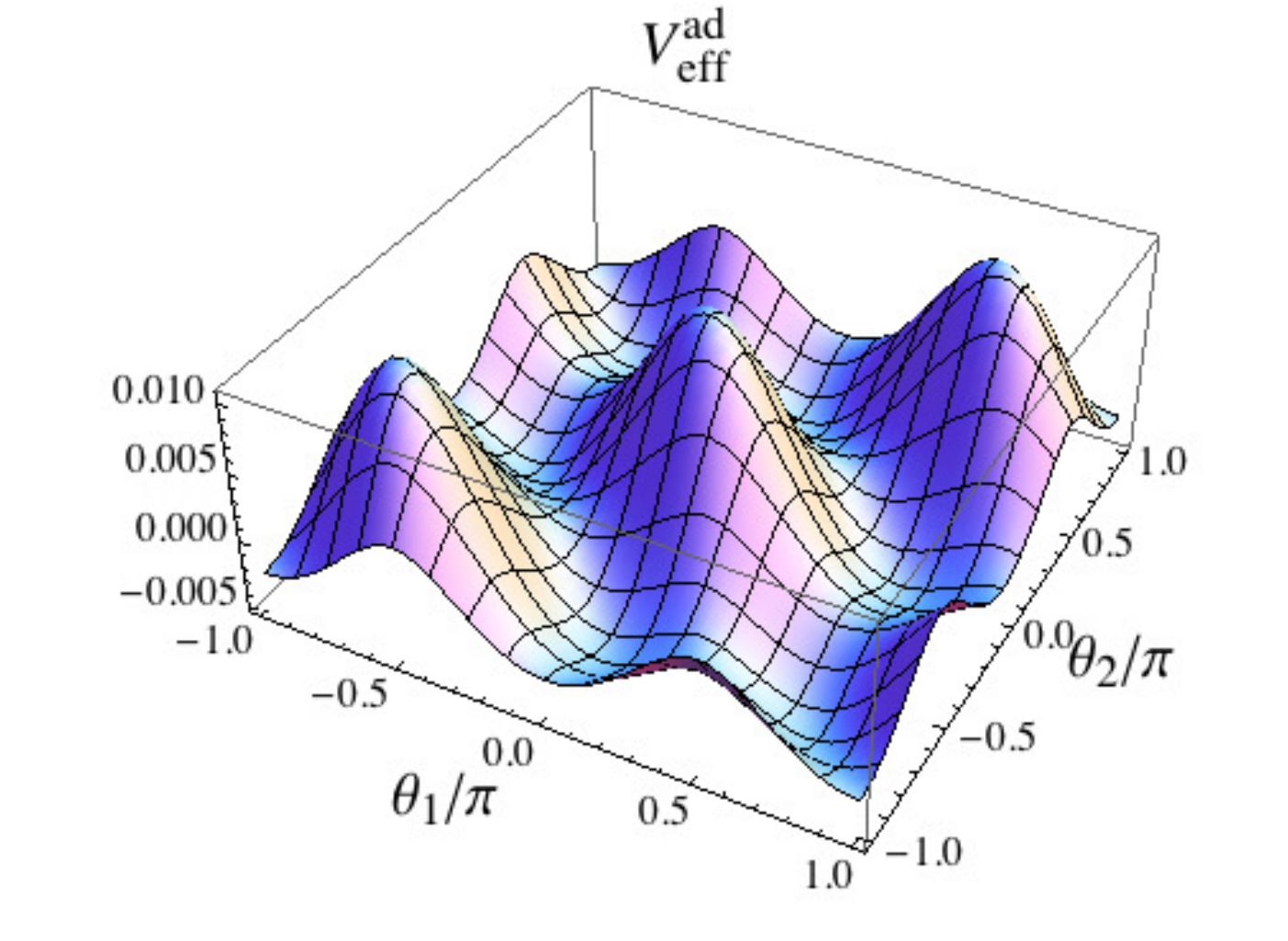}
  \caption{Three contributions to the effective potential $V_{\rm
      eff}^{\rm g+gh}(\theta_H)$, $V_{\rm eff}^\ad(\theta_H)$ and 
    $V_{\rm eff}^\fund(\theta_H)$ are plotted for the case with $d=4$,
    $m_\fund = m_\ad =0$ and $ \alpha_\fund = \alpha_\ad =0$.
    $R$ is normalized to unity.}
  \label{fig:Vg-Vad-Vfd}
\end{figure}
Note that, in eq.~(\ref{effV2}), both the momentum
integrations and the infinite sums of KK modes yield ultraviolet (UV) 
divergences which are independent of $\theta_H$ and $\alpha_{\fund,\ad}$. 
Through the calculations summarized in Appendix~\ref{UsefulFormulae}, 
we obtain expressions of each contribution
\begin{eqnarray}
V_\eff^{\rm g+gh}(\theta_H) &=&
(d-2) \sum_{j,k=1}^3 V(\theta_j - \theta_k, 0),
\nonumber
\\
V_\eff^\fund (\theta_H)
&=& 
-2^{\lfloor d/2 \rfloor} 
\sum_{j=1}^{3} V(\theta_j + \alpha_\fund, m_\fund),
\nonumber
\\
V_\eff^{\rm ad} (\theta_H)
&=&  
-2^{\lfloor d/2 \rfloor} 
\sum_{j,k=1}^{3} V(\theta_j - \theta_k + \alpha_\ad, m_\ad),
\label{effV2b}
\end{eqnarray}
where
\begin{eqnarray}
V(\theta, m) &=& \frac{\Gamma(d/2)}{\pi^{d/2} (2\pi R)^{d-1}} h_d(\theta, m) ~, \cr
\noalign{\kern 10pt}
h_d(\theta, m) &=& 
\sum_{k=1}^{\infty} \frac{1 - \cos k\theta  }{k^d} B_{d/2}(2\pi k m R ) ~, \cr
\noalign{\kern 10pt}
B_{d/2} (x) &\equiv& \frac{x^{d/2} K_{d/2}(x)}{2^{\frac{d}{2}-1} \Gamma(d/2)},
\quad
B_{d/2}(0) = 1,
\label{effV3}
\end{eqnarray}
and $K_{d/2}(x)$ is the modified Bessel function of the second kind.

In Fig.~\ref{fig:Vg-Vad-Vfd}, $V_{\eff}^{\rm g+gh}(\theta_H)$, 
$V_{\eff}^{\rm ad}(\theta_H)$ and 
$V_{\eff}^\fund(\theta_H)$ are plotted for $m_\fund =
m_\ad=0$ and $\alpha_{\rm ad}=\alpha_\fund=0$.
In this case, $V_{\eff}^{\rm g+gh}(\theta_H)$ has degenerate global
minima at $A_{1,2,3}$, reflecting the $Z_3$ symmetry.
On the other hand, 
$V_{\eff}^\fund(\theta_H)$ has degenerate global minima at $A_{2}$ and $A_{3}$
while $V_{\eff}^\ad(\theta_H)$ has global minima at $C$, {\it i.e.}
at the all permutations of $(0,\twothird\pi,-\twothird\pi)$.

In the case with adjoint fermion and $\alpha_{\rm ad}=0$, 
the effective potential can be rewritten, 
neglecting terms independent of $\theta_H$,
in terms of the trace of $W_3^k$ in eq.~(\ref{eq:PolyakovFund}) giving:
\begin{equation}
\sum_{i,j}^3 h_d(\theta_i-\theta_j, m) \propto
-\sum_{k=1}^{\infty} \frac{|\Tr W_3^k|^2}{k^d} B_{d/2}(2\pi k m R ) 
\equiv
-\sum_{k=1}^{\infty} c_k(mR) |\Tr W_3^k|^2
\label{eq:veff_P}
\end{equation}
which gives a direct interpretation as the sum of 
contributions coming from paths wrapping $k$ times in the 
compact direction (see also~\cite{AnberUnsal2013}). The sum is dominated by the first term ($k=1$) for every values of $mR$
(the worst case being $mR=0$ where $c_1/c_2 = 2^d$), and it looks like a self interaction of the Polyakov loop, respecting
the center symmetry. Indeed, terms like this appear in various
dimensionally reduced models with center stabilization terms, and are known as
double-trace deformations (see {\it e.g.}~\cite{Ogilvie:2012is}). In the
limit $mR\rightarrow 0$ ({\it i.e.} $c_1 \rightarrow 1$), the combined
potential of gluons and adjoint fermions has the minimum for $|P_3| = 0$
(a phase with confinement) as we show in the next section. The opposite limit
$mR\rightarrow \infty$ corresponds to the pure gauge case, $c_1
\rightarrow 0$. We use these observations later in the discussion on
 the phase diagram found on the lattice.

\subsection{Vacuum in presence of fermions}\label{perturbativeVeff}

In the presence of fermions, $V_\eff(\theta_H)$ exhibits a rich structure.
Let us consider a model with $N_\fund$($N_\ad$) fundamental (adjoint)
fermions which is described by eq.~(\ref{VeffWithFermions}).
For simplicity, we restrict ourselves to the case where
$m_\fund = m_\ad$ and $\alpha_\fund= \alpha_\ad$. $N_\fund=4$ and $N_\ad=2$ 
are the minimal numbers of flavors for the standard staggered formalism for the 
lattice fermion used in the simulations.
Therefore, in the following we briefly summarize the behavior 
of $V_{\eff}(\theta_H)$ for $(N_\fund, N_{\rm ad}) = (0,2)$ and $(4,0)$ 
for $d=4$, {\it i.e.} $R^3 \times S^1$ compactification.

\subsubsection{Adjoint fermions : $(N_\fund,N_{\rm ad}) = (0,2)$}
\label{VeffAdjointSec}

Let us begin with the case of adjoint fermions under the periodic
boundary condition $\alpha_\ad =0$.
At one-loop level, $V_\eff(\theta_H)$ depends on the mass in the product $m_\ad R$.
The global minimum of $V_\eff(\theta_H)$ changes position according to
the following pattern:
\begin{eqnarray}
A_{1,2,3}\ \ &{\rm for}&\ \ 0.499 \le m_{\rm ad} R ~,
\nonumber \\
B_{1,2,3}\ \ &{\rm for}&\ \ 0.421 \le m_{\rm ad} R \le 0.499 ~,
\nonumber \\
C\ \ &{\rm for}&\ \ 0 \le m_{\rm ad} R \le 0.421 ~.
\label{mRregions}
\end{eqnarray}
The fact that there are coexisting phases $A$ and $B$ ($B$ and $C$)
at the transition point $m_\ad R=0.421(0.499)$ implies that
the transition is of first order. 

\begin{figure}[h]
  \centering
  \includegraphics[width=0.8\textwidth]{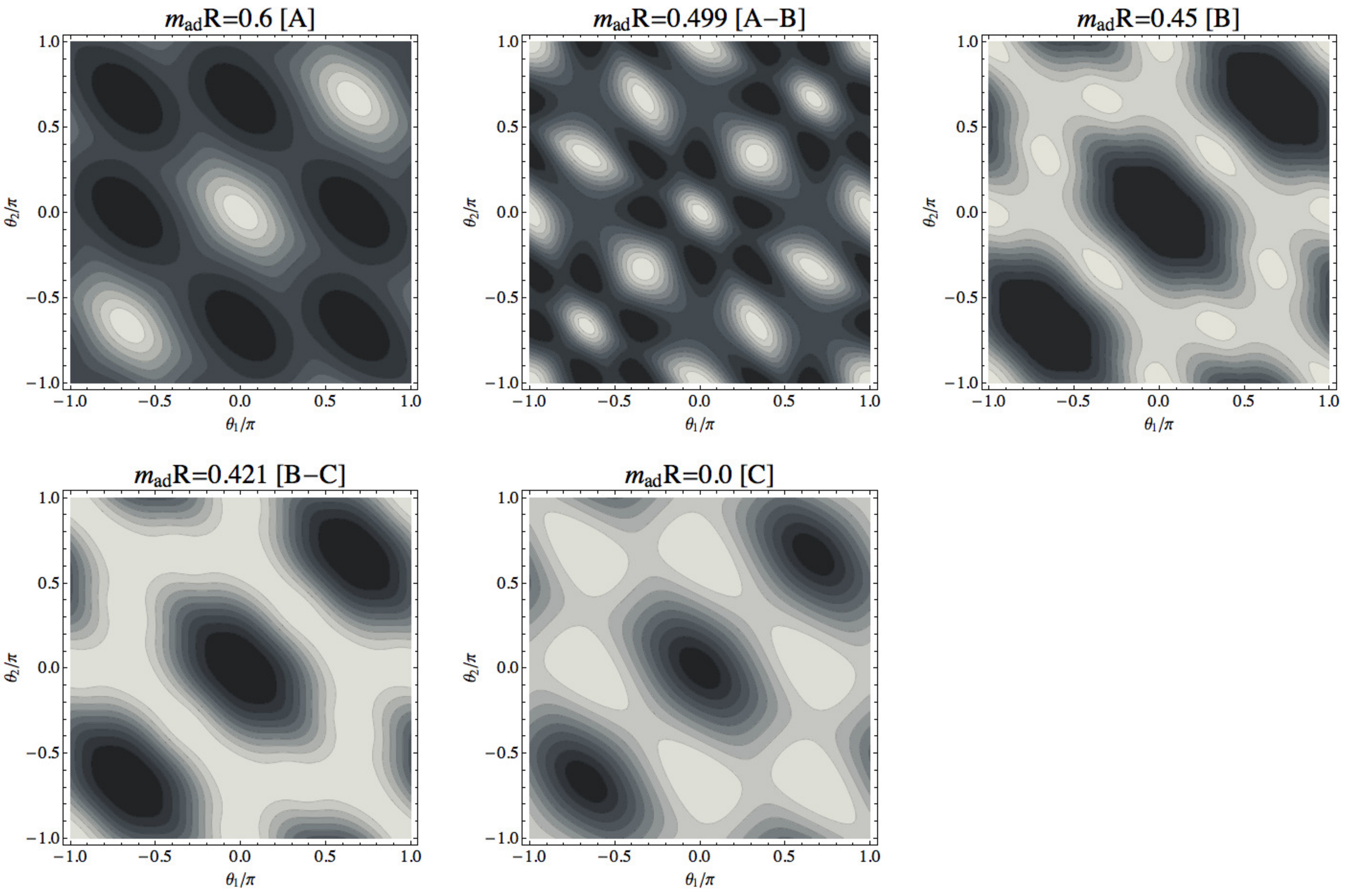}
  \caption{Effective potential for the case of $N_\ad = 2$ adjoint fermion 
    with periodic boundary condition ($\alpha_{\rm ad}=0$) for 
    the values of $m_{\rm ad} R$ in $d=4$. They are corresponding to 
    the $A$ phase, the $A$-$B$ transition point, the $B$ phase, the $B$-$C$
    transition point and the $C$ phase, respectively.
    Lower values of $V_{\eff}$ are indicated by lighter colors.}
  \label{fig:ad-mass}
\end{figure}

In Fig.~\ref{fig:ad-mass}, contour plots of $V_\eff(\theta_H)$ are
displayed in the order $m_\ad R$ is decreasing to cover the phases
$A$, $B$ and $C$ and transitions $A$-$B$ and $B$-$C$
according to eq.~(\ref{mRregions}).
One notes that 
the $A$-$B$ transition is more prominent than the $B$-$C$ transition
because the barrier separating two minima in the potential is much
higher for the former.

We find that the location of the global minimum depends on the value of
$\alpha_\ad$ as well. In particular, for $\alpha_\ad=\pi$,
the effective potential is identical to that at finite temperature $T =(2\pi R)^{-1}$ 
and the SU(3) symmetry remains unbroken.
For massless fermions $m_\ad=0$, the global minima of 
the effective potential are located at
\begin{eqnarray}
A_{1,2,3}\ \ &{\rm for}\ \ &  0.416\pi \le |\alpha_{\rm ad}| \le \pi ~,
\nonumber\\
B_{1,2,3}\ \ &{\rm for}\ \ & 0.319\pi \le |\alpha_{\rm ad}| \le 0.416 \pi ~,
\nonumber\\
C\ \ &{\rm for}\ \ & |\alpha_{\rm ad}| \le 0.319\pi ~.
\end{eqnarray}

\subsubsection{Fundamental fermions : $(N_\fund, N_{\rm ad}) = (4,0)$}\label{VeffFdm}

\begin{figure}[t]
  \centering
  \includegraphics[width=0.8\textwidth]{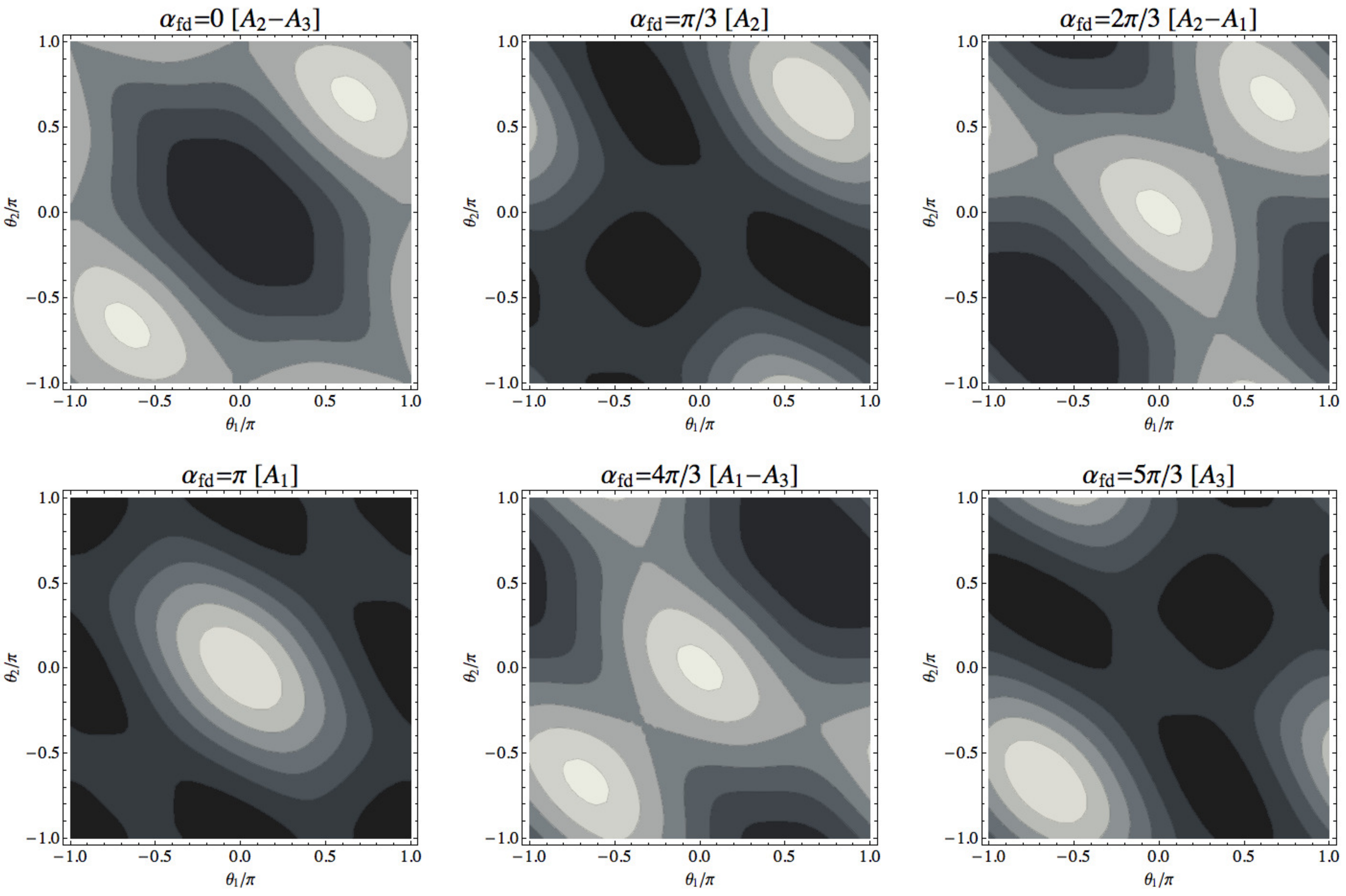}
  \caption{The effective potential with four massless fundamental fermions
    in the $(\theta_1/\pi,\theta_2/\pi)$ plane.
    Boundary condition of the fermions are changed from $\alpha_\fund=0$
    to $\alpha_\fund=5\pi/3$.
    We plot the three phase transitions 
    $A_2$-$A_3$, $A_2$-$A_1$ and $A_1$-$A_3$, and three phases $A_1$, $A_2$ and $A_3$.
    Lower values of $V_{\eff}$ are indicated by lighter colors.}
  \label{fig:fd-bc}
\end{figure}

\begin{figure}[h]
  \centering
  \includegraphics[width=0.45\textwidth]{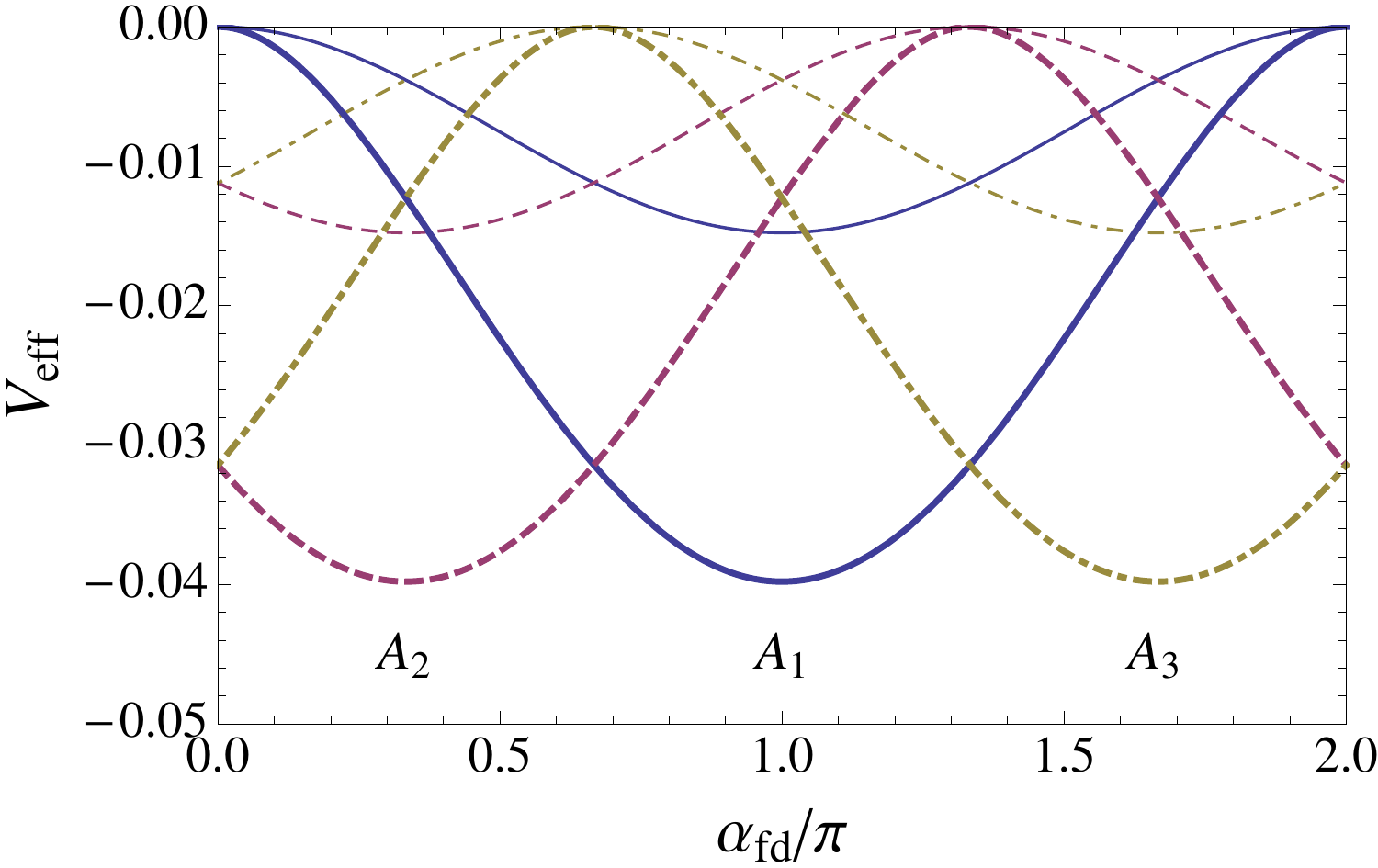}
\caption{$V_{\eff}$ for $(N_\fund,N_{\rm ad}) = (4,0)$ versus $\alpha_\fund$.
  Solid, dashed and dot-dashed lines correspond to values at $A_1$, $A_2$
  and $A_3$, respectively. Thick [thin] lines are for $m_\fund=0$ [$m_\fund R=0.4$].}
\label{fig:f-mass}
\end{figure}

In the presence of the fundamental fermions, the global $Z_3$ symmetry
\begin{eqnarray}
\theta_j \to \theta_j + \frac{2}{3}\pi ~, \quad j=1,2,3.
\label{Z3}
\end{eqnarray}
is broken.
The boundary condition parameter $\alpha_\fund$ plays the role of
selecting one of the $Z_3$ related minima.
We find that the fermion mass $m_\fund$, on the other hand, has only a small
effect on the location of the global minimum
unless $m_\fund R$ is large enough for the effect of fermions to be negligible.
Contour plots of $V_\eff (\theta_H)$ with $d=4$ and $N_\fund =4$ 
are displayed for various values of $\alpha_\fund$ in Fig.~\ref{fig:fd-bc}.
The global minimum is found at 
\begin{eqnarray}
A_2 \ \ &{\rm for}\ \ & 0 \le \alpha_\fund \le \frac{2\pi}{3} ~,
\nonumber \\
A_1\ \ &{\rm for}\ \ & \frac{2\pi}{3} \le \alpha_\fund \le \frac{4\pi}{3} ~,
\nonumber \\
A_3\ \ &{\rm for}\ \ & \frac{4\pi}{3} \le \alpha_\fund \le 2 \pi ~.
\label{VeffResultsFdm}
\end{eqnarray}
Therefore, the phase changes $A_2 \go A_1 \go A_3 \go A_2$ as $\alpha_\fund$
increases from $0$ to $2\pi$.
In Fig.~\ref{fig:f-mass},
we plot the corresponding $V_{\eff}(\theta_H)$ a function
of $\alpha_\fund$ for two different values of $m_\fund R$. 
One observes that the line of global minima has a period of $2\pi/3$
and non-analyticity at $\alpha_\fund=2n\pi/3,\ \ (n=0,1,2,\dots)$,
and the transition is expected to be of first order.
This is known as Roberge-Weiss phase structure~\cite{Roberge1986} 
which we discuss in Sec.~\ref{FundamentalSection}.

\subsection{Scalar (Higgs) masses}\label{sec:HiggsMass}

As discussed in Sec.~III, non-vanishing $\theta_j$ in (\ref{vevAy}) can induce
symmetry breaking where zero modes $\phi(x) \equiv A_y^{(0)}(x)$  play the role of the Higgs field
in $R^{d-1}$.  
With the normalization in eq.\ (\ref{expansion1}),  $- \int dy \, {\rm Tr}\, F_{\mu y}^2 $ yields
the canonically normalized kinetic term for $\phi(x)$.
There are eight real scalars; $\phi(x)  = \sum_{a=1}^8 \phi_a(x) T^a$ 
where $T^a$'s are $SU(3)$ generators.  Some of them are absorbed by $A_\mu^{(0)} (x)$, 
which makes vector fields massive in the broken symmetry sector.  
The rest of the $\phi_a$'s remain physical.  They are massless at the tree level, but acquire
finite masses at the quantum level.  In the application to electroweak interactions, they correspond
to the physical neutral Higgs boson.  

The masses of these physical scalar fields are related to $V_\eff (\theta_1, \theta_2)$ in the 
previous subsections.  $V_\eff (\theta_1, \theta_2)$ is the effective potential for $\phi_3(x)$ 
and $\phi_8(x)$ as well.  The relationship between $\{ \theta_j \}$ and $\{ \phi_a \}$ are given by
\beeq
(\theta_1, \theta_2, \theta_3) =
g \sqrt{\frac{\pi R}{2}} ~ \bigg( \phi_3 + \frac{\phi_8}{\sqrt{3}}  , 
 - \phi_3 + \frac{\phi_8}{\sqrt{3}}  , - \frac{2 }{\sqrt{3}}\phi_8  \bigg) ~.
 \label{Higgs1}
\eneq
The mass eigenstates are determined by diagonalizing the mass matrix
$\partial^2 V_\eff / \partial \phi_a \partial \phi_b$ $(a,b=3,8)$ at
the minimum of $V_\eff$.
One easily finds $\phi_3$ and $\phi_8$ are the eigenstates whose
masses are given by
\beqn
m_3^2 &=& \frac{g^2 \pi R}{2} \Big( \frac{\dd}{\dd \theta_1} - \frac{\dd}{\dd \theta_2} \Big)^2
~ V_\eff (\theta_1, \theta_2) \Big|_\min ~, \cr
\noalign{\kern 10pt}
m_8^2 &=& \frac{g^2 \pi R}{6} \Big( \frac{\dd}{\dd \theta_1} + \frac{\dd}{\dd \theta_2} \Big)^2
~ V_\eff (\theta_1, \theta_2) \Big|_\min 
\label{Higgs2}
\eeqn
where the second derivatives are evaluated at the global minimum  of $V_\eff$.
Note that although the form of the effective potential $V_\eff (\phi_a)$ is 
gauge-dependent in general, the masses of the scalar fields $\phi_a$, which are related 
to the curvatures of $V_\eff$ at its  global minimum, are gauge-invariant quantities.   
In particular, the expression (\ref{Higgs2}) is valid to $O(g^2)$.

At this stage it is instructive to see how the formulas would change or remain invariant 
when a different boundary condition were adopted.  Suppose that the boundary condition matrix $V$ in 
\eqref{eq:BoundCond_1} is given by
\beeq
V =  
\begin{pmatrix} e^{i \delta_1} \cr & e^{i \delta_2} \cr  && e^{i \delta_3} \end{pmatrix} , \quad
\sum_{j=1}^3 \delta_j =0 ~.
\label{newBC2}
\eneq
With this boundary condition the  spectra of the KK modes of $A_M, \psi_\fund, \psi_\ad$
are given by the formula  \eqref{spectrum1} where $\theta_j$ is replaced by $\theta_j - \delta_j$.
As a consequence the effective potential is given by
\beeq
V_\eff (\theta_1, \theta_2 ; \delta_1, \delta_2) = 
V_\eff (\theta_1 - \delta_1, \theta_2 - \delta_2 ; 0, 0) ~,
\label{newVeff1}
\eneq
where $V_\eff (\theta_1 - \delta_1, \theta_2 - \delta_2 ; 0, 0)$ is $V_\eff (\theta_1, \theta_2)$
given by \eqref{VeffWithFermions} and \eqref{effV2b}.
The location of the global minimum of $V_\eff$ is given by 
$\theta_j^\min (\delta_1, \delta_2) = \theta_j^\min(0,0) + \delta_j$.
The masses $m_3$ and $m_8$ in \eqref{Higgs2} remain invariant.

As an example, consider the case with $d=4, N_\fund =0, \alpha_\ad =0$.  The effective potential 
is given by
\beqn
V_\eff(\theta_1, \theta_2) &=&  \frac{1}{2 \pi^5 R^3} 
\Big[ h_4(\theta_1 - \theta_2, 0) 
    +  h_4(2\theta_1 + \theta_2, 0)  +  h_4(\theta_1 + 2\theta_2, 0)  \cr
\noalign{\kern 10pt}
&&\hskip -.5cm
- 2 N_\ad \big\{ h_4(\theta_1 - \theta_2, m_\ad) +  h_4(2\theta_1 + \theta_2, m_\ad)   
+  h_4(\theta_1 + 2\theta_2, m_\ad) \big\} \Big] ~.
\label{effV6}
\eeqn
Here $h_4(\theta, m)$ is defined in (\ref{effV3}). 
The masses are given by
\beqn
m_3^2 &=& \frac{g^2}{4 \pi^4 R^2} \Big[ 
\big\{ 4 h_4^{(2)} (\theta_1 - \theta_2, 0) 
    +  h_4^{(2)}(2\theta_1 + \theta_2, 0)  +  h_4^{(2)} (\theta_1 + 2\theta_2, 0) \big\} \cr
\noalign{\kern 10pt}
&& - 2 N_\ad \big\{ 4 h_4^{(2)} (\theta_1 - \theta_2, m_\ad) 
+ h_4^{(2)}(2\theta_1 + \theta_2, m_\ad)  
+ h_4^{(2)} (\theta_1 + 2\theta_2, m_\ad) \big\}\Big]_\min ~, \cr
\noalign{\kern 10pt}
m_8^2 &=& \frac{3 g^2}{4 \pi^4 R^2} \Big[ 
\big\{   h_4^{(2)}(2\theta_1 + \theta_2, 0)  +  h_4^{(2)} (\theta_1 + 2\theta_2, 0) \big\} \cr
\noalign{\kern 10pt}
&& \hskip 1.cm
- 2 N_\ad \big\{  h_4^{(2)}(2\theta_1 + \theta_2, m_\ad)  
+ h_4^{(2)} (\theta_1 + 2\theta_2, m_\ad) \big\}\Big]_\min ~, 
\label{Higgs3}
\eeqn
where $h_4^{(2)} (\theta, m) = d^2 h_4 (\theta, m) / d\theta^2$.

The corresponding mass spectrum for each phase is as follows.

\bigskip

\noindent
\underline{A: SU(3) symmetric}

\vskip 5pt
In this phase all $\phi_a$ are physical, and all $m_a$'s are degenerate.
\beeq
m_a^2 = \frac{3g^2}{2 \pi^4 R^2} \big\{ h_4^{(2)} (0,0) - 2 N_\ad  h_4^{(2)} (0,m_\ad) \big\} 
\quad (a=1, \cdots, 8) ~.
\label{HiggsM1}
\eneq

\noindent
\underline{ B: SU(2) $\times$ U(1) symmetric}

\vskip 5pt
In this phase $\phi_4, \phi_5, \phi_6, \phi_7$ are absorbed by the corresponding vector fields.
$\phi_1, \phi_2, \phi_3, \phi_8$ are physical.
\beqn
m_1^2 && =m_2^2 =m_3^2 \cr
\noalign{\kern 10pt}
&&= \frac{g^2}{2 \pi^4 R^2} \Big[ 2h_4^{(2)} (0,0) + h_4^{(2)} (\pi ,0) 
- 2 N_\ad  \big\{ 2 h_4^{(2)} (0,m_\ad)  + h_4^{(2)} (\pi ,m_\ad) \big\}  \Big] ~, \cr
\noalign{\kern 10pt}
m_8^2&& =
\frac{3g^2}{2 \pi^4 R^2} \Big[  h_4^{(2)} (\pi ,0) - 2 N_\ad   h_4^{(2)} (\pi ,m_\ad)   \Big] ~.
\label{HiggsM2}
\eeqn
Notice that $m_{1,2,3} \not= m_8$.
\vskip 5pt

\noindent
\underline{C: U(1) $\times$ U(1) symmetric}

\vskip 5pt
In this phase $\phi_1, \phi_2,\phi_4, \phi_5, \phi_6, \phi_7$ are absorbed by the corresponding 
vector fields. $\phi_3, \phi_8$ are physical.
\beeq
m_3^2  = m_8^2 =
\frac{3g^2}{2 \pi^4 R^2} \Big[ h_4^{(2)} (\twothird \pi ,0)  
- 2 N_\ad  h_4^{(2)} (\twothird \pi ,m_\ad)  \Big] ~.
\label{HiggsM3}
\eneq

\begin{figure}[h]
  \centering
  \includegraphics[width=0.45\textwidth]{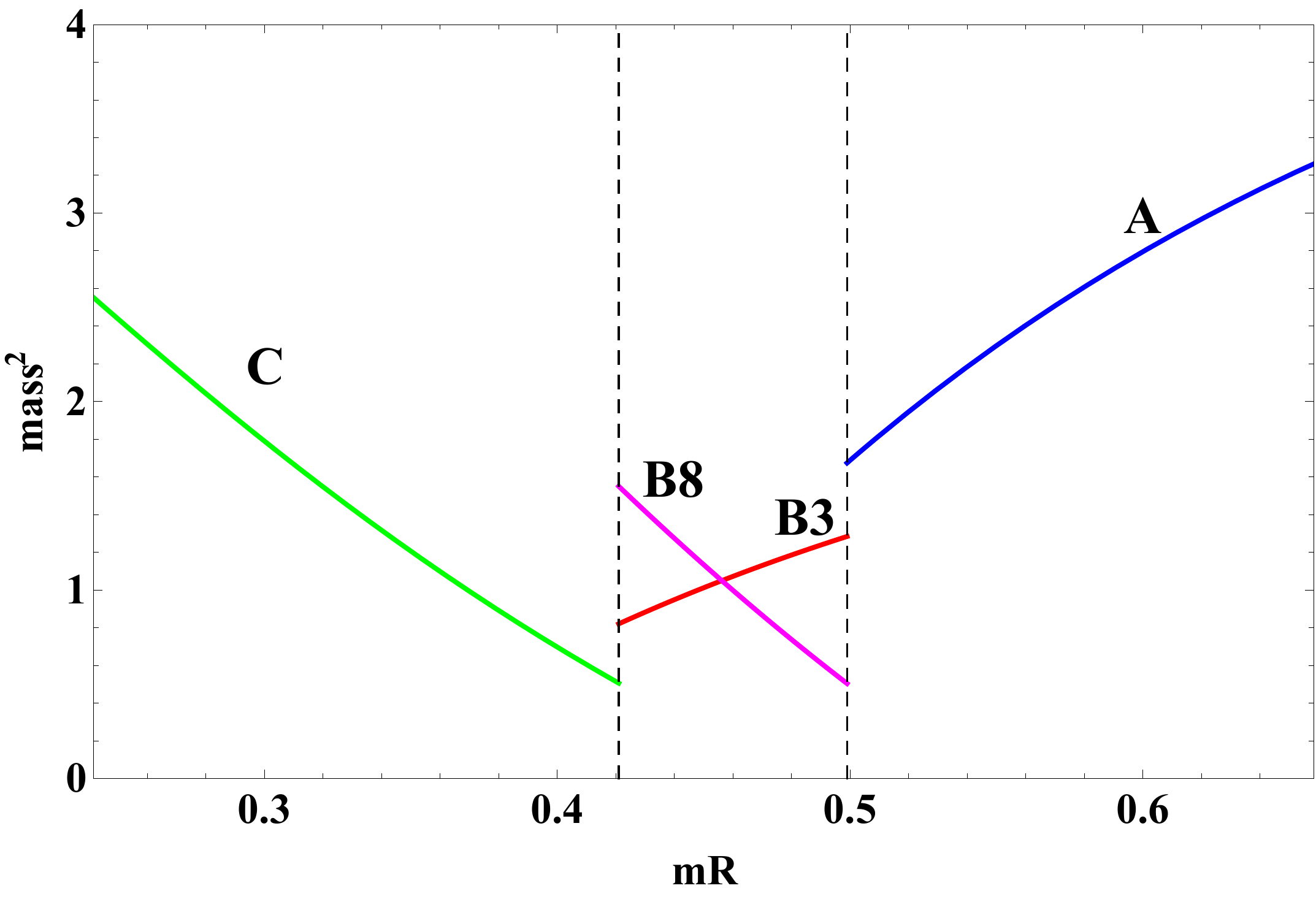}
\caption{ The dependence on the compactification parameter $mR$ of scalar
 ``Higgs'' mass squared in phases A,~B and C for $(N_{\rm fd},N_{\rm ad})=(0,2)$.
For each phase, the value in units of $g^2/2 \pi^4 R^2$ is plotted. 
B3 and B8 are the masses for the $m_3^2$ and $m_8^2$ in eq.~(\ref{HiggsM2}), respectively.}
\label{fig:scalar-mass}
\end{figure}

In Fig.~\ref{fig:scalar-mass} we plot the dependence of the masses on the parameter $mR$ in the A,~B and C phases.
 We note here that $m_3$ and $m_8$ are expected to cross at some value of $mR\sim 0.46$ in the B phase and that $m_3=m_8$ in the
C-phase grows as $mR\rightarrow 0$ ({\it i.e.} $\beta \rightarrow \infty$ in terms of lattice parameters).

\section{Lattice results\label{sec:LatticeResults}}

 Following the general remarks in Sec.~\ref{GeneralRemark}, we 
 present our lattice study with the adjoint (fundamental) fermions
 in Secs.~\ref{AdjointSection} (\ref{FundamentalSection}).
 We separately discuss, in Sec.~\ref{sec:Directmeasure}, the connection 
 to the perturbative prediction by the analysis of the eigenvalue 
 distribution. 

\subsection{General remarks}\label{GeneralRemark}

The lattice action used through the entire work is the standard Wilson 
gauge action and standard staggered fermions (in the fundamental and adjoint representation).

We compute Polyakov loops $P_3$ and $P_8$
on the $16^3\times 4$ volume gauge configurations
sampled with the weight 
$e^{-S_g - S_f}$, where the lattice actions $S_g$ and $S_f$ are 
given in eqs.~(\ref{eq:LatticeGauge}) and (\ref{eq:FermionAction}), 
respectively. 
By comparing the distribution of $P_3$ on the complex plane and 
Fig.~\ref{P3sketch}, one can distinguish which phase is realized.
This qualitative analysis is also done in comparing the 
eigenvalue distribution and the vacua 
(Figs.~\ref{fig:ad-mass} and \ref{fig:f-mass})
obtained from the perturbative analysis. 
The transition points are determined by the susceptibility
\beq
\chi_\Omega = N_x^3 \left( \langle \Omega^2 \rangle - \langle \Omega
\rangle^2 \right)\label{definition_of_chi}
\eeq
of the observable $\Omega\in \{|P_3|,P_8\}$ which
should scale with the lattice volume for first order phase transitions.
 In connection to the perturbative results, where the relevant parameter 
 is $m_\fund R$ or $m_\ad R$, increasing $\beta$ has the effect of 
decreasing those parameters, due to the running of the renormalized 
fermion mass in the lattice unit.
We estimate statistical errors by employing the jackknife method 
with appropriate bin sizes to incorporate any auto-correlations.

\subsection{Adjoint fermions}\label{AdjointSection}

\subsubsection{Phase structure}

In the numerical simulation for $(N_{\rm ad},N_\fund) = (2,0)$, 
we use bare masses $ m_\ad a = ma =0.05$ and 0.10, 
changing $\beta$ covering the range $5.3 \le \beta \le 6.5$.
Periodic boundary condition is used ($\alpha_\ad=0$) 
in the compact direction, which is different 
from the case with anti-periodic boundary conditions (finite temperature) 
where only the confined and deconfined phases are 
realized~\cite{Karsch:1998qj}.
To explore the phase structure in heavier mass region, we also examine 
bare masses $ m_\ad a = ma =0.50$ and $0.80$ 
for the range of $5.5 \le \beta \le 9.8$ and $5.5 \le \beta \le 20.0$,
respectively. As will be discussed, data with those masses require 
even more careful treatment.

\begin{figure}
  \centering 
  \includegraphics[width=0.45\textwidth]{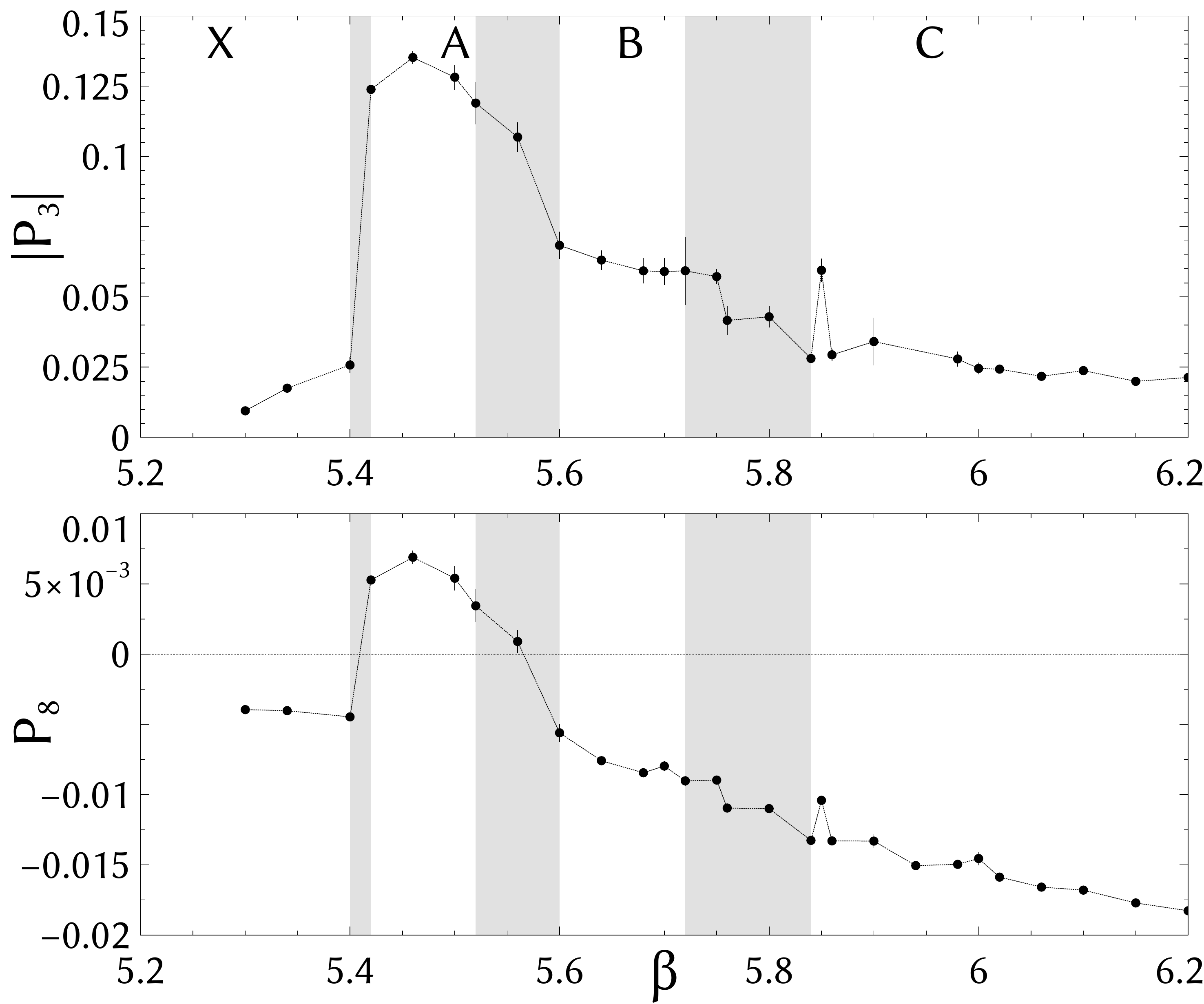}
  \includegraphics[width=0.45\textwidth]{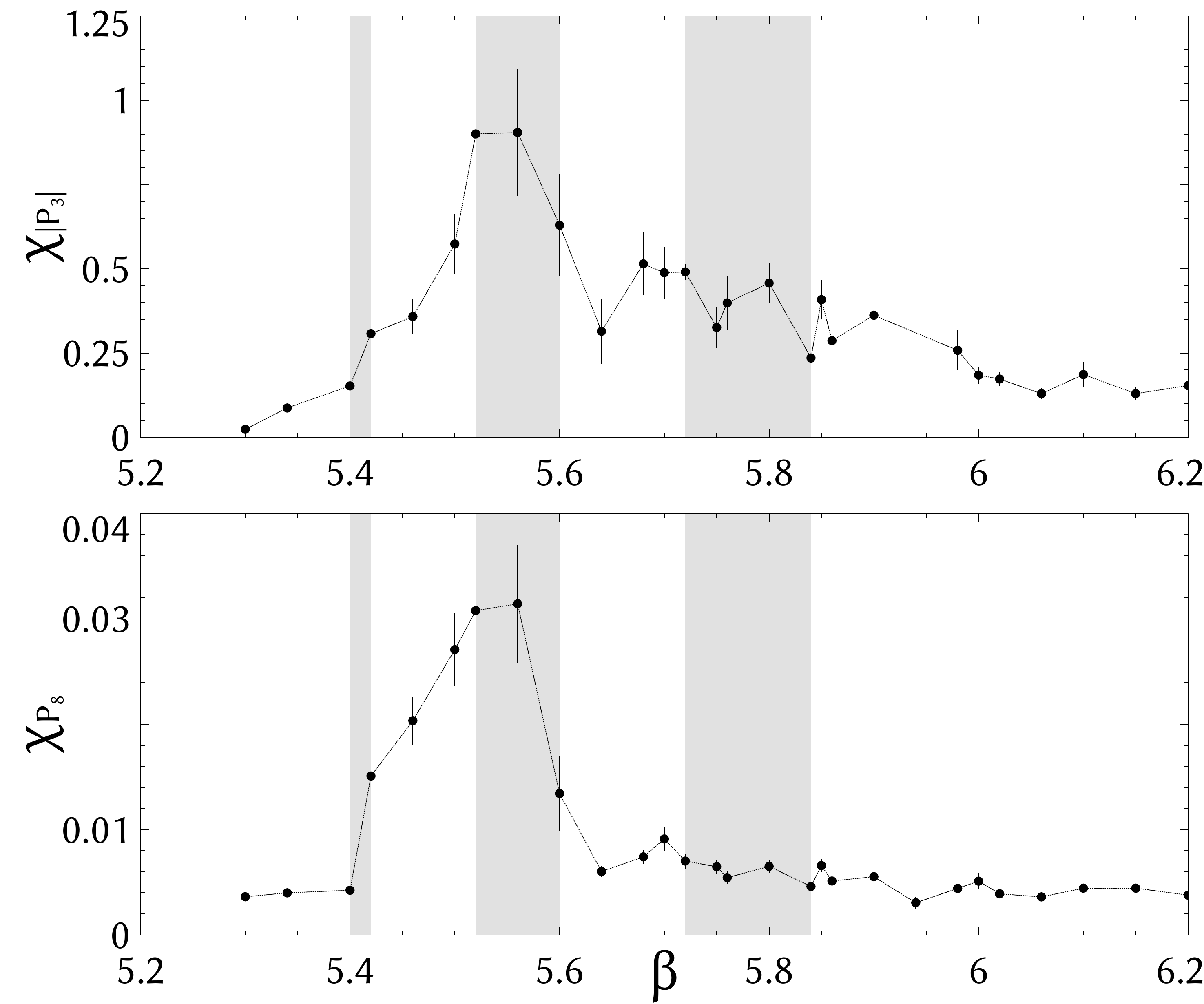}
  \caption{Left: $\beta$-dependences of $|P_3|$ (upper) and $P_8$
    (lower). The gray bands indicate the transition regions as listed in 
    Table~\ref{table-critical-beta}. Each phase is labeled in the upper
    panel, accordingly.
    Right: susceptibilities corresponding to the light panels.
    Results for $ma=0.05$ are shown.}
  \label{P3adj_vb-ma-0.05}
\end{figure}

\begin{figure}[t]
  \centering
  \includegraphics[width=0.45\textwidth]{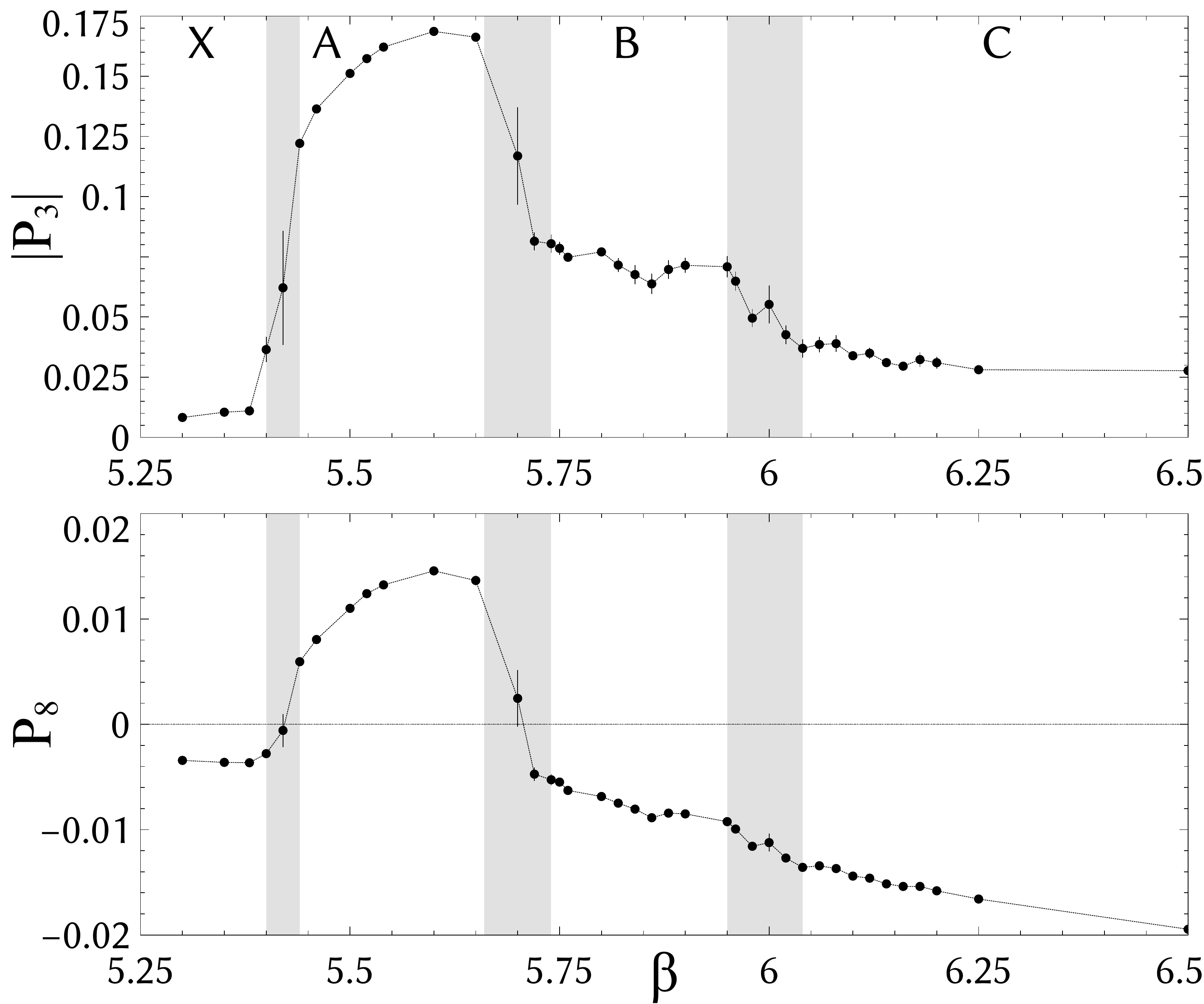}
  \includegraphics[width=0.45\textwidth]{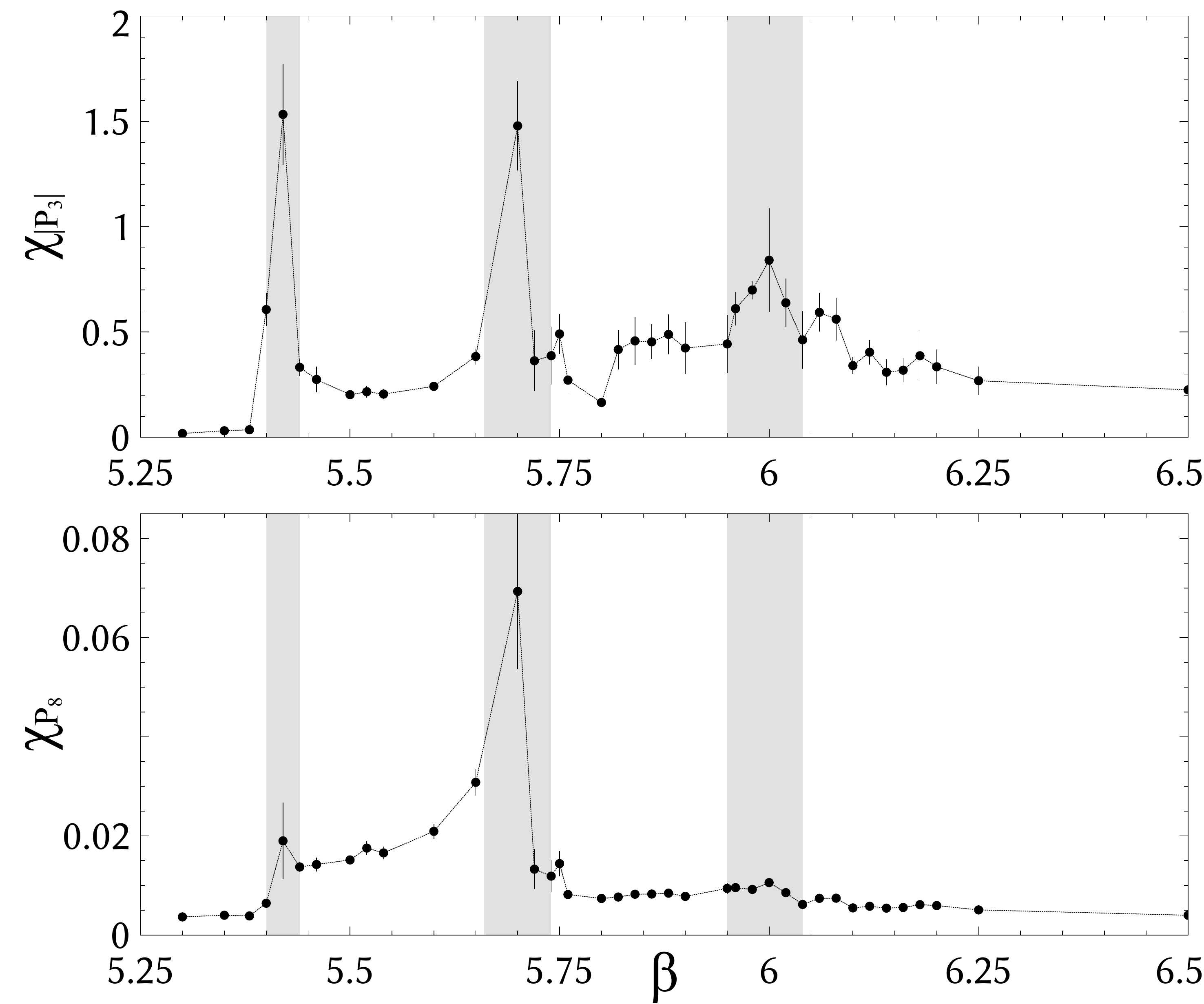}
  \caption{Same figure as Fig.~\ref{P3adj_vb} but for $ma=0.10$.}
  \label{P3adj_vb}
\end{figure}

\begin{figure}[h]
  \centering
  \includegraphics[width=0.45\textwidth]{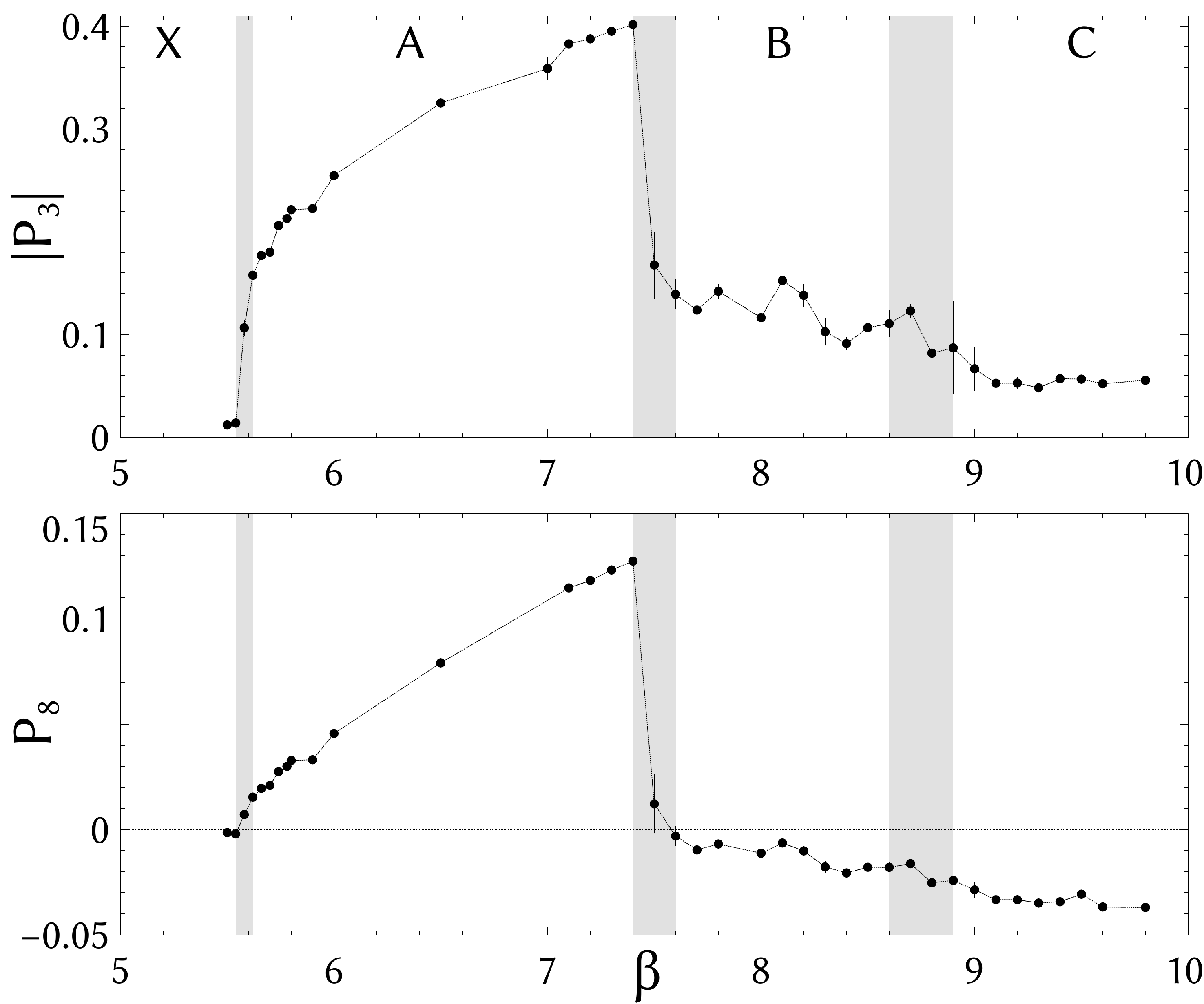}
  \ \
  \includegraphics[width=0.45\textwidth]{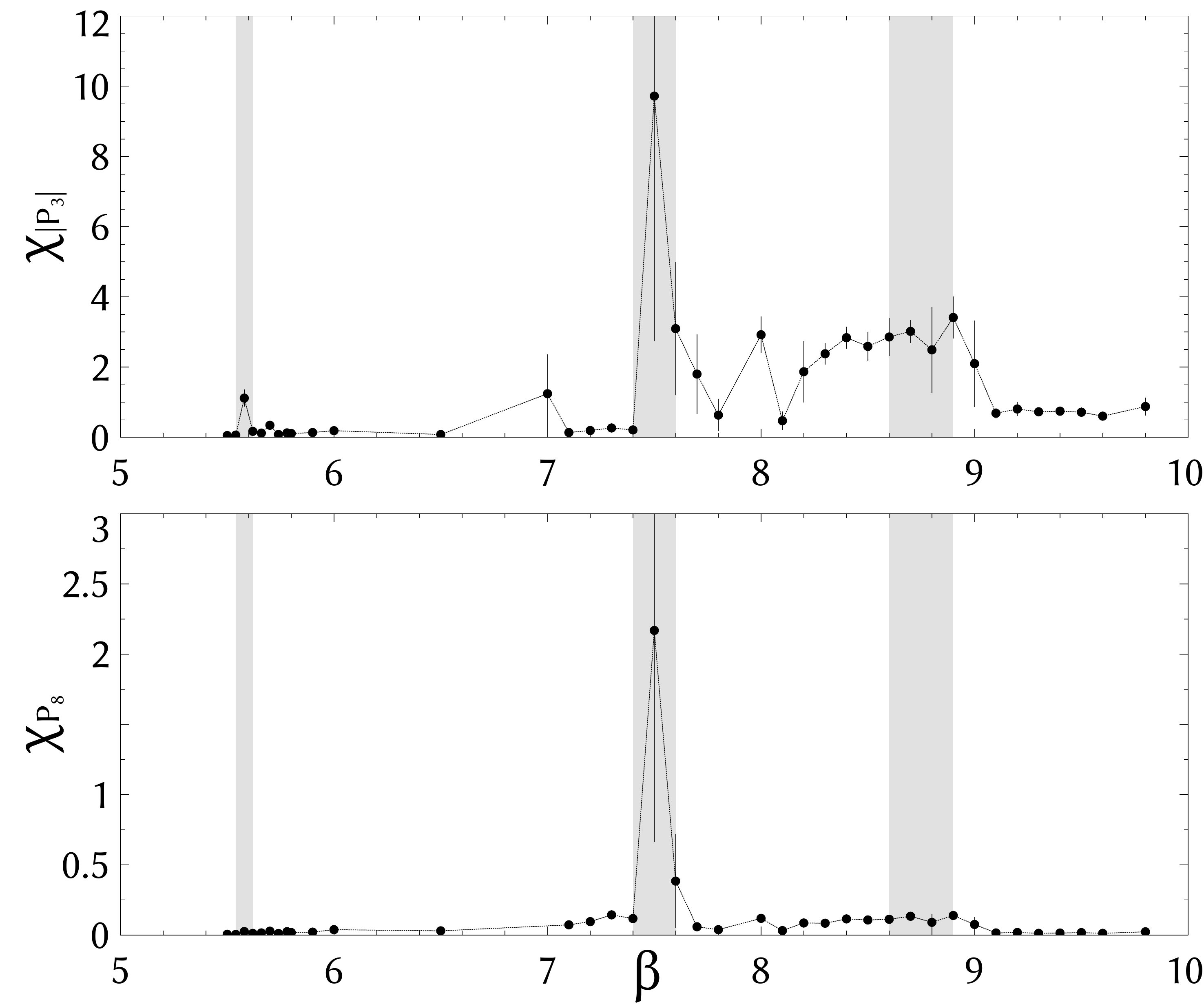}
 \caption{Same figure as Fig.~\ref{P3adj_vb} but for $ma=0.50$.}
 \label{P3adj_vb-ma-0.50}
\end{figure}

\begin{figure}[h]
  \centering
  \includegraphics[width=0.45\textwidth]{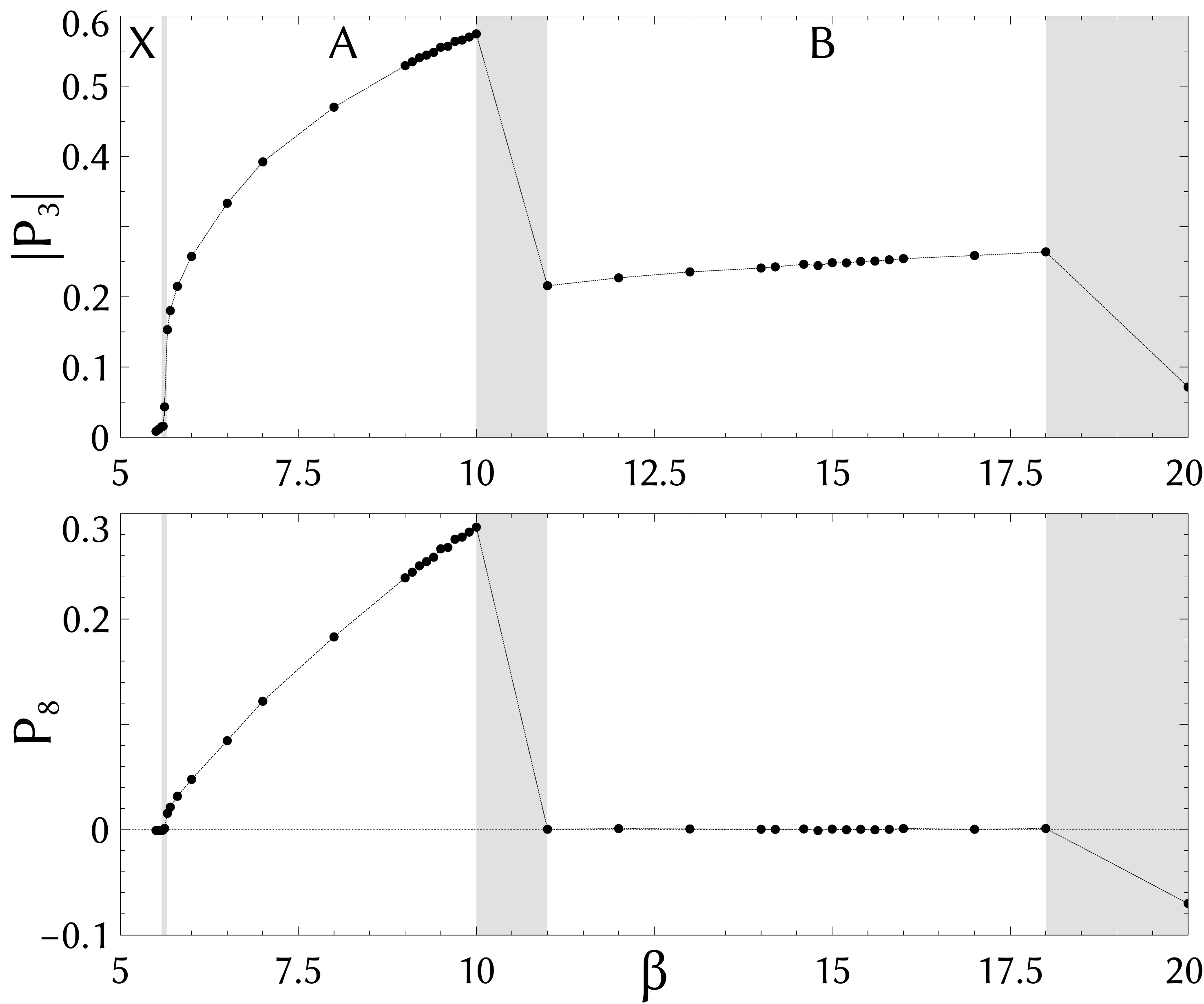}
  \ \
  \includegraphics[width=0.45\textwidth]{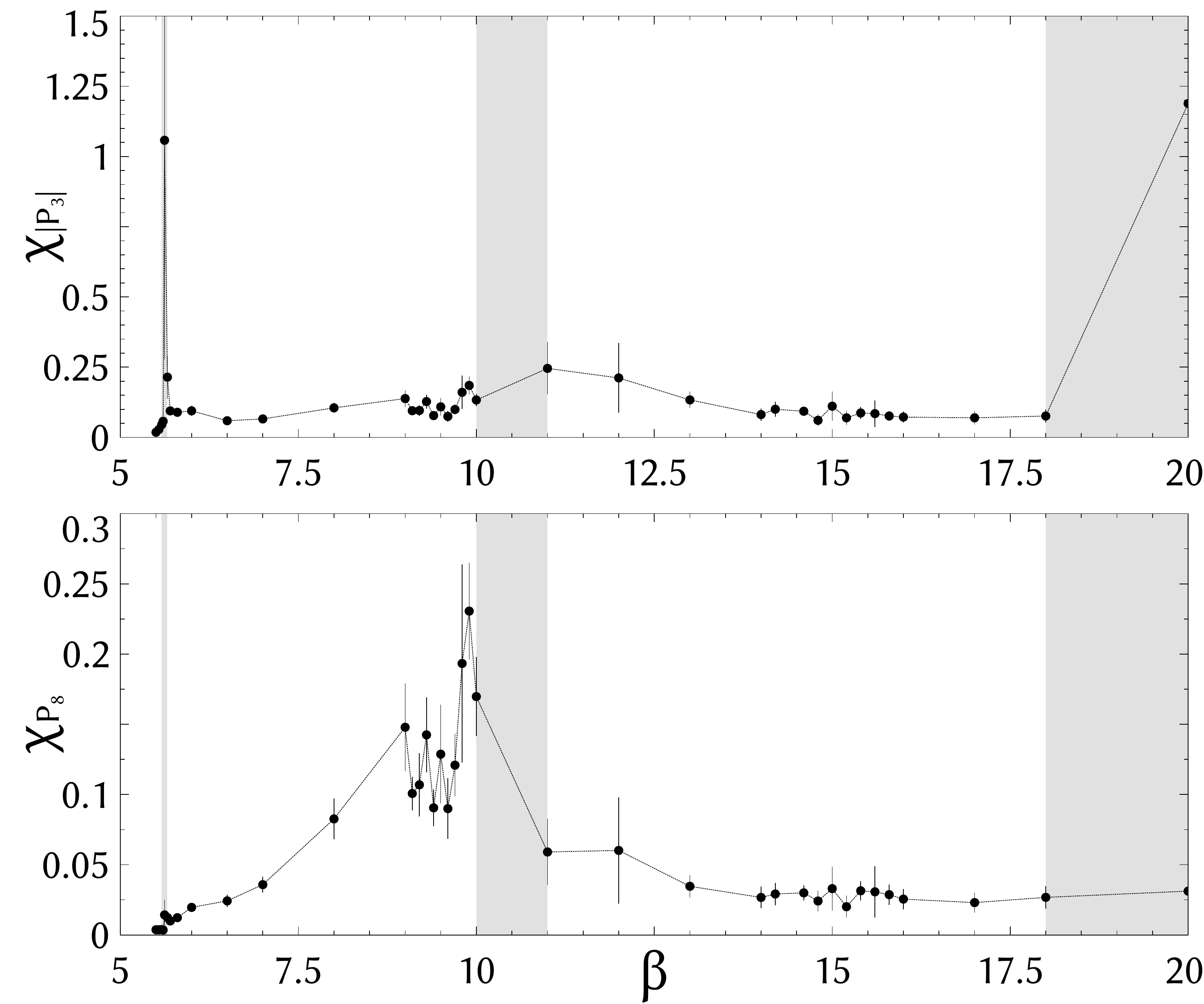}
  \caption{Same figure as Fig.~\ref{P3adj_vb} but for $ma=0.80$.}
  \label{P3adj_vb-ma-0.80}
\end{figure}

For each $ma$, after checking rough phase structure from the
distribution plot of $P_3$, we determine the transition points which we call
$\beta_{X/A}$, $\beta_{A/B}$ and $\beta_{B/C}$ for 
the $X$-$A$, $A$-$B$ and $B$-$C$ transitions, respectively (see
Figs.~\ref{P3adj_vb-ma-0.05},~\ref{P3adj_vb},~\ref{P3adj_vb-ma-0.50} 
and~\ref{P3adj_vb-ma-0.80}).
For this purpose, it is convenient to investigate $|P_3|$, $P_8$ and the
susceptibilities of them. 
 Since the data is at finite physical volume,
and the $P$ are not renormalized, 
their magnitudes could differ from the predictions summarized
in Table~\ref{table-phase}. Nevertheless, from the $|P_3|$ data, 
we can identify four regions for all the masses studied, that
respectively correspond to the observation of phases 
$X$, $A$, $B$ and $C$. The peaks of the susceptibility at the
$B$-$C$ transition is milder than the first two
as explained in the qualitative discussion of potential barrier in 
Sec.~\ref{VeffAdjointSec} for the behavior of $V_{\eff}(\theta_H)$.
It is also interesting to see that $P_8$ becomes zero at the $X$-$A$ transition
($\beta=5.42$) and the $A$-$B$ transition ($\beta=5.70$).
The values at these points change from $-1/8$ to $1$ and from $1$ to $0$
according to the analytical prediction summarized in Table~\ref{table-phase}.
Since, in general, the change of the sign in observables is not affected by 
lattice artifacts, the zeros of $P_8$ give a reliable way of locating the
transition points.
However, this idea is only applicable to the transition where 
the sign of $P_8$ changes. Contrary to the analytic prediction, $P_8$
takes a 
negative value in the $B$-phase for reasons which we will discuss later.
Therefore, there is no possibility to have another zero around $\beta=6.0$ 
for the $B$-$C$ transition. As seen in the figure, $\chi_{P_8}$ is less
sensitive to the transitions than others, hence remains in a complemental role.


 Looking over $P_8$ for all bare masses, we see 
the $X$-$A$ transition point depends on $ma$ only mildly. This is explained 
by the fact that the existence of the adjoint fermions does not affect the
${\rm Z}_3$ symmetry because the adjoint gauge link is invariant under the
associated global ${\rm Z}_3$ transformation.
On the other hand, for larger $ma$, 
the $A$-$B$ transition occurs at the larger $\beta$ at which,
in the perturbative
language, the value of $m_\ad R = \hat{m}(\beta) \times N_y/2\pi$ remains 
in the same level by the running of the renormalized mass
$\hat{m}(\beta)$ in the lattice unit.  
Accordingly, $P_8$ in the $B$-phase approaches zero as expected analytically 
for the perturbative region. In particular, at $ma=0.80$, 
$P_8$ is consistent with zero in the $B$-phase and the $B$-$C$ 
transition is detected at the point where $P_8$ starts to deviate from zero.
However, for such large $\beta$ value, the physical 
lattice size is exponentially small. 
For further discussion on the properties of these transitions,
a more detailed study on the finite size scaling has to be done. 
 
\begin{figure}[h]
 \centering
 \includegraphics[clip=true,width=0.65\textwidth]{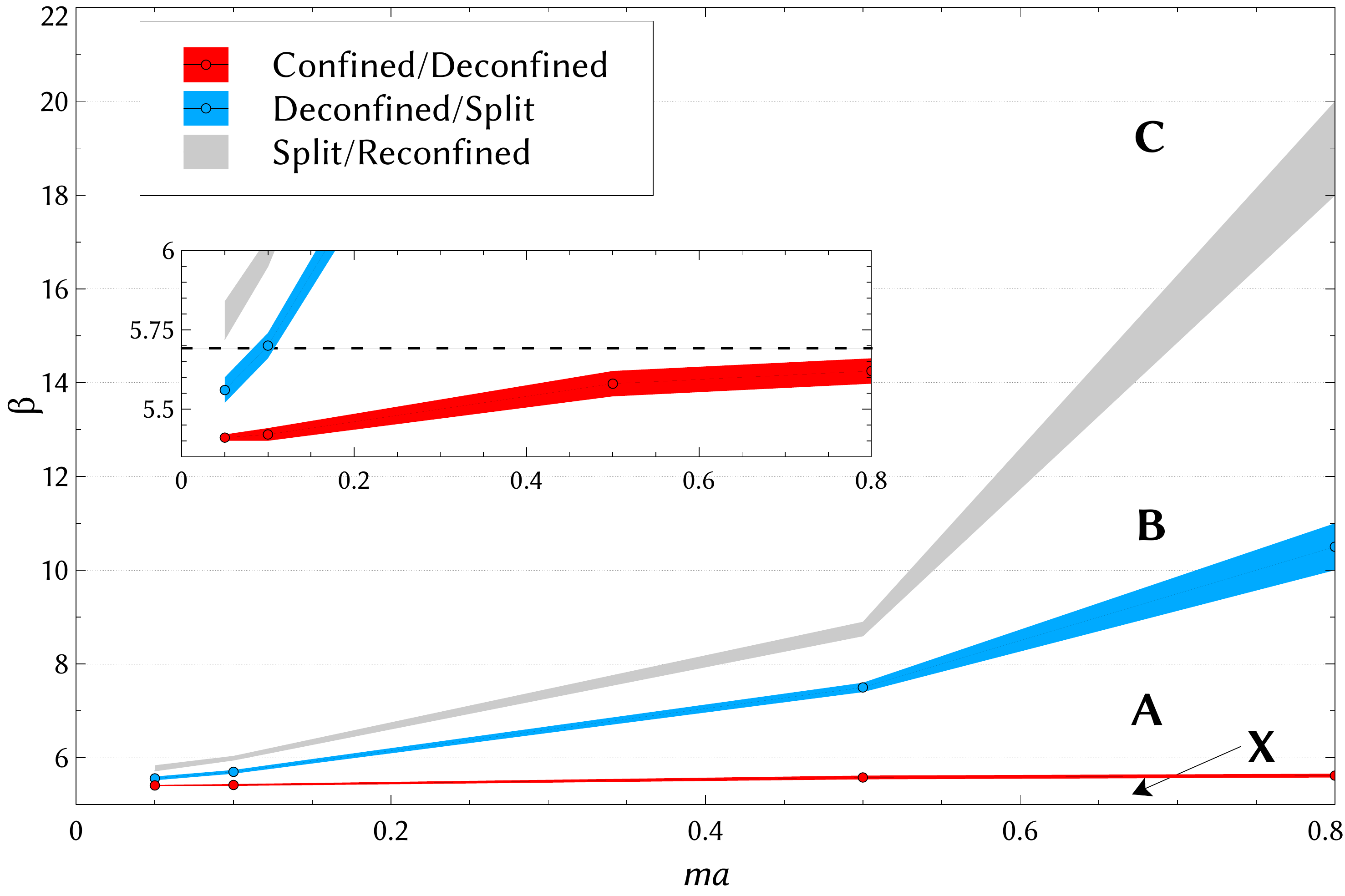}
 \caption{Phase diagram for the $N_\ad =2$ adjoint fermion system with periodic 
   boundary condition in the compact dimension. 
   In the window, the $X$-$A$ transition line is compared with the 
   pure gauge case (dashed line)~\cite{Fukugita1989}. }
 \label{fig:phase-diagram}
\end{figure}

With a caveat for the heavy mass region, we summarize the critical
values of $\beta$ for each mass in Table~\ref{table-critical-beta}.
Based on this result, the phase diagram on the $\beta$-$ma$ plane 
is depicted in Fig.~\ref{fig:phase-diagram}. In the magnified plot in
the inset, we show the approach of the $X$-$A$ transition point 
to the confined-deconfined transition transition point $\beta=5.692(20)$
 (dashed line) 
for the pure gauge case~\cite{Fukugita1989}.  
Because the $B$-$C$ transitions are hard to observe clearly from the Polyakov
loops or the susceptibilities at this volume,
we estimate the interval where the transition occurs as follows.
The lower boundary of the interval is the highest $\beta$ where, by
inspecting the eigenvalue distribution, we can still clearly identify the
$B$-phase. 
The upper boundary comes accordingly from the lowest $\beta$ where 
 we are certainly in the $C$-phase.
Due to the subjective character of the data, we do not quote any error: 
it is an identification of the region where the transition is occurring.

\begin{table}[bth]
\caption{Critical values of $\beta$ for each $ma$.\label{table-critical-beta}}
\centering
\begin{tabular}{|c|c|l|l|}
\hline
 $ma$ & $\beta_{X/A}$ &\ \ \ \ $\beta_{A/B}$ &\ \ \ \ \ \ \ $\beta_{B/C}$ \\
 \hline
 0.05 &\ \ 5.41(1)\ \ &\ \ \,5.56(4)   & \, [5.72,\, 5.84 ] \\
 0.10 &\ \ 5.42(2)\ \ &\ \ \,5.70(4)   & \, [5.95,\, 6.04 ] \\
 0.50 &\ \ 5.58(4)\ \ &\ \ \,7.50(10)  & \, [8.60,\, 8.90 ] \\
 0.80 &\ \ 5.62(4)\ \ &\ 10.50(50)  & [18.00, 20.00 ] \\
\hline
\end{tabular}
\end{table}

\begin{figure}[h]
 \centering
 \includegraphics[clip=true,width=0.45\textwidth]{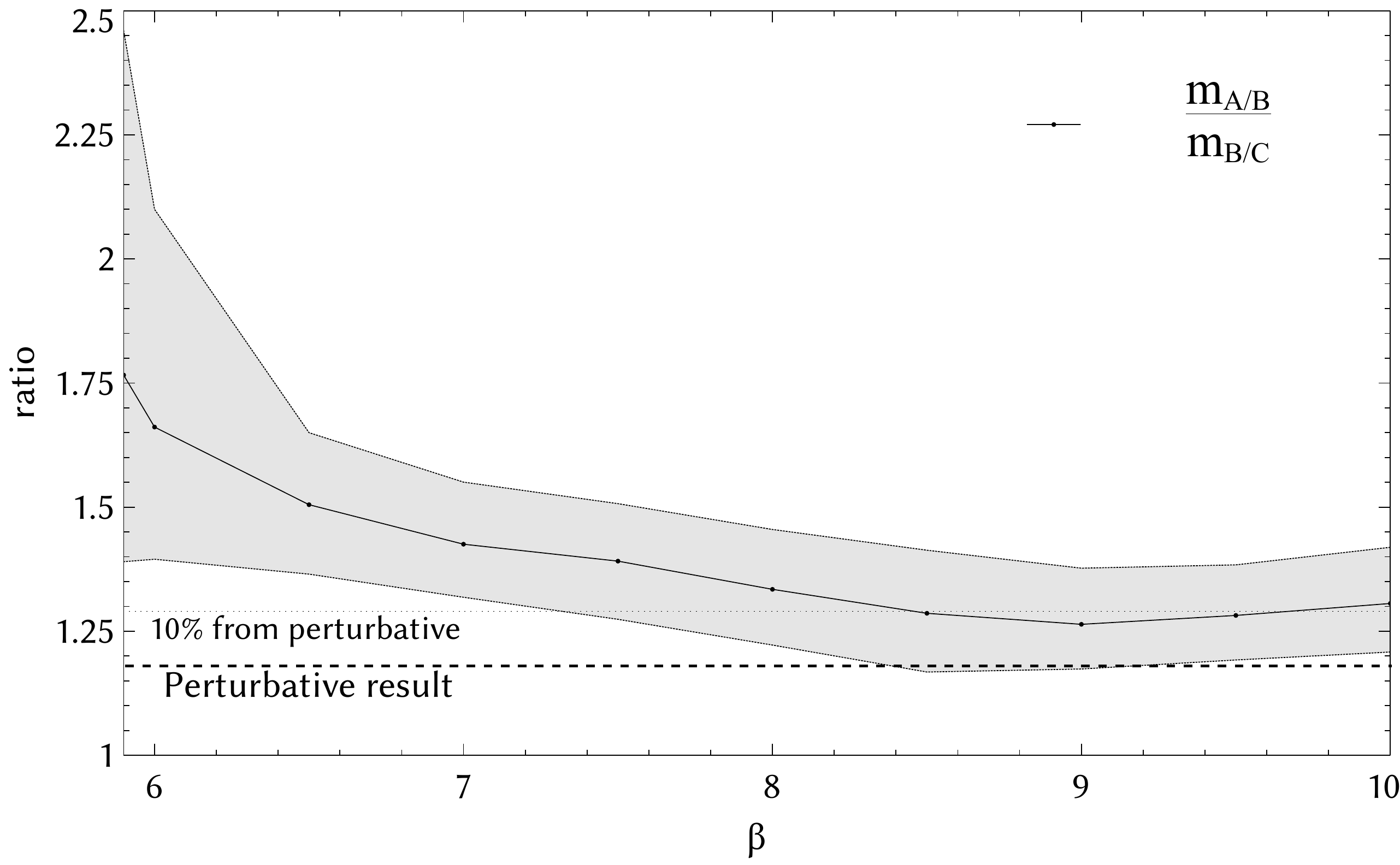}
 \caption{The ratio of mass-parameter at the phase transitions
 compared with the perturbative prediction. Measurements are done in
 steps of 0.5 in $\beta$.}
 \label{fig:masses-ratio}
\end{figure}

The phase diagram depicted in Fig.~\ref{fig:phase-diagram} can be compared with the perturbative 
expectations in Sec.~\ref{perturbativeVeff}. 
First we consider ``trajectories'' with constant $\beta$ 
on which only the adjoint quark masses change.
 If the perturbative parameter $m_{\rm ad}$ and the lattice parameter $ma$
are related by a multiplicative factor, which only depends on $\beta$,
then we can compare the ratio $m_{\rm A/B}/m_{\rm B/C}$ of these bare
parameters at the phase transitions
 to the prediction from eq.~(\ref{mRregions}) {\it i.e. $\sim 1.18$} (constant).
In Fig.~\ref{fig:masses-ratio} we plot the result of this analysis. The
intercepts of the constant $\beta$ lines to the phase transition ones
are obtained after spline interpolation. The errors are pure statistical
not including the systematic of the interpolation method. 
For $\beta \gtrsim 8$, the lattice results seem to reach a constant value 
which is $\approx 10\%$ larger than the perturbative one.

It should be noted that
since the effective potential is written in terms
of $\Tr W_3$ in eq.~(\ref{eq:veff_P}) and, as already explained, can be approximated by its
first term, our model is then related to a simpler one with the A, B and
C phases~\cite{Bialas:2004gx, Ogilvie:2012is, Myers:2007vc}.
By this comparison, it is inferred that the truncation of the adjoint
fermion part to the first term in eq.~(\ref{eq:veff_P}) is enough to
reproduce the phase structure.

\subsubsection{Eigenvalues of the Wilson line} \label{sec:Directmeasure}

The measured values of $P_3$ and $P_8$ are comparable with the predictions for 
the Hosotani mechanism listed in Table~\ref{table-phase}.
In order to further clarify the connection of these phases with the perturbative 
effective potential predictions of Sec.~\ref{VeffAdjointSec}, we present here the main result of the paper: 
the density plots for the eigenvalues of the Wilson line wrapping around the
compact dimension (cf. eq.~(\ref{eq:Wilson1})).
These observables are the fundamental degrees of freedom in the perturbative description. 
We demonstrate also that the lattice non-perturbative calculation matches the perturbative results in
the weak coupling limit. In the strong coupling region, we show qualitative 
agreement with the $V_{\rm eff}$ and its phase structure.

We recall here that on the lattice the Wilson lines are given by 
\beq
W_3^{\rm latt}(x)= \prod_{y=1}^{N_y}U_{(x,y),4}, \ \ 
W_8^{\rm latt}(x)= \prod_{y=1}^{N_y}U^{(8)}_{(x,y),4}.
\eeq
The eigenvalues of the Wilson line, eq.~(\ref{ABphase1}), are
independent of the gauge transformations, 
and their degeneracies classify 
the pattern of gauge symmetry breaking as explained in 
Sec.~\ref{sec:sym-breaking}.
The three complex eigenvalues are denoted by $\lambda=e^{i \theta_1}, e^{i \theta_2}$ and $e^{i \theta_3}$.
We constrain each phase within the interval $-\pi\le
\theta_1,\theta_2,\theta_3 \le \pi$.

\begin{figure}[t]
 \centering
  \includegraphics[clip=true, width=0.3\columnwidth]{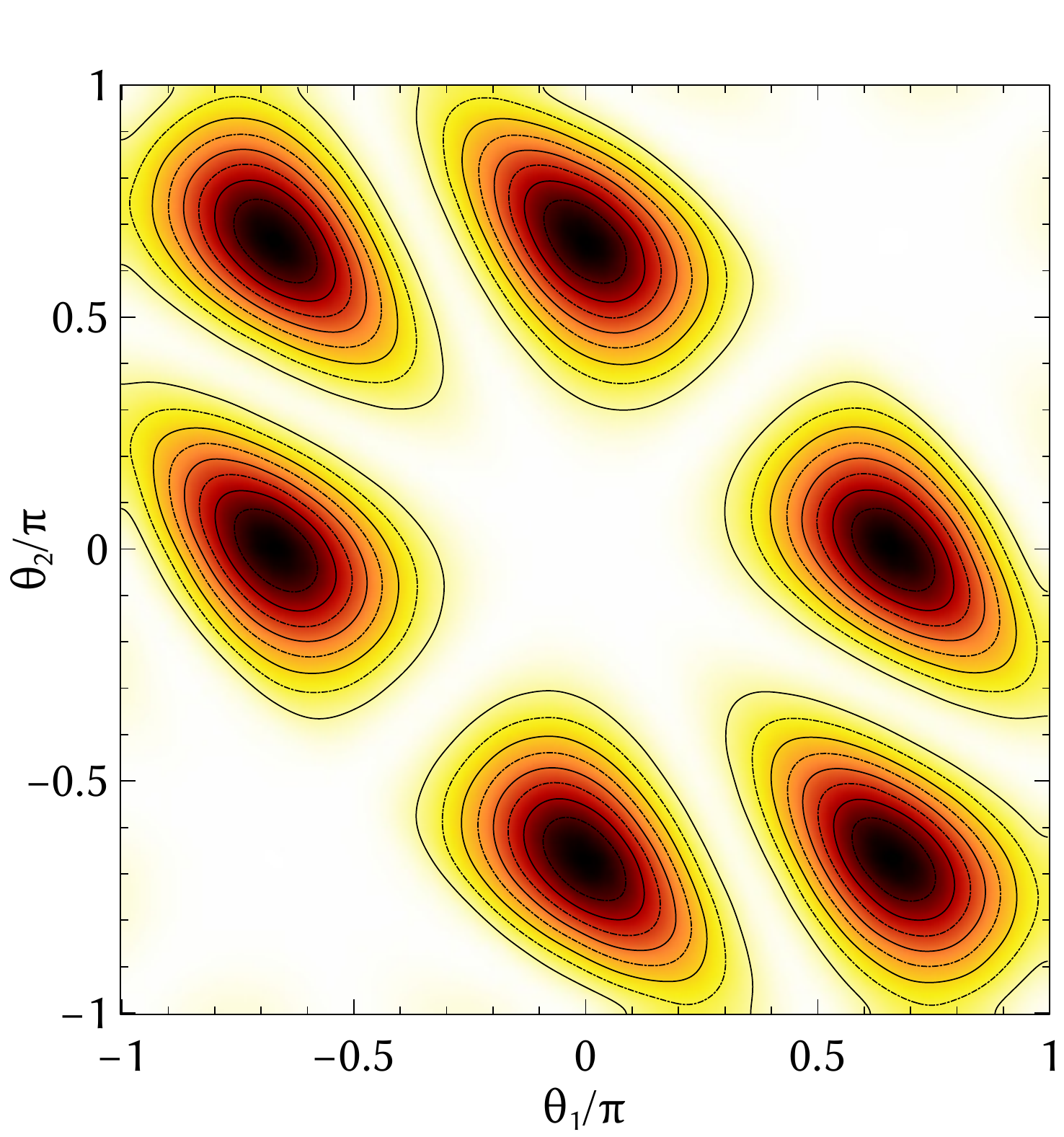}
 \caption{Haar measure density plot. Darker colors denote the highest density regions.}
 \label{fig:Haar-measure}
\end{figure}

A direct comparison of the density plots of the Polyakov
loop with the perturbative result for $V_ {\rm eff}$ must take into account that 
a complete degeneracy for the eigenvalues can be never measured directly. 
This is easily explained by the Haar measure for SU(3) (that can be derived using the Weyl integration formula, see e.g. \cite{SimonsLieGroups})
\beq
\prod_{i>j} \sin^2 \frac{\theta_i - \theta_j}{2} 
= \prod_{i>j}  \frac{1}{4}|e^{i\theta_i} - e^{i\theta_j}|^2
\label{eq:Haar-measure}
\eeq
which forbids such configurations. See the density plot in 
Fig.~\ref{fig:Haar-measure}, 
where the eigenvalue triplets of the Polyakov loop is obtained by 
random numbers constrained by eq.~(\ref{eq:Haar-measure}).
This measure term gives a strong repulsive force for the eigenvalues 
that must be subtracted to get the non-perturbative $V_ {\rm eff}$.

\begin{figure}[ht]
 \centering
  \includegraphics[clip=true, width=0.45\columnwidth]{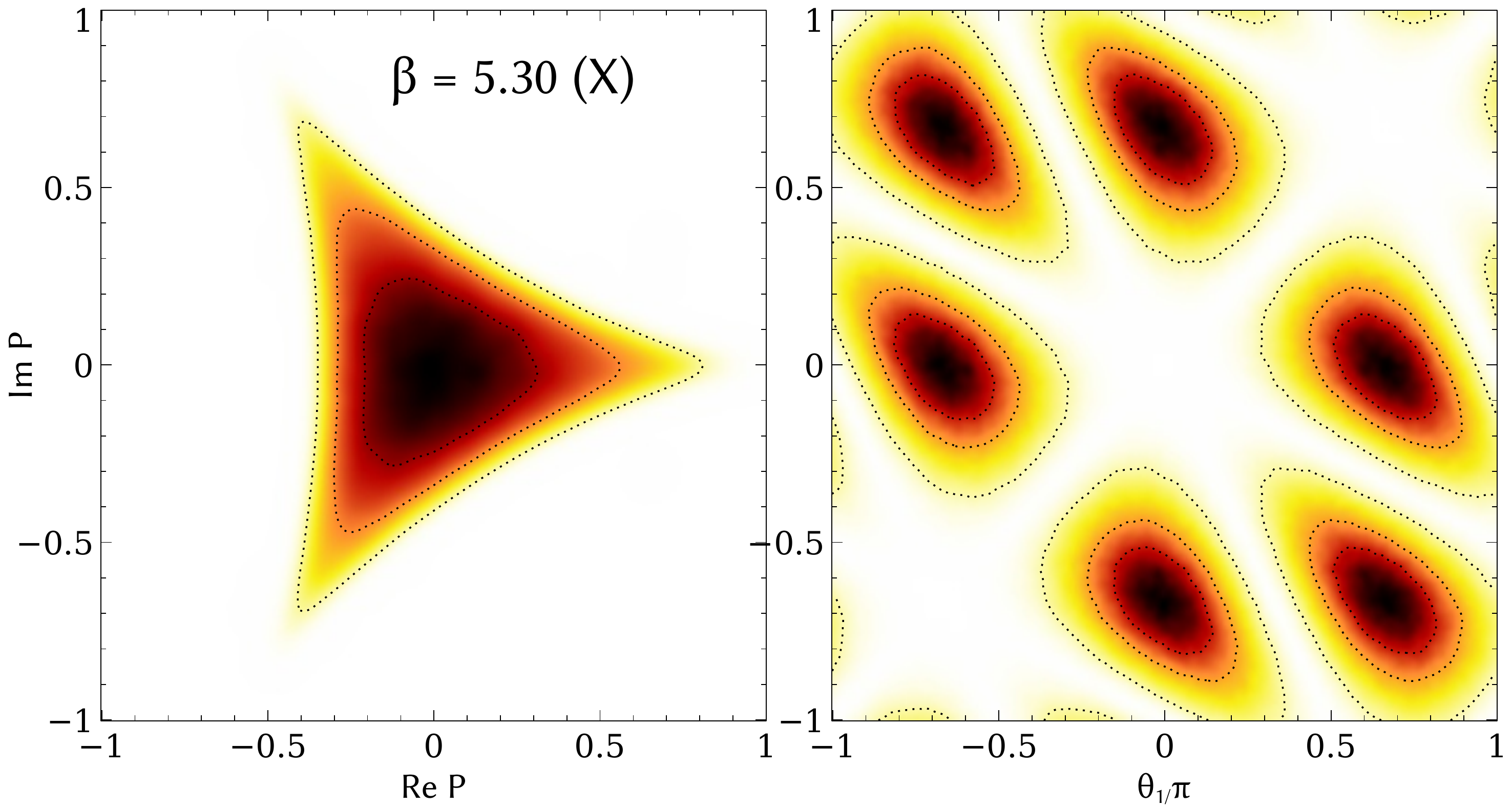}
  \includegraphics[clip=true, width=0.45\columnwidth]{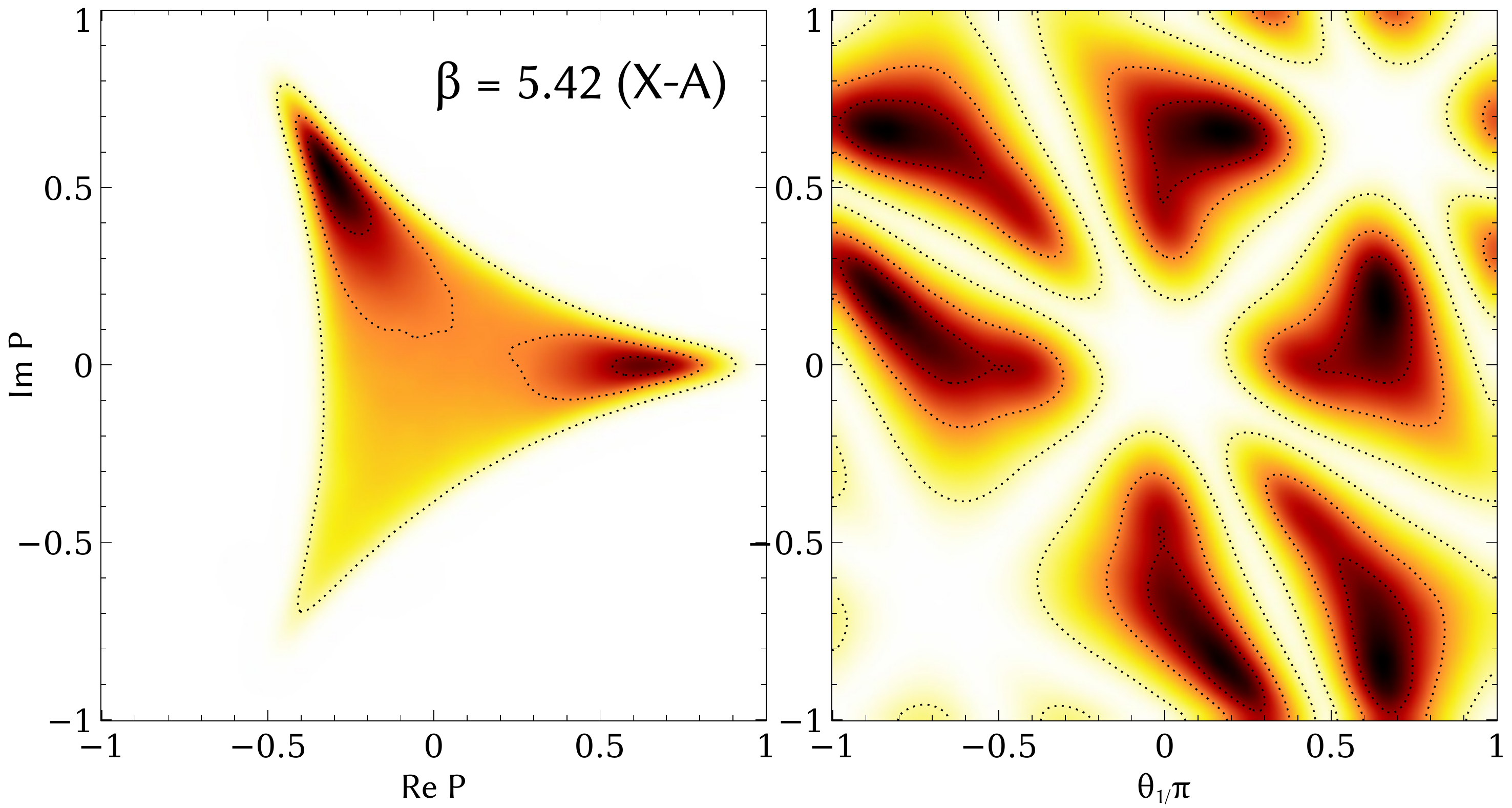}\\
  \includegraphics[clip=true, width=0.45\columnwidth]{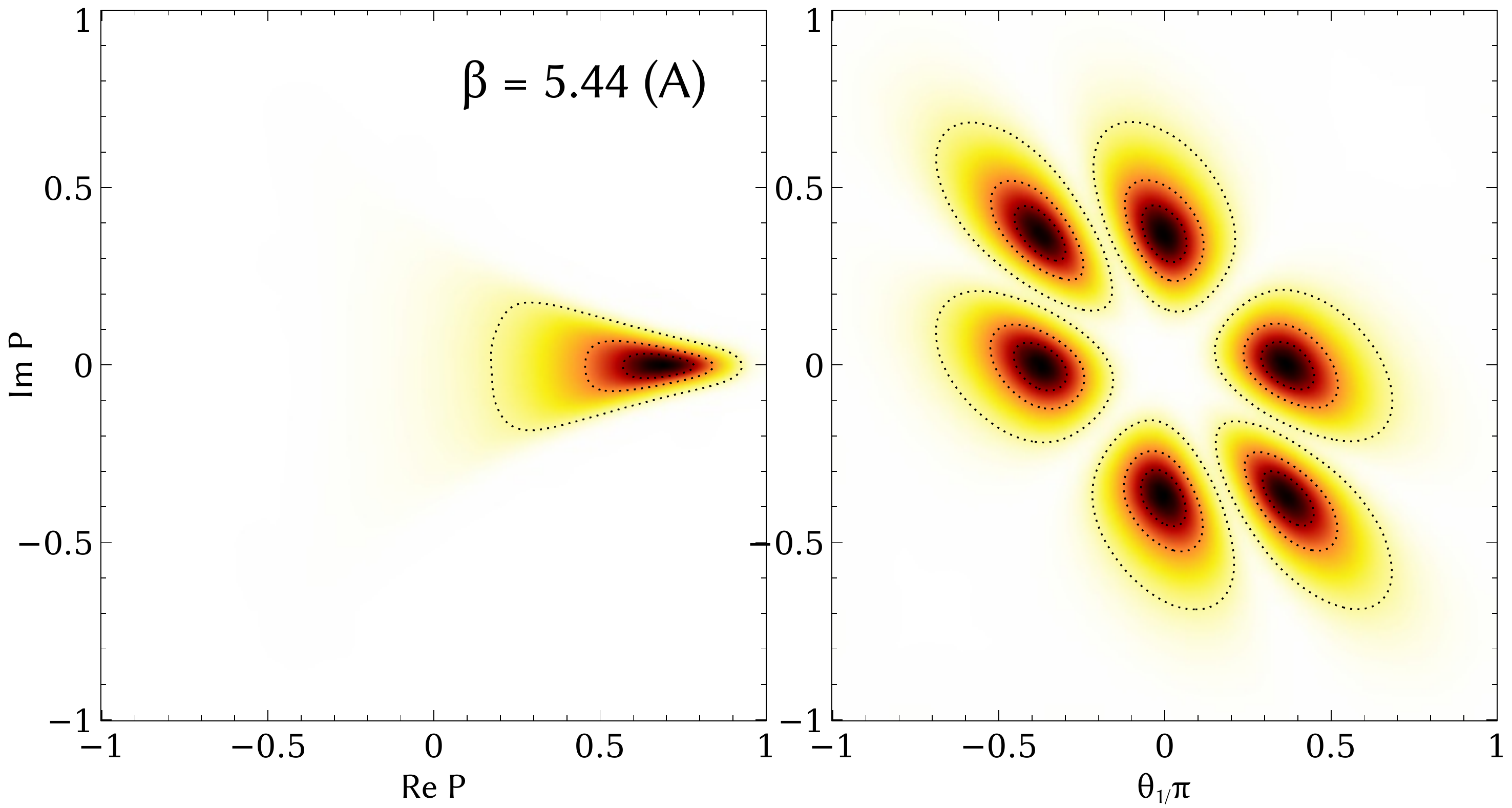}
  \includegraphics[clip=true, width=0.45\columnwidth]{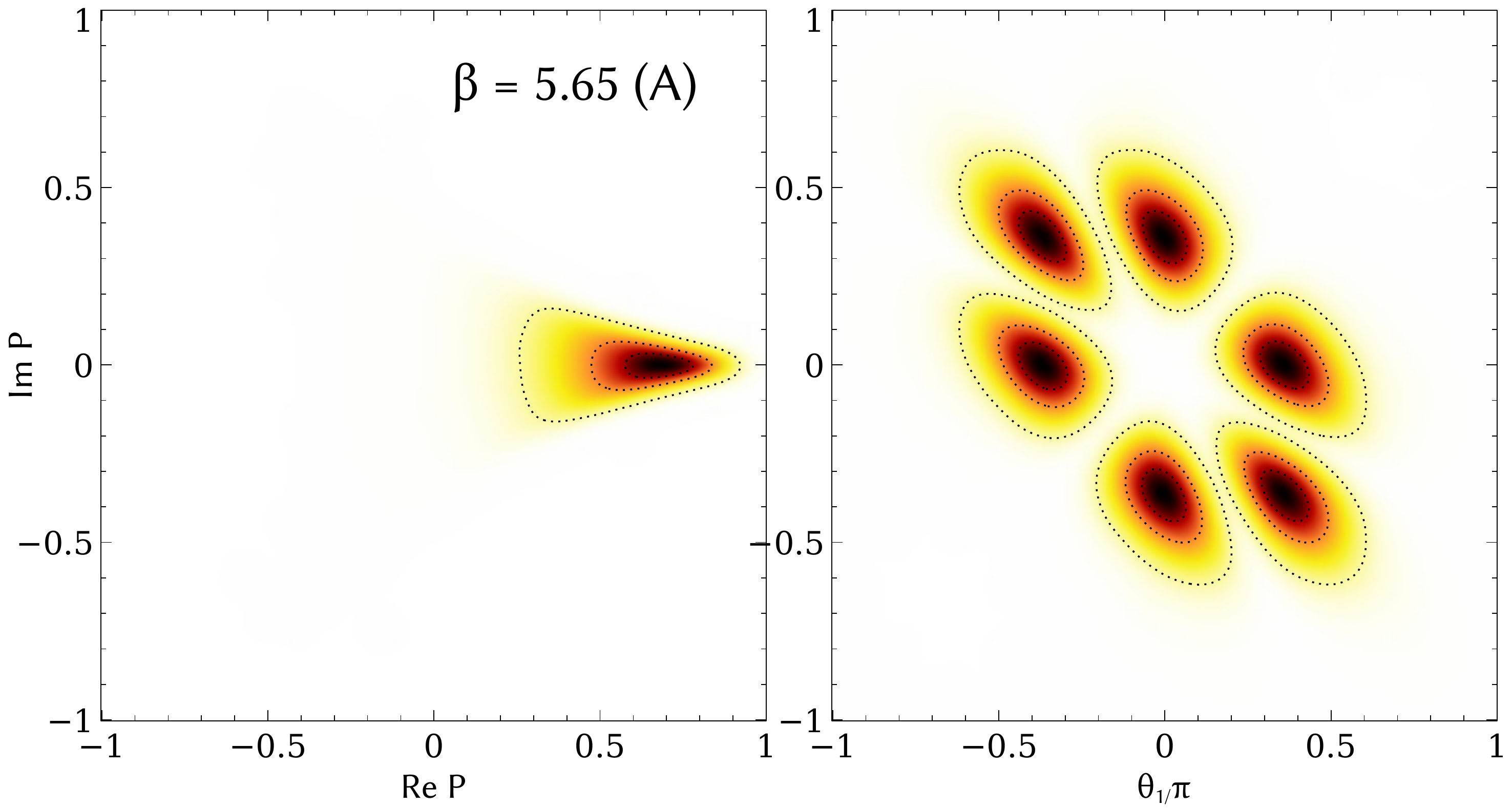}\\
  \includegraphics[clip=true, width=0.45\columnwidth]{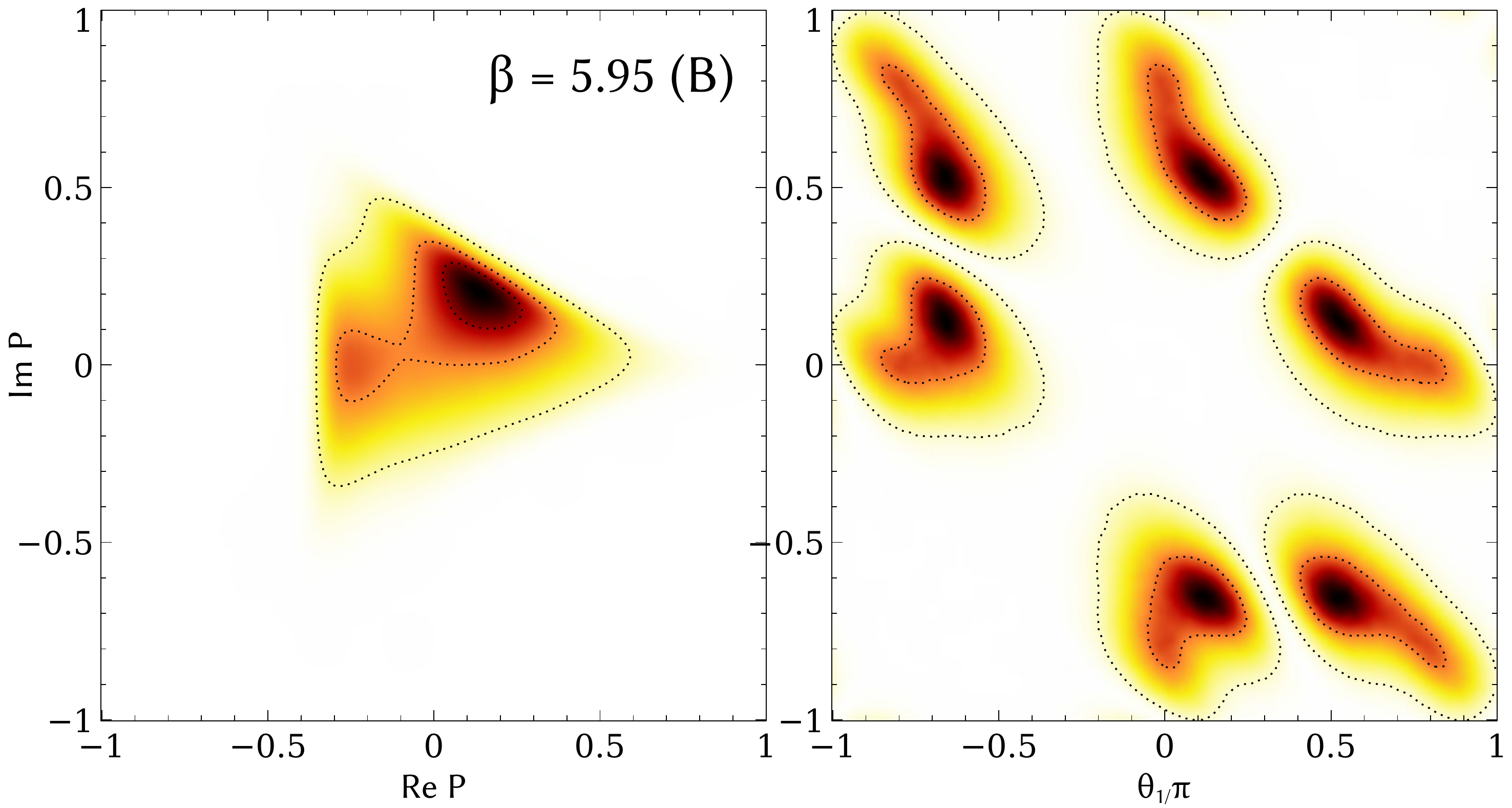}
  \includegraphics[clip=true, width=0.45\columnwidth]{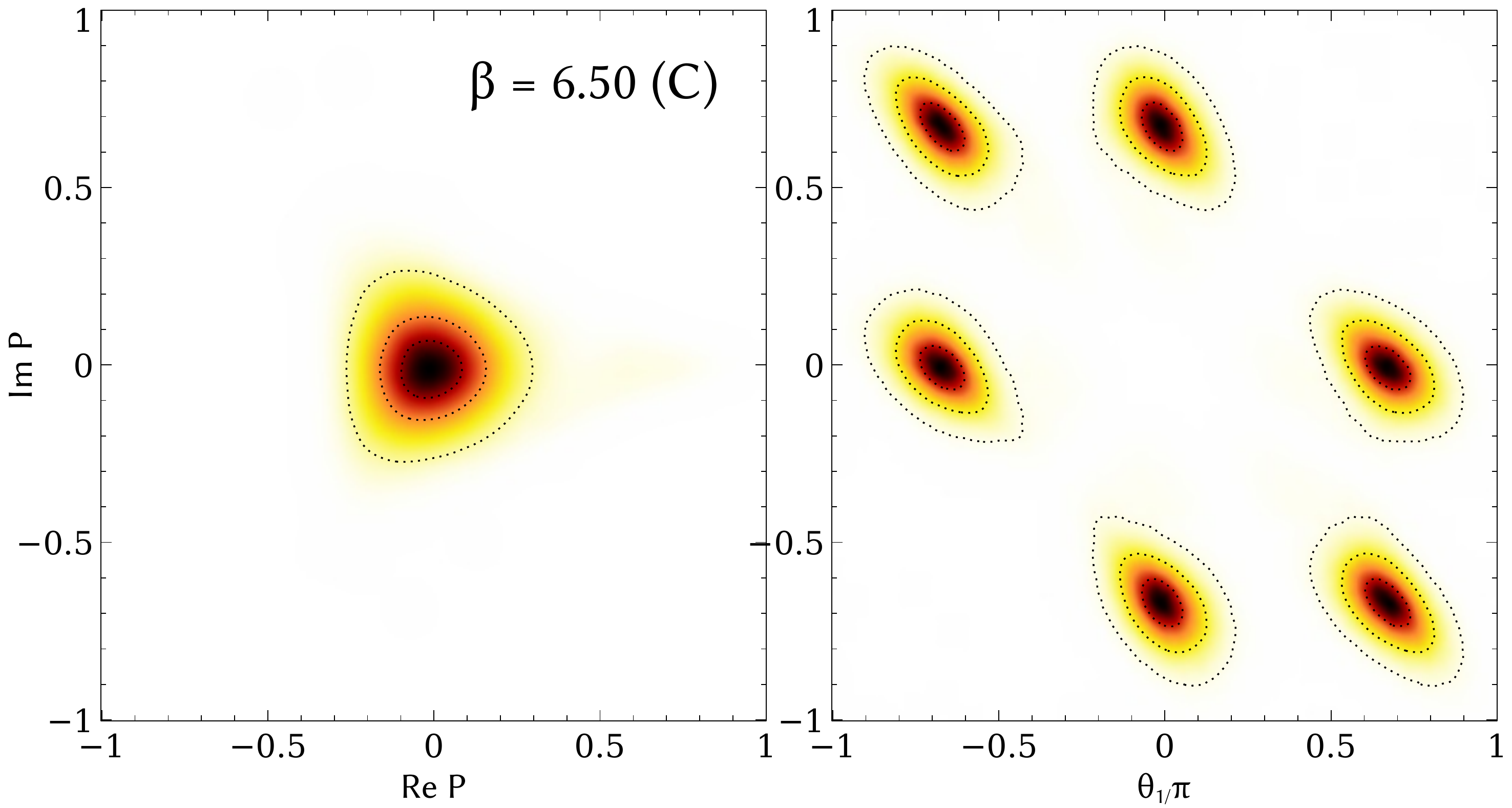}
 \caption{Density plots of the Polyakov loop 
  (in the complex plane) and its eigenvalues 
  (in the $\theta_1/\pi$ - $\theta_2/\pi$ plane) are shown 
  side-by-side at several $\beta$'s. All possible pairs
  $\{(\theta_i,\theta_j), i\neq j \}$ are included in the plots. 
  Darker colors denote the highest density regions.}
 \label{fig:density-plots}
\end{figure}

The results of our investigations are shown in the panels of Fig.~\ref{fig:density-plots}. 
These plots come in couples and each one of them displays the density plots for the Polyakov loop
 $P_3$ itself (left) and for the phases $(\theta_1,\theta_2)$ of its
 eigenvalues (right). 
Smearing is applied to the configuration before measurements 
(5 steps of stout smearing \cite{smearing} 
with the smearing parameter $\rho = 0.2$) to filter the ultraviolet
modes that are essentially just lattice noise and are not relevant 
for $V_{\rm eff}$.
By this technique, the gauge configuration is smoothed out by averaging 
the links over the nearest-neighbors, in a gauge invariant way. 
Several successive steps of smearing can be applied gradually increasing 
the radius of involved neighbors. 
The final result is a configuration where the ultraviolet oscillations 
of the gauge field at the level of the lattice spacing are highly suppressed. 
It is typically used to clear propagator signals, or obtain information on topological objects. 
Since we are only interested in
locating the minima of the potential, any fluctuations of the Polyakov
loop induced by the coarseness of the lattice around that minima are not
relevant at this level. We find that smearing is essential to extract
useful information from the configurations generated.
Data for the 2D density plots are also smoothed by a gaussian filter with 
a radius of 5 nearest-neighbors for clarity in the presentation.
The panels in Fig.~\ref{fig:density-plots}, from left to right, top to
bottom, show the change of distributions in passing 
the $X$-$A$-$B$-$C$ phases.
Modulo the Haar measure contribution, we observe a good correspondence 
between the perturbative shape of the potential and the location of the 
maximum of the densities of the measured Polyakov loop eigenvalues.

\begin{figure}[t]
 \centering
  \includegraphics[clip=true, width=0.24\columnwidth]{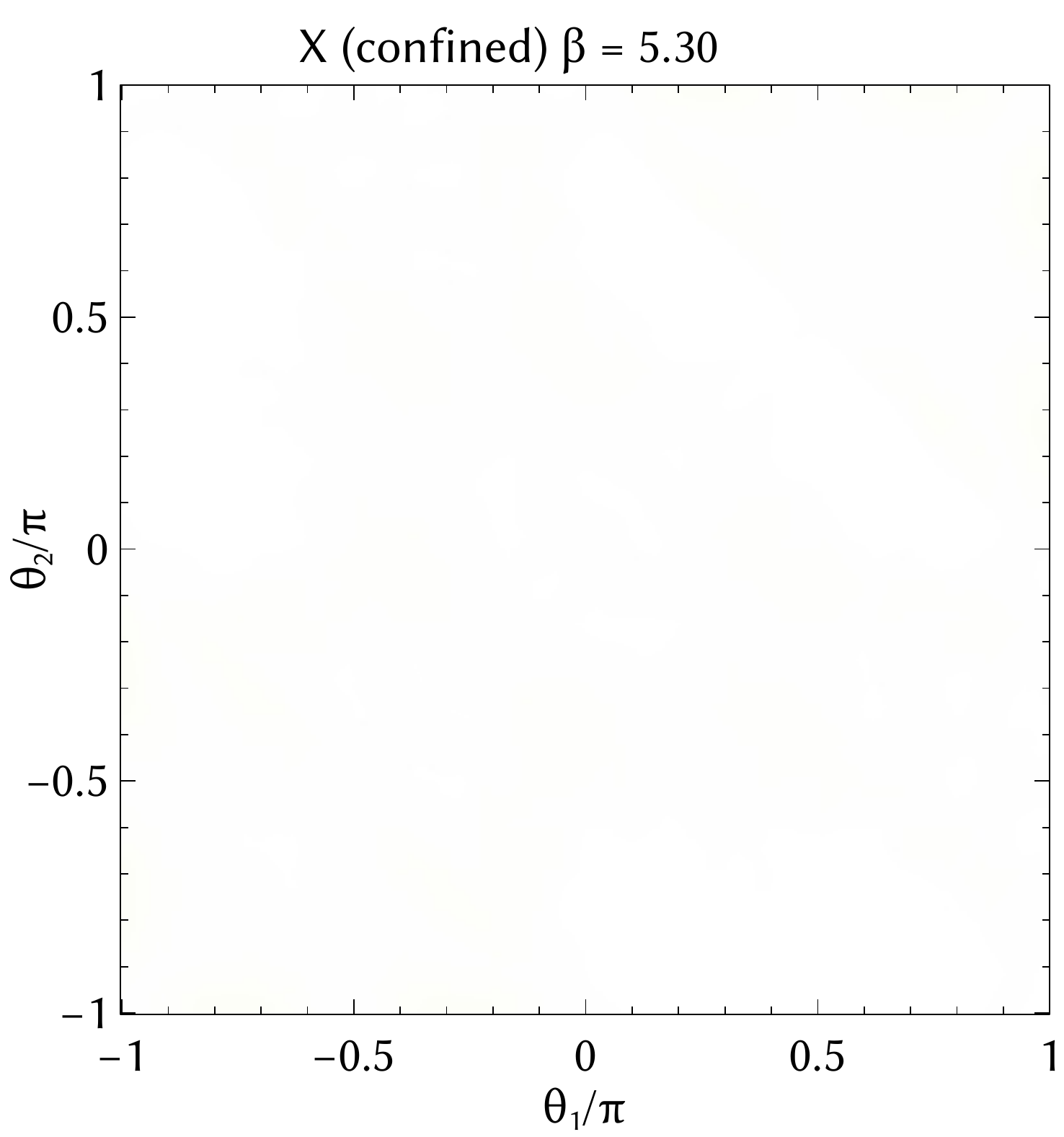}
  \includegraphics[clip=true, width=0.24\columnwidth]{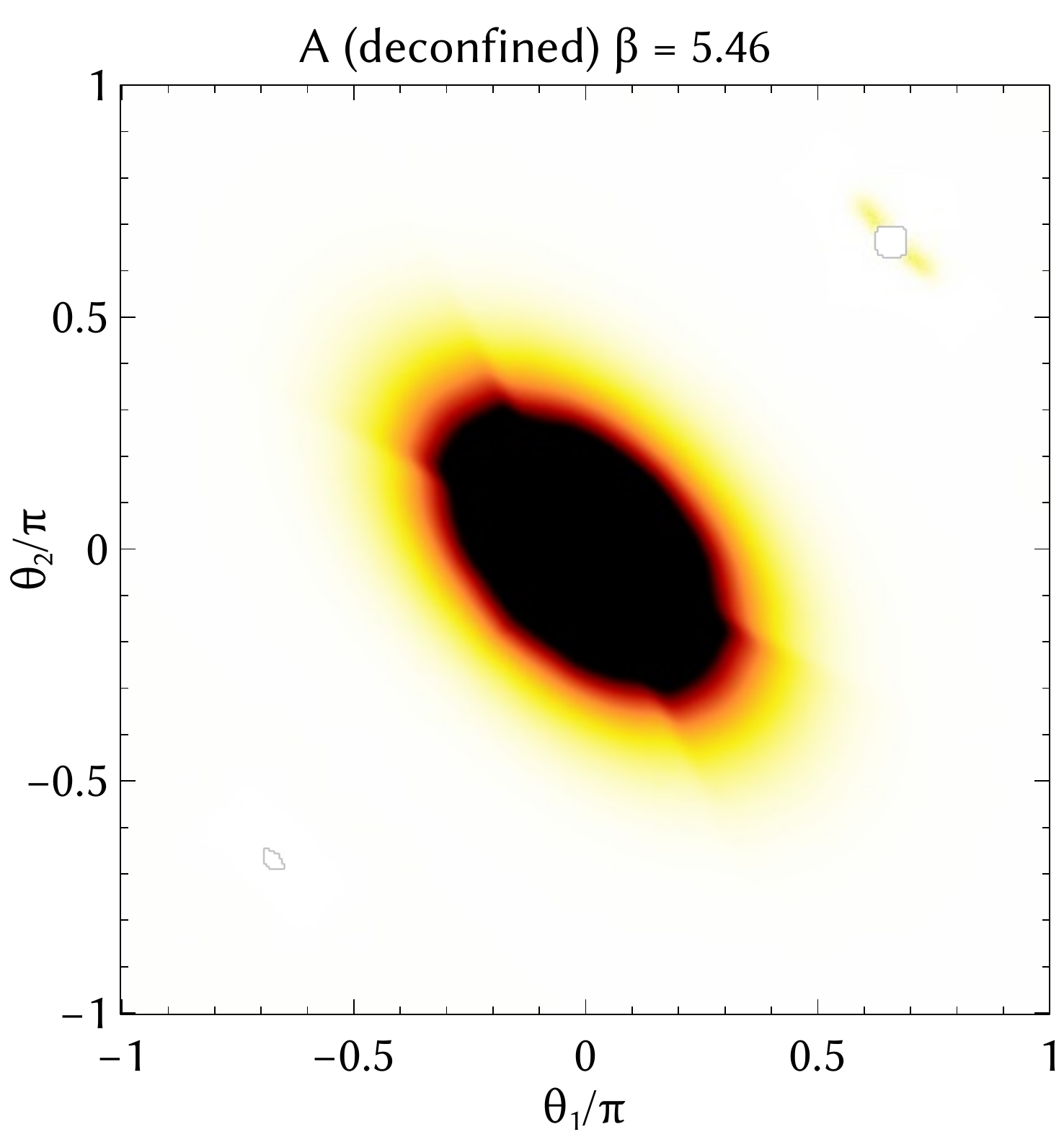}
  \includegraphics[clip=true, width=0.24\columnwidth]{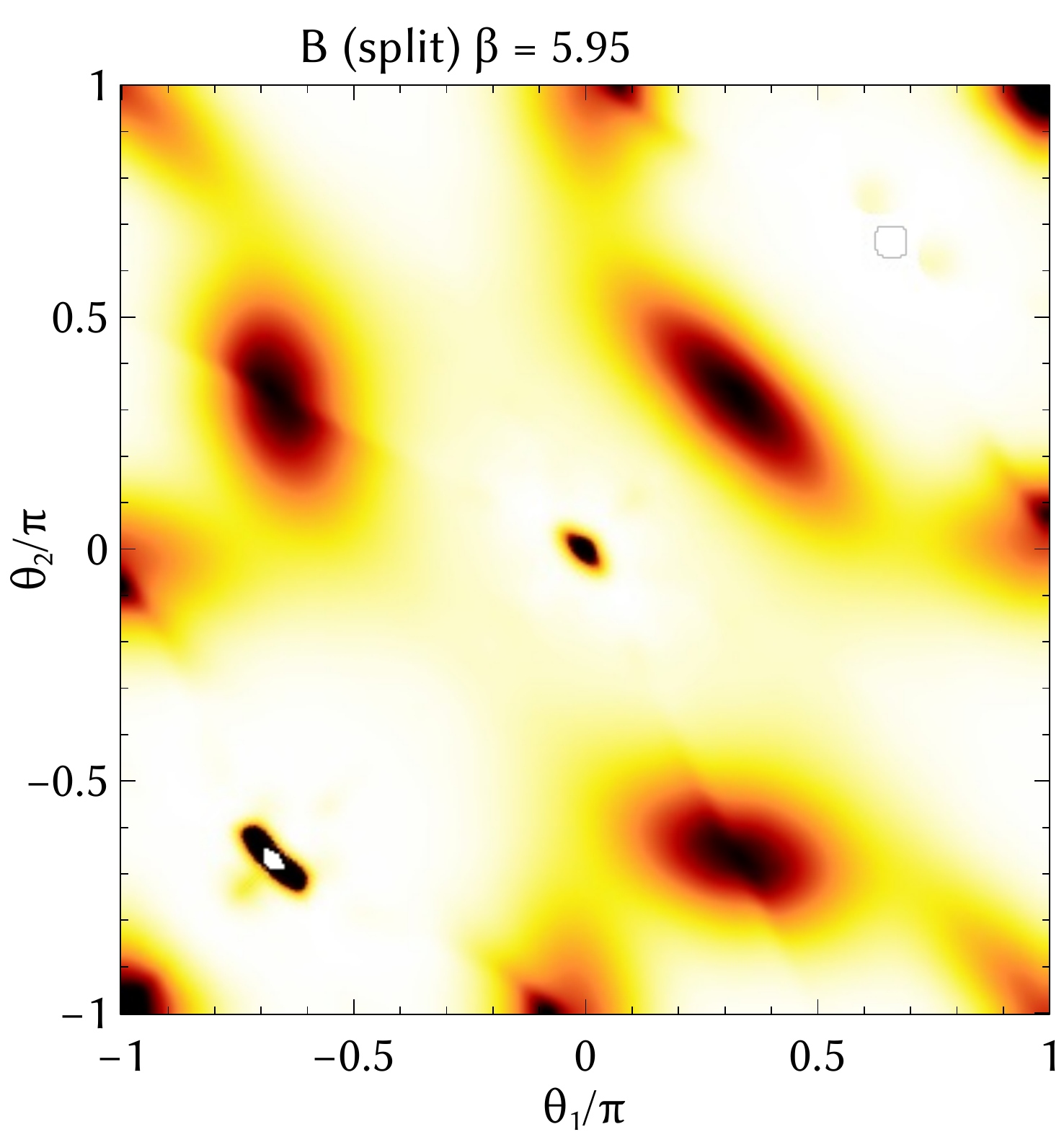}
  \includegraphics[clip=true, width=0.24\columnwidth]{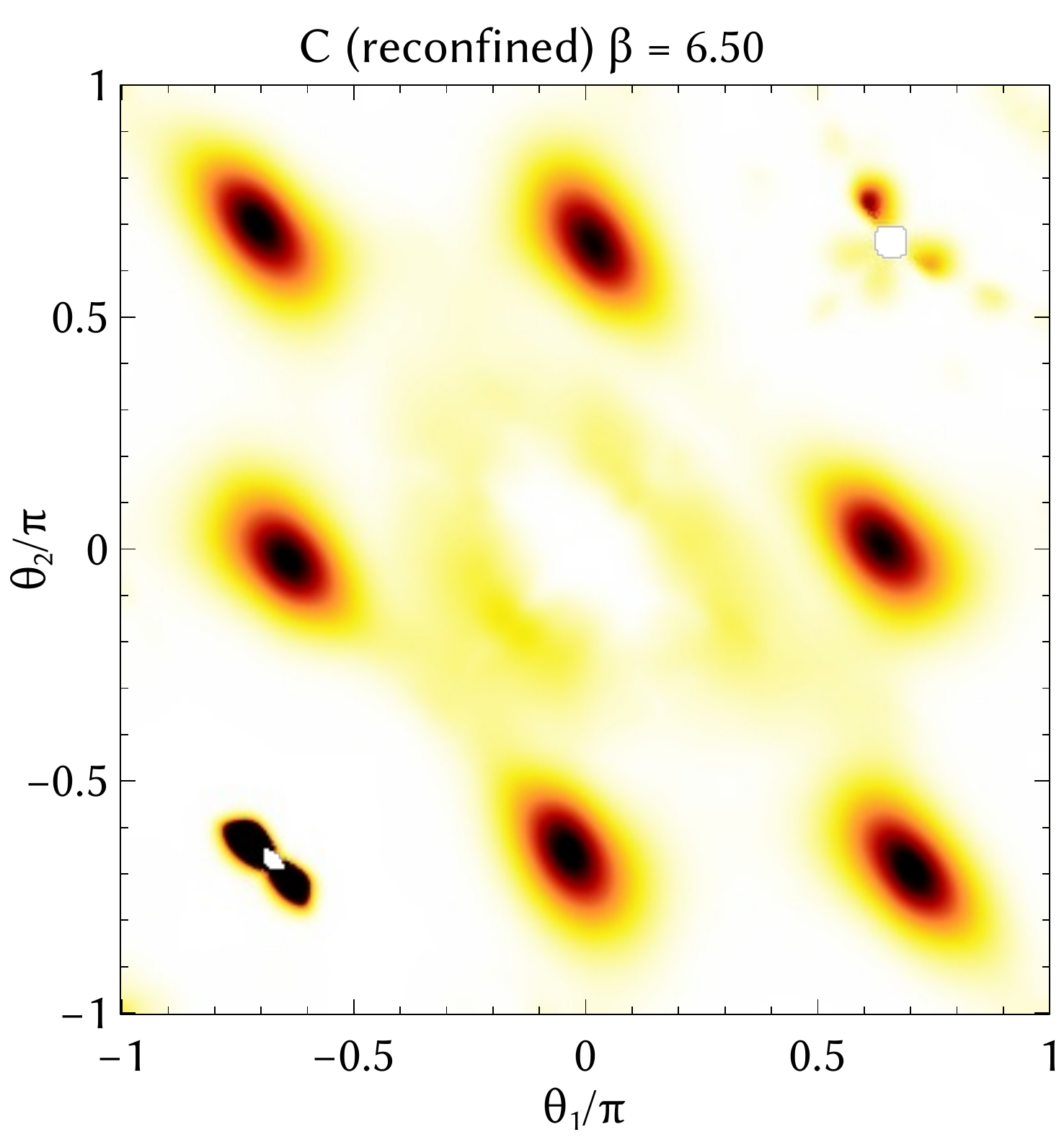}
 \caption{Density plots at several $\beta$'s for the Polyakov Loop
  eigenvalues (in the $\theta_1/\pi$ - $\theta_2/\pi$ plane), 
  as in Fig.~\ref{fig:density-plots}. 
  Here the original data is divided by a similar density plot of the 
  Haar measure distribution. 
  From left to right, the panels correspond to the $X$, $A$, $B$ and 
  $C$ phases.
  The first panel is white as a result of the calculation.
  Darker colors denote the highest density regions.}
 \label{fig:normalized-density-plots}
\end{figure}

To strengthen our view, we perform another analysis to 
eliminate the contribution of the Haar measure seen in 
Fig.~\ref{fig:Haar-measure} from the density plots. We generated a random 
ensemble of O($5\cdot 10^6$) eigenvalues distributed according to the 
Haar measure (Fig.~\ref{fig:Haar-measure}) and, using the same
normalization as the lattice data plots of Fig.~\ref{fig:density-plots}, 
it is now easy to isolate the effective potential
contribution from the kinetic term of the group measure. The result is
plotted in Fig.~\ref{fig:normalized-density-plots}.
The two-dimensional bins of the histogram have always a finite density 
of eigenvalues, somewhere very small, so that we are never dividing by zero. 
Although the procedure introduces some artifacts and more noise due to the
subtle cancellations caused by the Haar measure \footnote{The most problematic points are the minima of the Haar measure
  where the numerically calculated histogram has almost zero occupation number and a perfect cancelation is required to get the signal.
Around these points we can get divergent results. In the B and C phases these are essentially harmless since 
we know by theoretical arguments that there should be no signal there. It becomes more dangerous 
in the A phase where the peak of the distribution is poorly determined.}, it 
is useful from the qualitative point of view.
At this stage of the work we would like to show that the effective potential
has the features anticipated by the perturbative calculation and can be compared with 
Fig.~\ref{fig:ad-mass}. Notice that the density plots derived in this way are
proportional to $e^{- \int d^3 x \,  V_{\rm eff}}$. Location of the minima (maxima of density) 
is clearly not affected by this monotonic transformation.
The plots, from left to right, are respectively the $X$, $A$,
$B$, and $C$ phases ($\beta = 5.30, 5.46, 5.95$ and $6.50$). The distribution 
in the $X$ (confined) phase is almost a constant {\it i.e.} unity so the plot 
is a white image, which is a manifestation of a uniform random
distribution of the eigenvalues in the two dimensional plane. The $A$
and $B$ phases confirm the conjecture of Fig.~\ref{fig:density-plots}, 
{\it i.e.} that the discrepancy with perturbative prediction
is only due to the Haar measure contribution. 
In the region around $\theta_1 = \theta_2 = (0, 2\pi/3,-2\pi/3)$, 
the Haar measure density is close to zero in a wide area and very 
precise lattice data is needed in order to have a perfect cancellation, 
sampling also the low density regions, so we 
see some artifacts as a result.
As a further remark we underline that the $C$ phase shows a completely
different behavior from the confined one although the Polyakov loop
is centered around zero. The eigenvalues are now not
distributed in a random fashion but located in peaks around the $Z_3$
symmetric values $\theta_H = (0, 2\pi/3,-2\pi/3)$ (again some artifacts
appear), with maximal repulsion between them.
There have been recently studies~\cite{Poppitz:2012nz,AnberUnsal2013} using
semi-classical models that relate this phase to a weak coupling region where confinement is realized by
abelian degrees of freedom ($U(1) \times U(1)$ remaining). Future works
will be devoted to the quantitative tests of these ideas.
In conclusion, by removing the gauge group kinetic term from our data, although at
the price of introducing some noise, we are able to confirm the
nice agreement of lattice data with the perturbative potential.
All the four predicted phases are clearly reproduced by the data, which
is a strong indication of the realization of the Hosotani mechanism
in 3+1 dimensions even at the non-perturbative level.
For further confirmation of the Hosotani mechanism on the lattice we need a direct measurement of
the particle spectrum from lattice observables to see if the symmetry breaking 
is taking place.  This task is left for future investigation.  Here we would like to point out another
correspondence between the density plots of the Polyakov loop eigenvalues and
the scalar ($A_y$) mass spectrum in the continuum perturbation theory.

\begin{figure}[h]
 \centering
  \includegraphics[clip=true, width=0.49\textwidth]{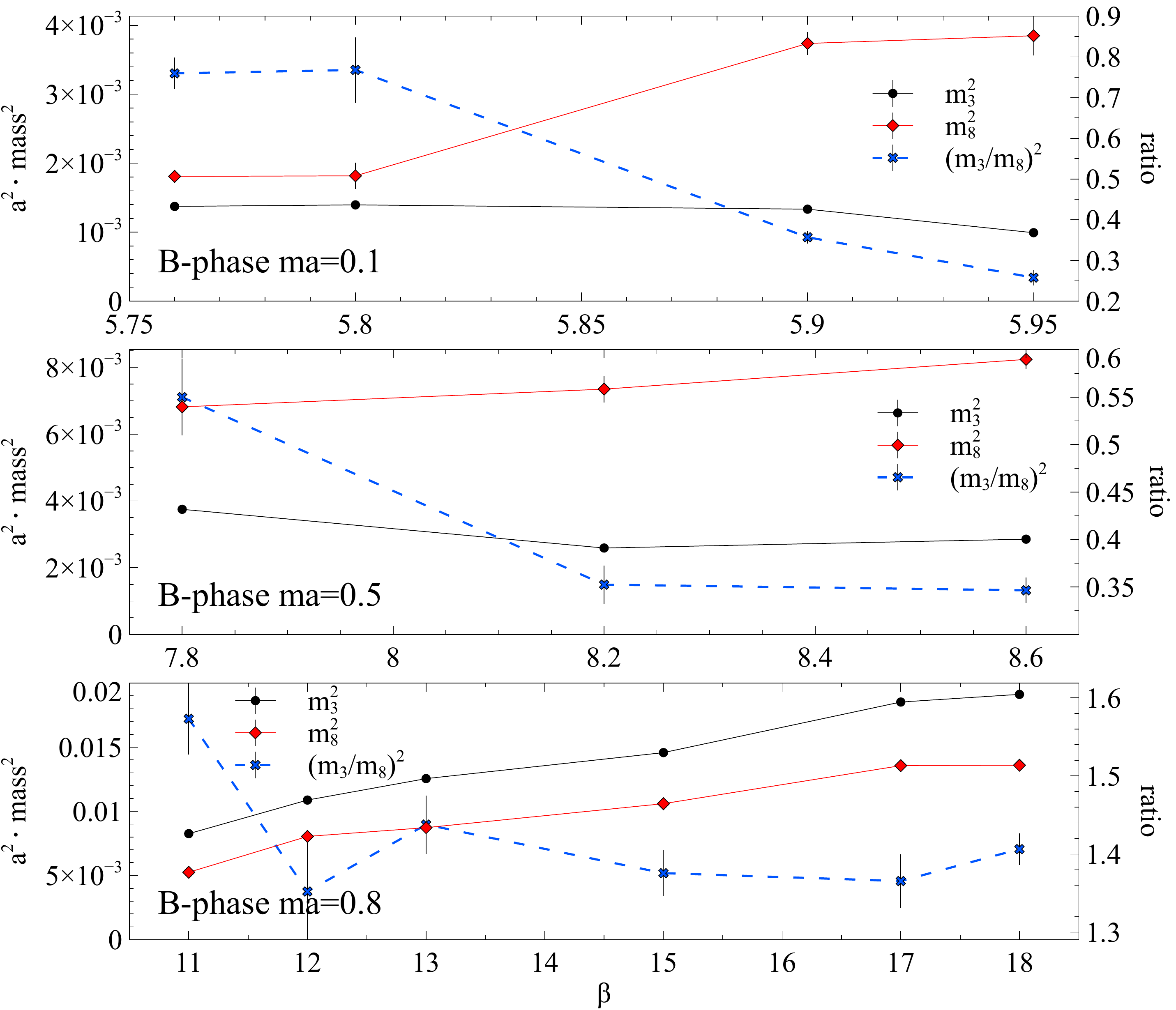}
  \includegraphics[clip=true, width=0.49\textwidth]{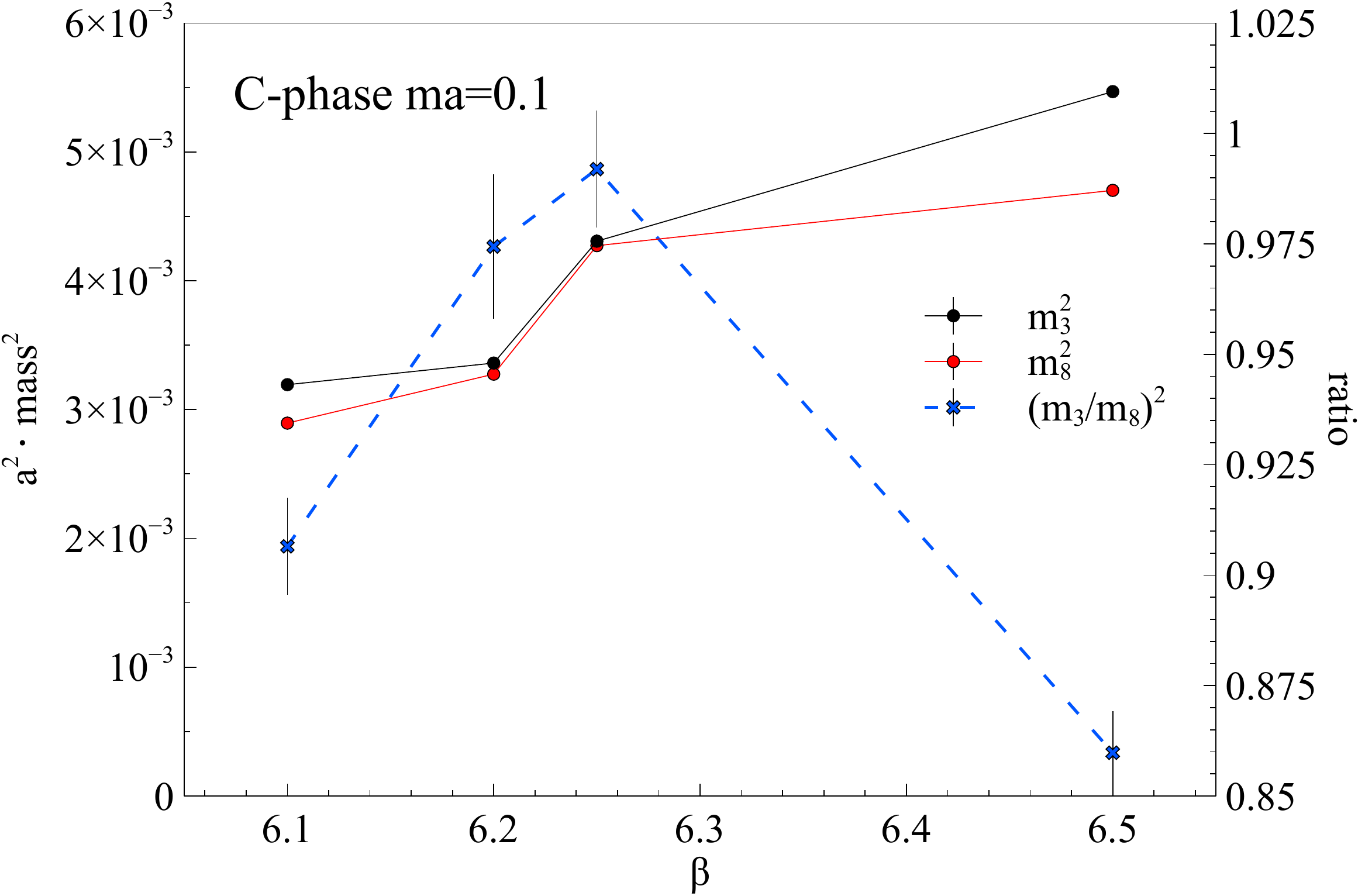}
 \caption{Results for the masses $m_3^2$ and $m_8^2$ in lattice units (left B-phase, right C-phase). The ratio is also plotted as blue line 
and the axis on the right indicates the scale}
 \label{fig:masses_m3_m8}
\end{figure}
From the normalized plots we can estimate the values of the
masses $m_3^2$ and $m_8^2$ of Sec.~\ref{sec:HiggsMass}
in each phase by fitting the peaks of the density plots to a Gaussian 
curve after converting the variables $(\theta_1,\theta_2)$ to 
$(\phi_3, \phi_8)$ along eq.~(\ref{Higgs1}). 
In the continuum theory the distribution density is controlled by  $e^{- \int d^3 x \, V_\eff (\phi)  }$
where  $V_\eff (\phi) \simeq 
\onehalf \{ m_3^2 (\phi_3- \phi_3^\min)^2 + m_8^2 (\phi_8 -\phi_8^\min)^2 \}$
around each minimum  of $V_\eff (\phi) $.  
With the $SU(3)$ invariance taken into account, the distribution density $\rho (\theta_1, \theta_2)$
is given by
\beqn
&&\hskip -1.cm
\int d\theta_1 d\theta_2 \, \rho (\theta_1, \theta_2)
= \int [dU] \, e^{- (vol) V_\eff (\theta_1, \theta_2)} \cr
\noalign{\kern 10pt}
&&\hskip -1.cm
= \int d\theta_1 d\theta_2 \,  \prod_{j<k} \sin^2 \onehalf (\theta_j - \theta_k) 
e^{- (vol) V_\eff (\theta_1, \theta_2)}  ~.
\label{densitydistribution}
\eeqn
In other words, $ \rho (\theta_1, \theta_2)$ divided by the Haar measure (\ref{eq:Haar-measure}) 
can be fitted with a Gaussian distribution around the minima of $V_\eff (\theta_1, \theta_2)$
(this assumes that the deviation from the Gaussian  are negligible around the maxima, 
and we estimated this  by the quality of the fits, always with $\chi^2 \sim 1$ and even smaller for 
many fitting points).
In this way we can obtain a qualitative comparison with the one-loop
results in eqs.~(\ref{HiggsM1}), (\ref{HiggsM2}) and (\ref{HiggsM3}).
The left panels of Fig.~\ref{fig:masses_m3_m8} 
show the result for B-phase. 
We observe the mass ratio $m_3^2/m_8^2$ deviates from $1$ and its
non-trivial dependence on the bare parameters:
 $m_3^2/m_8^2>1$ if $ma>0.5$ and $m_3^2/m_8^2<1$ for lighter masses. 
On the other hand, in the C-phase, as seen in the right panel of the
figure, we obtain the degenerate mass 
$m_3^2(\simeq m_8^2)$ which is increasing for $mR\rightarrow 0$ ($\beta\rightarrow
\infty$) as expected. For the A-phase, although the non-perfect cancellation
obscures a clear peak, we can fairly conclude that $m_3 \simeq m_8$ from 
the tails of the distributions. 
From the one-loop calculation, one expect that 
the mass ratio should cross in the B-phase passing from the A/B
transition to the B/C transition. 
We could not find a direct evidence of this crossing with our current
data but we observe an inversion of the ordering in the highest mass
region. 
In summary, we find a match of the mass (non)degeneracy pattern to
the perturbative prediction in this analysis.

\subsection{Phase structure with fundamental fermions}
\label{FundamentalSection}

As a further test of the perturbative prediction in
Sec.~\ref{sec:PertResults}, we study 
the dependence of $P_3$ and $P_8$ on the boundary phase $\alpha_\fund$ 
for several values of $\beta$ in the presence of fundamental fermions. 
As explained in
Sec.~\ref{LatticeGeneral}, we introduce $\alpha_\fund$ through 
the boundary condition (\ref{alpha_lattice}).
This setup is formally equivalent to finite
temperature QCD with an imaginary chemical potential $\nu= \pi
+\alpha_\fund$.
Roberge and Weiss~\cite{Roberge1986} have already shown that  
the corresponding partition function with SU($N$) gauge
symmetry is periodic in $\nu$ as 
\begin{eqnarray}
  Z(\nu)=Z(\nu+2\pi /N)
\end{eqnarray}
and there are discontinuities (first order lines) at $\nu =2n\pi/N$ with 
$n=0,\cdots, N-1$.
These discontinuities exist in a region of high $\beta$
down to some endpoints which require a non-perturbative study 
to be located. 
Several numerical simulations on the lattice, for example~\cite{Forcrand2002,dElia2003},
have determined these points as well as the phase structure 
in the $(\beta,\alpha_\fund)$ plane, {\it i.e.} the transition lines
which were associated with chiral phase transitions and the breaking of
the approximate $Z_3$ symmetry for $\alpha_\fund$.

We carry out a numerical simulation with $(N_{\rm ad},N_\fund) = (0,4)$.
The basic setup is the same as in ref.~\cite{dElia2003} 
except for the bare fermion mass being fixed to $ma=0.10$ in our case.
Since we are interested in the symmetries of the Polyakov loop,
we determine the locations of the transition points 
by the peak points of $\chi_{|P_3|}$.
Other technical matters related to this simulation are
briefly summarized in Appendix~\ref{TechnicalMatter}.

\begin{figure}[t]
 \centering
  \includegraphics[width=0.335\textwidth,clip]{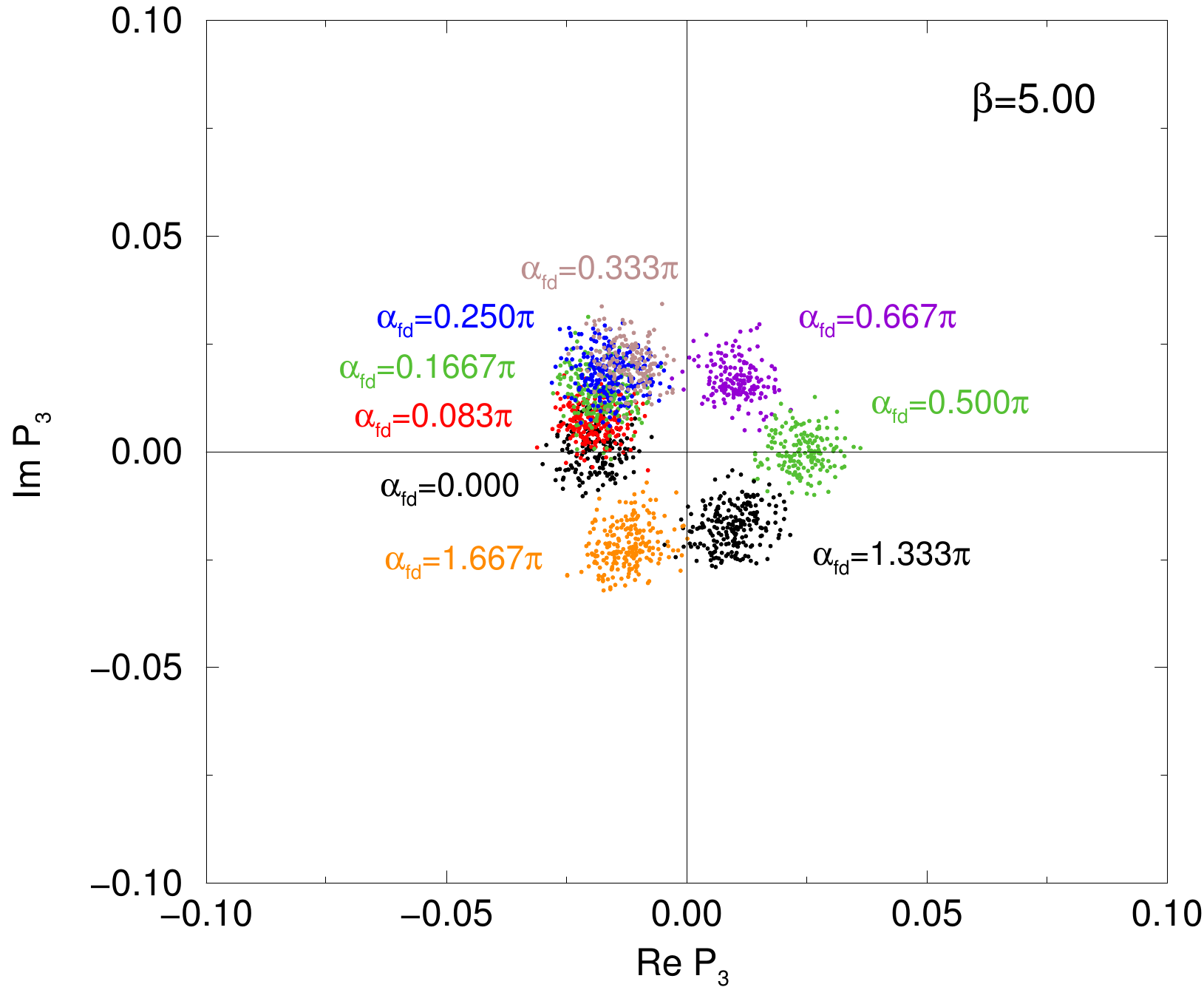}
  \includegraphics[width=0.3\textwidth,clip]{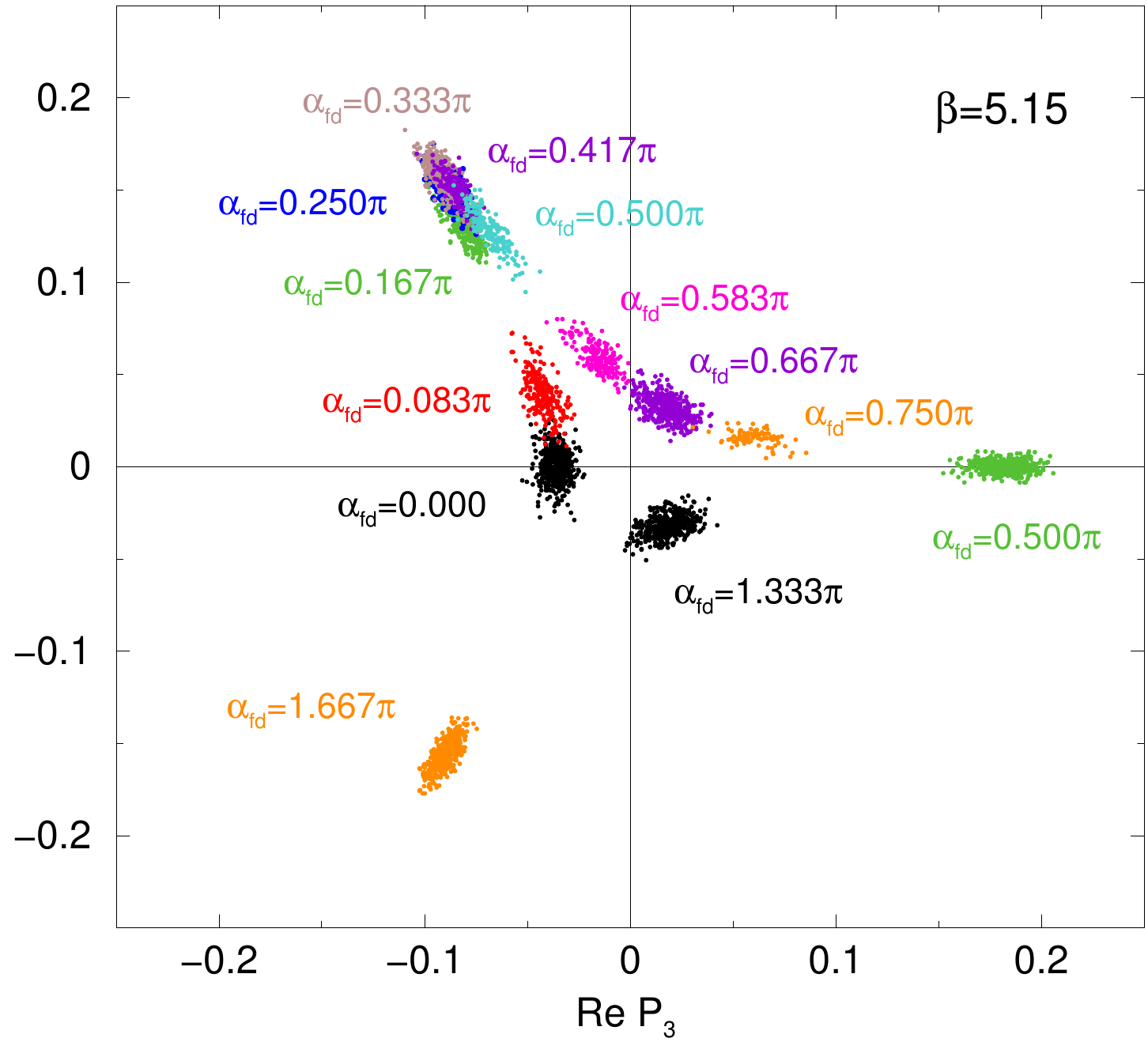}
  \includegraphics[width=0.3\textwidth,clip]{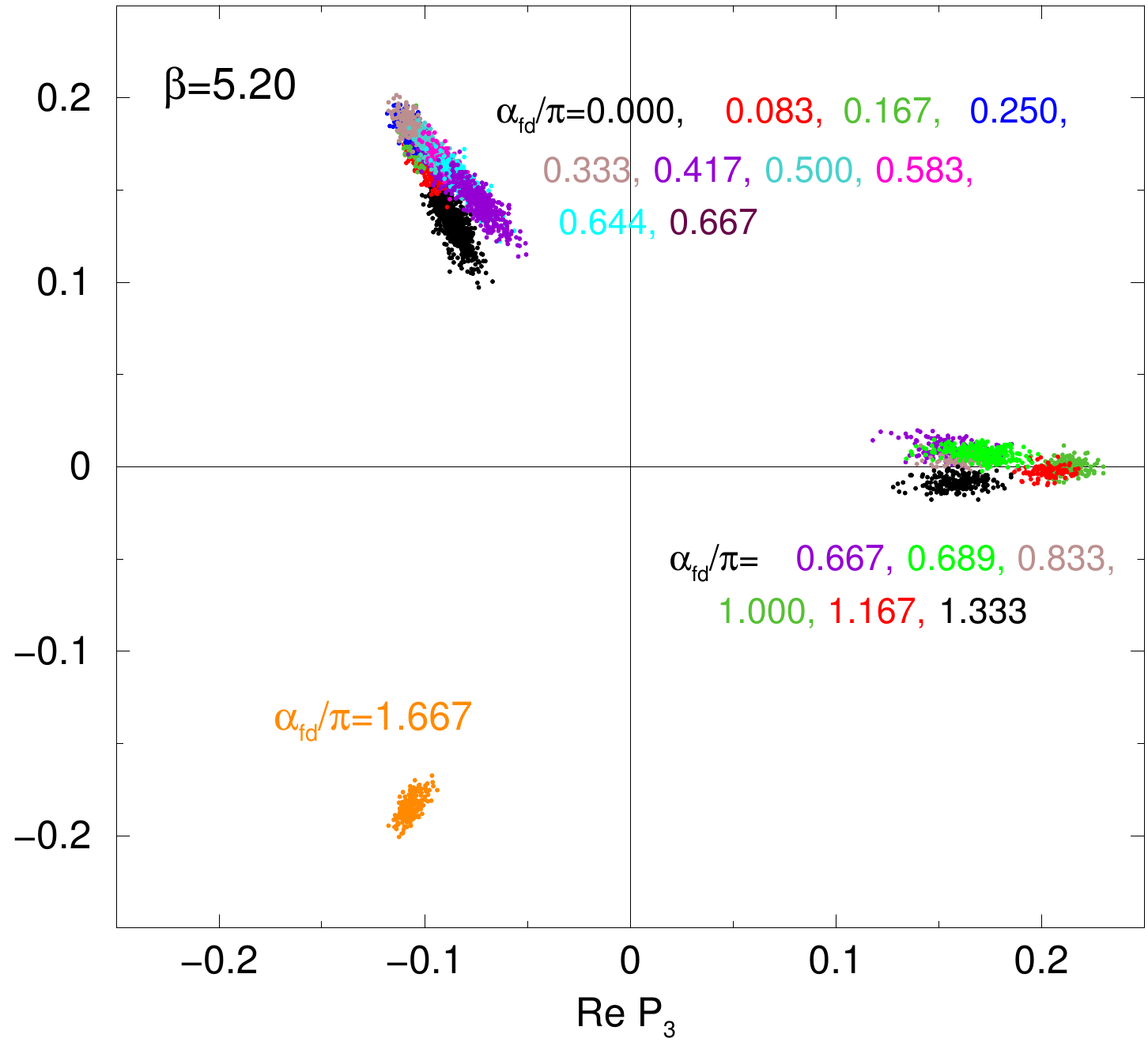}
  \caption{Distributions of $P_3$ obtained on gauge ensembles
    with a variation of $\alpha_\fund$ for $\beta=5.00$ (left), 
    $5.15$ (center) and $5.20$ (right). The degrees of $\alpha_\fund$ used in
    the calculation are indicated with the corresponding data. 
    Overall, data points with same degrees of $\alpha_\fund$ are 
    indicated by same colors.}
  \label{fdm_deconfined}
\end{figure}

\begin{figure}[t] 
 \centering
  \includegraphics[width=.4\columnwidth,clip]{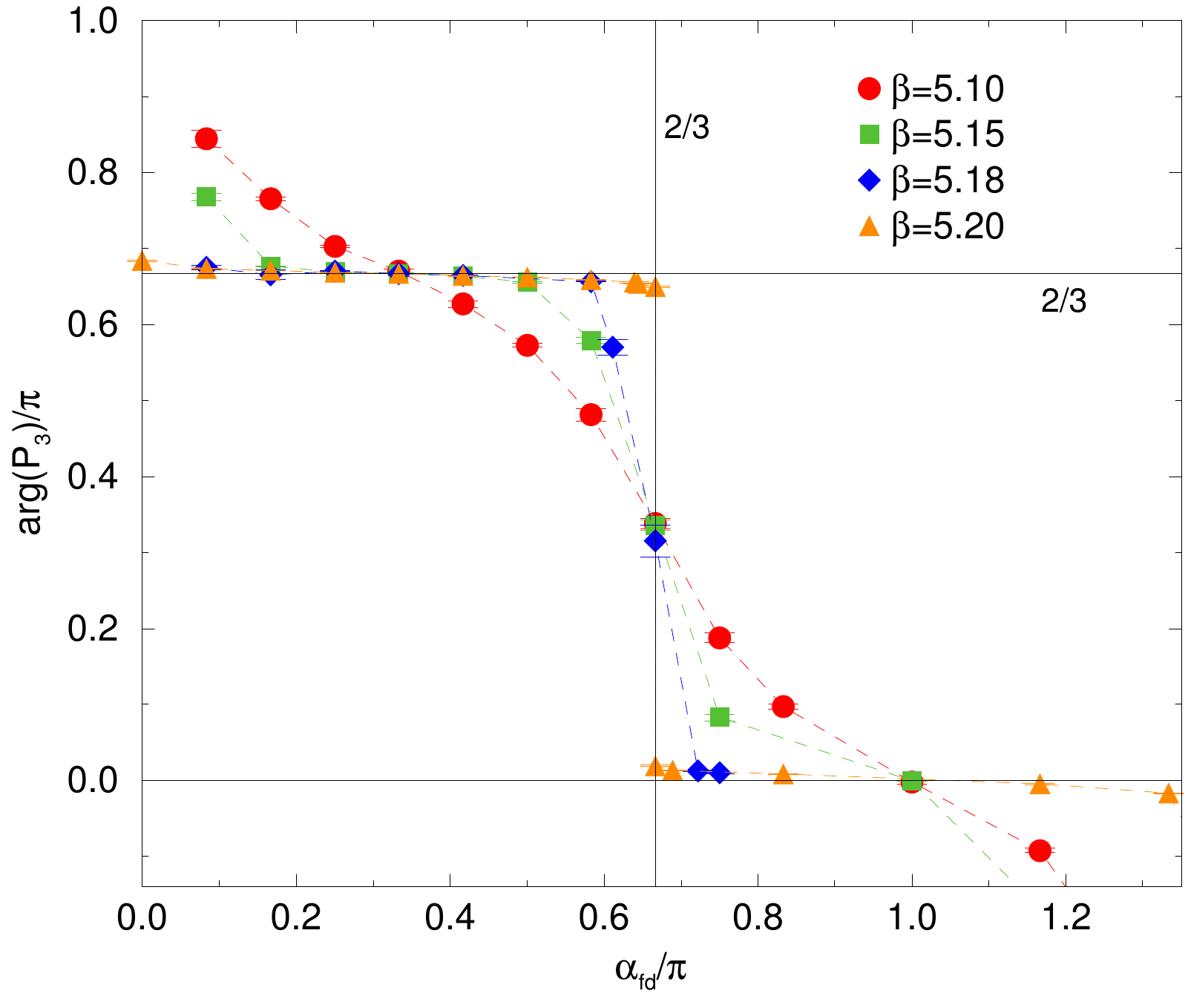}
  \includegraphics[width=.45\columnwidth,clip]{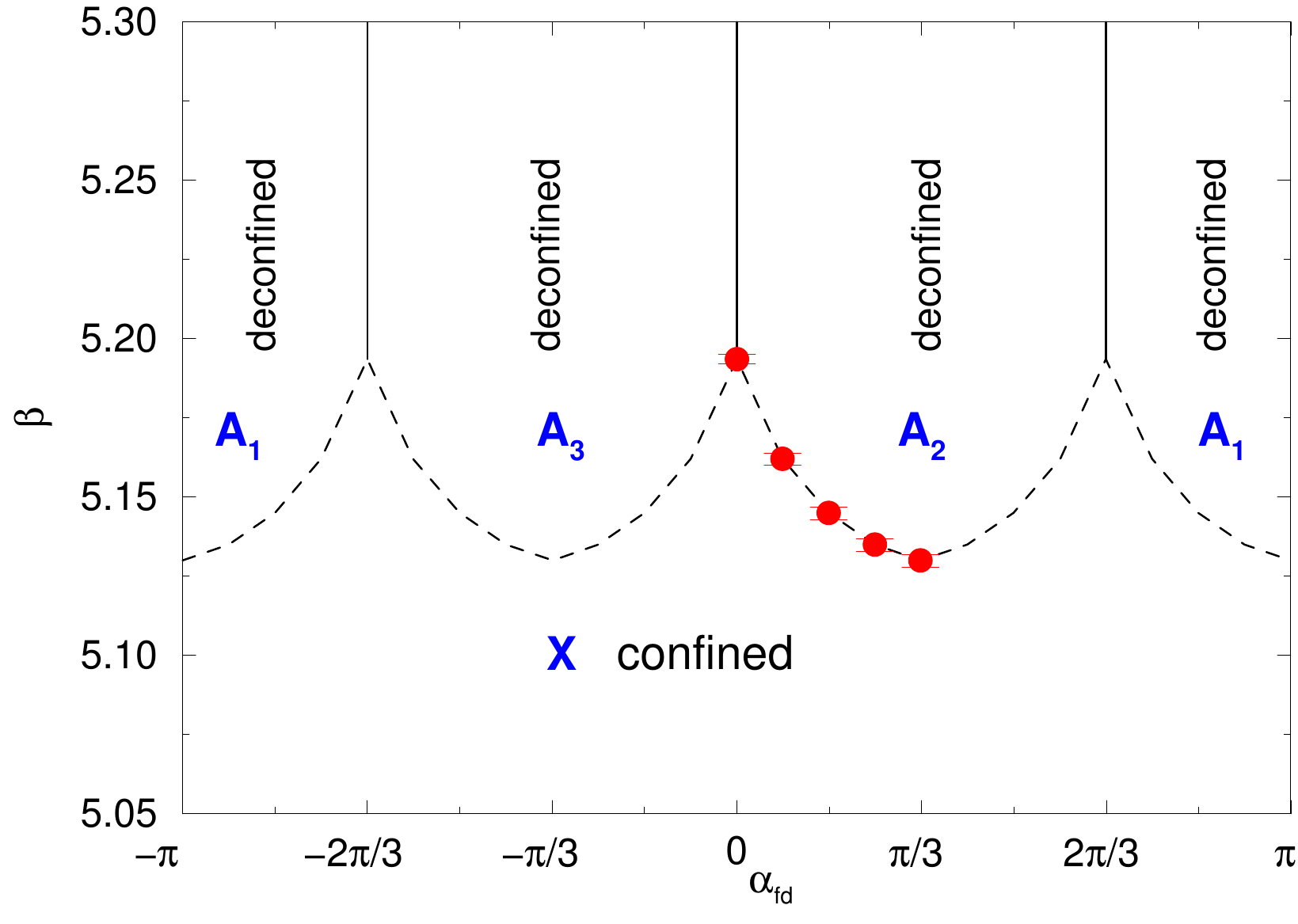}
  \caption{(Left) ${\rm arg}(P_3)$ as a function of $\alpha_\fund$ with various values of $\beta$.
  (Right) Phase diagram of the $N_\fund=4$ fundamental fermion case. 
    Solid lines indicate discontinuity separating the $A$ phase 
    in the period of $2\pi/3$. Dashed lines are the critical line drawn by
    connecting the actual data points (red squares) and its copy based on 
    the symmetry and periodicity explained in the text.}
  \label{fig:fdm_theta_va}
\end{figure}

We compute the Polyakov loop in a region of
the $(\beta,\alpha_\fund)$ plane which covers the known phase structure.
The resulting distributions of $P_3$ are shown in
Fig.~\ref{fdm_deconfined} for 
$\beta=5.00$ (left), $5.15$ (center) and $5.20$ (right). 
In each panel, we cover the range of $\alpha_\fund$ from $0$ to $5\pi/3$.
For $\beta=5.00$, the sizable shift of the data from the origin is 
caused by the non-zero value of $(ma)^{-1}$, which breaks the 
${\rm Z}_3$ symmetry. We observe a continuous change of  
${P_3}$ as a function of $\alpha_\fund$ in this case.
This behavior does not change even at $\beta=5.15$. 
On the other hand, the discontinuity of $P_3$ 
around $\alpha_\fund=2\pi/3$ is clearly visible for $\beta=5.20$. 
In particular, we find that the data at $\alpha_\fund=2\pi/3$ is
in the $A_1$ phase or the $A_2$ depending on the initial configuration 
in HMC. This is the indication of the non-analyticity of $\theta$.
For a better illustration of this behavior, in Fig.~\ref{fig:fdm_theta_va}, 
we plot the phase ${\rm arg}(P_3)$ as a function of $\alpha_\fund$.
As seen in the figure, the transition from a continuous 
behavior to a discontinuous one around $\alpha_\fund=2\pi/3$ 
becomes more evident for increasing $\beta$.
From the location of the peaks of $\chi_{P_3}$ we can draw the phase diagram of Fig.~\ref{fig:fdm_theta_va}.
We note that our data do not differ significantly
from the results of {\it e.g.} ref.~\cite{dElia2003} with $ma =0.05$.
It suggests that no significant mass dependence of the phase structure
is expected.
Because the perturbative region is realized at large $\beta$,
there would be a split of phases into three classes $A_1$,$A_2$ and $A_3$
as described in eq.~(\ref{VeffResultsFdm}) and Figs.~\ref{fig:fd-bc} 
and~\ref{fig:f-mass}.

\section{Discussion}\label{Discussion}
In this paper we explored the Hosotani mechanism of symmetry breaking 
in the SU(3) gauge theory on the  $16^3 \times 4$ lattice. 
The Polyakov loop, its eigenvalue phases, and the susceptibility were measured
and analyzed  in models with periodic adjoint fermions and with fundamental fermions 
with general boundary conditions.
Among the four phases appearing in the SU(3) model with adjoint fermions~\cite{Cossu},
the $A$, $B$, and $C$ phases are interpreted as the SU(3), 
SU(2)$ \times $U(1), and U(1)$ \times $U(1) phases classified from the location of the 
global minimum of the effective potential of the AB phases.  
We confirmed natural correspondence between the effective potential evaluated
in perturbation theory on $R^3 \times S^1$ and the distribution of phases of eigenvalues
of the Polyakov loop in the lattice simulations.
The correspondence was seen in the model with fundamental fermions with varying 
boundary conditions as well.

The next issue to be settled is the particle spectrum.  If the SU(3) symmetry is broken,
asymmetry in the particle spectrum must show up in two-point correlation functions
of appropriate operators.   This can be explored in 4D lattice simulations.
It is important from the viewpoint of phenomenological applications.  We would like
to explain why and how the $W$ and $Z$ bosons become massive in the scheme of
the Hosotani mechanism non-perturbatively.

In the end, we would need a five-dimensional simulation to have realistic gauge-Higgs 
unification models of electroweak interactions. The continuum limit of the 5D lattice
gauge theory has been under debate in the literature.   
Furthermore, we will need chiral fermions in four dimensions.  
It is customary to start from gauge theory on orbifolds in
phenomenology, however.  
Lattice gauge theory on orbifolds needs further refinement.
We would like to come back to these issues in future.

\acknowledgments

We thank E. Itou for the great contribution in all phases of this work.
We would like to thank J.~E.~Hetrick for his enlightening comment which 
prompted us to explore the Hosotani mechanism on the lattice.
We also thank M.~D'Elia  and Y.~Shimizu for providing helpful information, 
H.~Matsufuru for his help in developing the simulation code, and 
K.~J.~Juge for careful reading of the draft.
Numerical simulation was carried out on Hitachi SR16000 at YITP, Kyoto University,
and Hitachi SR16000 and IBM System Blue Gene Solution at KEK 
under its Large-Scale Simulation Program (No.~T12-09 and 12/13-23).
This work was supported in part 
by scientific grants from the Ministry of Education and Science, 
Grants No.\ 20244028, No.\ 23104009 and  No.\ 21244036.
G.~C and J.~N are supported in part by
Strategic Programs for Innovative Research (SPIRE) Field 5.
H.~H is partly supported by NRF Research Grant 2012R1A2A1A01006053 (HH) of 
the Republic of Korea.

\appendix
\section{Useful formulas for $V_\eff (\theta)$}\label{UsefulFormulae}
We need to evaluate the following building block
\begin{eqnarray}
V(\theta, m) &\equiv& \frac{1}{2} \sum_{n=-\infty}^{\infty} 
\int \frac{d^{d-1}p}{(2\pi)^{d-1}} \ln \left[p^2 +
 \frac{1}{R^2} \left(n + \frac{\theta}{2\pi} \right)^2
 + m^2
\right],
\end{eqnarray}
in terms of which $V_\eff$ is written as in Eq.~(\ref{effV2b}).
We use the technique of the zeta regularization to write
$V(\theta,m) = - \zeta'(0)/2$,
where $\zeta(s)$ is the generalized zeta function defined by
\begin{eqnarray}
\zeta(s) &=& \frac{1}{\Gamma(s)} 
\sum_{n=-\infty}^\infty \int_0^\infty dt\, t^{s-1} 
\nonumber\\&& 
\times \int \frac{d^{d-1}p}{(2\pi)^{d-1}}
\exp 
\left\{ - t\left[p^2 +
	    \frac{1}{R^2} \left(n + \frac{\theta}{2\pi} \right)^2
	    + m^2\right] 
\right\}.
\end{eqnarray}
Here $\Gamma(s)$ is the Gamma function. 
Performing integration with $p$ and using  Poisson's summation formula,
\begin{eqnarray}
\sum_{n=-\infty}^{\infty} \exp\left[-t \left( \frac{2\pi n + \theta}{2\pi R}\right)^2 \right]
&=& \frac{2\pi R}{\sqrt{4\pi t}} \sum_{\ell =-\infty}^{\infty}
\exp\left(- \frac{(2\pi R)^2 \ell^2}{4t} + i \ell \theta\right),
\end{eqnarray}
we obtain
\begin{eqnarray}
\zeta(s) &=& \frac{\pi^{\frac{d-1}{2}}}{(2\pi)^{d-1}\Gamma(s)}
\frac{2\pi R}{\sqrt{4\pi}}
 \sum_{n=-\infty}^{\infty}
e^{in\theta}
\int_0^\infty dt \, t^{s - \frac{d}{2} -1} \exp\left[- \frac{(2\pi R)^2 n^2}{4t} - t m^2\right],
\end{eqnarray}
The $n=0$ part, though divergent, can be dropped  as it is $\theta$-independent.

Using a formula
\begin{eqnarray}
\int_0^\infty dt \, t^{-\nu-1} \exp\left[-t m^2 - \frac{(2\pi R)^2 n^2}{4t}\right]
&=& 
\frac{2^{1+\nu}}{(2\pi R)^{2\nu} n^{2\nu}} (2\pi R n m)^{\nu} K_{\nu}(2 \pi R n m),
\end{eqnarray}
and taking $s\to0$ limit for $\zeta'(s)$,
we obtain
\begin{eqnarray}
V(\theta,m) &=& - \frac{1}{2^{\frac{d}{2}-1}\pi^{d/2}(2\pi R)^{d-1}}
\sum_{n=1}^\infty \frac{\cos n\theta - 1}{n^d} \tilde{B}_{d/2}(2 \pi R n m).
\end{eqnarray}
where $\tilde{B}_{\delta}(x) \equiv x^\delta K_\delta(x)$. 
Normalizing by a factor $\tilde{B}_{d/2}(0) \equiv \lim_{x\to0} \tilde{B}_{d/2}(x) = 2^{\frac{d}{2}-1}\Gamma(d/2)$,
we finally obtain
\begin{eqnarray}
V(\theta,m) &=& - \frac{\Gamma(d/2)}{\pi^{d/2}(2\pi R)^{d-1}}
\sum_{n=1}^\infty \frac{\cos n\theta -1 }{n^d} B_{d/2}(2 \pi R n m) 
\label{effV5}
\end{eqnarray}
which has been used in (\ref{effV3}).


It would be useful to give another expression of $V(\theta,m)$, which
has been obtained in Refs. \cite{HH2011, Falkowski2007}; 
\beeq
V(\theta, m)
= \myfrac{1}{\Gamma(\tfrac{d-1}{2}) (4\pi)^{\frac{d-1}{2}} R^{d-1}}
\int_{mR}^\infty dt \, t \left(t^2 - (mR)^2\right)^{\frac{d-3}{2}} \ln \left[1 + \frac{\sin^2(\theta/2)}{\sinh^2(\pi t)}\right].
\label{effV4}
\eneq
In both expressions (\ref{effV5}) and (\ref{effV4}), the $\theta$-independent
constants have been chosen such that $V(0, m) =0$.

\section{Match to the weak coupling regime}

We also check the agreement of the eigenvalue distribution of the Wilson line
with the perturbative prediction in weak coupling regime, 
$\beta\rightarrow +\infty$ (that implies a compactification radius shrinking
to zero as well as the 3d volume -- with a ratio of 4 in our case), for $N_\fund=4$ fundamental fermions and $N_{\rm ad}=2$
adjoint fermions with periodic boundary conditions.
Based on the analysis of Sec.~\ref{sec:sym-breaking}, the perturbative 
predictions for $\Delta\theta_j$'s are
\begin{itemize}
\item Fundamental fermions with periodic boundary condition\\
In this case, the gauge symmetry is never broken. 
The true vacua of the effective potential are realized at 
$({\theta}_1,{\theta_2},{\theta_3})=(-\frac{2\pi}{3},-\frac{2\pi}{3},-\frac{2\pi}{3})$  
or $(\frac{2\pi}{3},\frac{2\pi}{3},\frac{2\pi}{3})$, see Fig.~\ref{fig:fd-bc}.  
In both vacua, the angle of the Wilson line phase ($\Delta \theta_i$) is  
\beq
(\Delta \theta_1, \Delta \theta_2, \Delta \theta_3)&=& (0,0,0).
\eeq

\item Adjoint fermions with periodic boundary condition\\
In this case, the SU($3$) gauge symmetry is broken to U($1$) $\times$ U($1$). 
The true vacua are realized at 
$({\theta}_1,{\theta_2},{\theta_3})=(0,\frac{2\pi}{3},-\frac{2\pi}{3})$ and 
the permutations, thus all eigenvalues of the Wilson line are not degenerate.
We expect that
\beq
(\Delta \theta_1, \Delta \theta_2, \Delta \theta_3)&=&
      (\tfrac{2\pi}{3},\tfrac{2\pi}{3}, \tfrac{4\pi}{3}), \ \ 
(0\le\Delta\theta_j < 2\pi) \label{DeltaThetaU1U1}
\eeq
with all possible permutations. In any case the peaks of the distribution are expected at $\pm \frac{2\pi}{3}$.
\end{itemize}

\begin{figure}[t]
 \centering
  \includegraphics[clip=true, width=0.68\columnwidth]{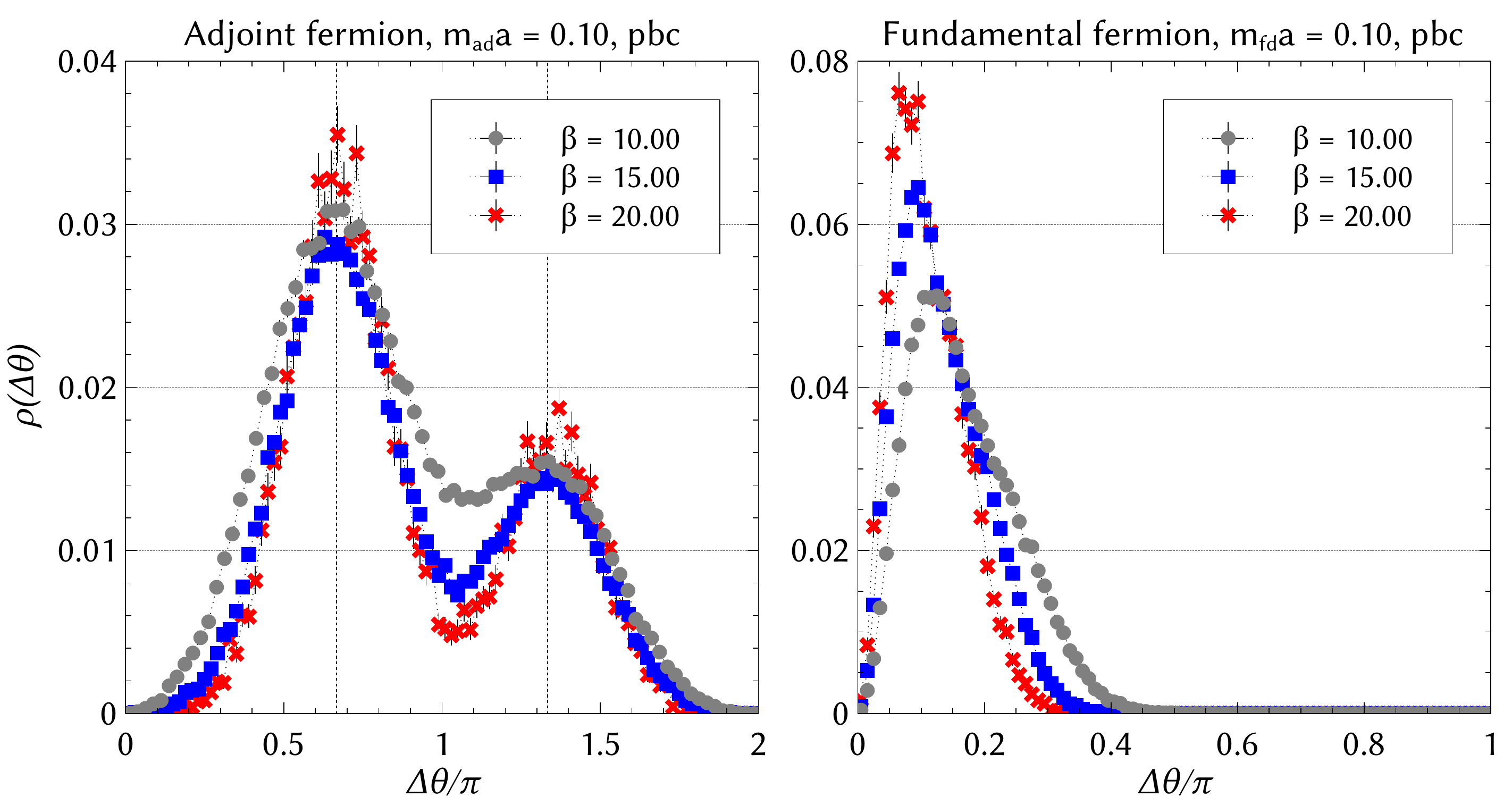}
  \caption{Density plots of $\Delta \theta_i$ ($i = 1,2,3$) for $ma=0.10$ with the periodic 
    boundary condition for adjoint and fundamental fermions, respectively. 
    In the vertical axis the total density fraction is reported. For
    the readability, in the adjoint fermions case, we draw two lines to
    identify the expected peak locations $\Delta\theta_i =2\pi/3$ and $4\pi/3$. }
  \label{fig:hist-dtheta-adj-fund}
\end{figure}

Figure~\ref{fig:hist-dtheta-adj-fund} shows the histograms 
of $\Delta \theta_1, \Delta \theta_2$ and $\Delta \theta_3$ densities.
The left and right panels show the results in the case of adjoint and
fundamental fermions, respectively.
In both cases, the fermion bare mass is fixed at $ma=0.10$ and the
boundary condition is periodic. 
We change the value of $\beta=10.0,15.0$ and $20.0$.

In the case of the presence of adjoint fermions, the distribution drift towards the double 
peaks expected around $2\pi/3$ and $4\pi/3$ with the ratio 2 to 1 as
indicated in (\ref{DeltaThetaU1U1}). 

In the case of fundamental fermions we find the position of the peaks 
approaches $\Delta \theta_i = 0$, including the effect of the Haar measure that suppresses
exactly the $\Delta \theta_i \simeq 0$ configurations.
The exponential shrinking of the 3d volume does not seem to affect the matching of our results with 
the 1-loop prediction.

\section{Lattice technical details}\label{TechnicalMatter}

In this section we describe some details of our lattice simulations. 

The algorithm to generate the configurations is the Hybrid Monte Carlo (HMC)~\cite{Duane1987,Gottlieb1987} for both adjoint and fundamental fermions.
We obtain a new configuration by $N_{MD}$ steps of evolution of the molecular dynamics trajectory of size $\tau = 1$.

In the case of adjoint fermions, the molecular dynamics integrator is the 
Omelyan~\cite{PhysRevE.65.056706} with Hasenbush 
preconditioning~\cite{Hasenbusch:2001ne}. 
Using $N_{\rm MD}=25$ and $\Delta \tau=0.04$,
we obtain an acceptance rate greater than 90\%. 
We accumulate 4,000 - 14,000 trajectories depending on the parameter set 
$(ma,\beta)$.
For the fundamental fermions case, by setting $N_{\rm MD}=50$ and $\Delta
\tau=0.02$ in the standard leapfrog integrator, we obtain acceptance 
rate $\mor 80\%$. Depending on the value of $(\beta,\alpha_\fund)$, 
we accumulate 2,500 - 110,000 trajectories depending on the significance
of signal.

The observables  $P_3$ and $P_8$ are computed every 10 generated 
configurations. For the error analysis, we employ the unbiased 
jackknife method in both cases.
As a reference we report the values of some plaquette values for few
values of $\beta$ and $ma$ in Table~\ref{table:reference}.

\begin{table}[bth]
\begin{center}
\renewcommand{\arraystretch}{1.4}
\caption{Plaquette and $|P_3|$ for some selected $\beta$ and $ma$ values
 for reference. Left: adjoint fermion case, Right: fundamental 
fermion case with $\alpha_{\rm fd}=0$.\label{table:reference}}
\vspace{0.5cm}
\begin{minipage}{7cm}
\begin{tabular}{|c|c|l|l|}
\hline
$\beta$ & $ma$ & plaquette & $|P_3|\times 10^2$ \\
\hline
5.30 & 0.1 & 0.5087(5) &  \ 0.88(4)\\
5.46 & 0.1 & 0.5671(1) &  13.4(2)\\
5.75 & 0.1 & 0.6081(1) & \  7.9(3)\\
5.95 & 0.1 & 0.6279(1) & \  7.1(4)\\
6.00 & 0.1 & 0.6325(1) & \  6.1(6)\\
6.50 & 0.1 & 0.6696(8) & \  2.6(3)\\
\hline
5.50 & 0.5 & 0.5272(4) & \  1.20(8)\\
6.00 & 0.5 & 0.6089(2) &  25.4(4)\\
7.00 & 0.5 & 0.6817(4) &  35(1) \\
8.00 & 0.5 & 0.7285(7) &  11(1) \\
9.00 & 0.5 & 0.7617(10)& \  6(2)  \\
\hline
\end{tabular}
\end{minipage}
\begin{minipage}{7cm}
\begin{tabular}{|c|c|l|l|}
\hline
$\beta$ & $ma$ & plaquette & $|P_3|\times 10^2$ \\
\hline
4.900 & 0.1 &0.4224(1)&  \, 1.73(2)\\
5.000 & 0.1 &0.4433(2)&  \, 1.97(3)\\
5.100 & 0.1 &0.4702(4)&  \, 2.83(3)\\
5.150 & 0.1 &0.4872(2)&  \, 3.71(4)\\
5.180 & 0.1 &0.4997(2)&  \, 4.88(5)\\
5.190 & 0.1 &0.5043(3)&  \, 5.53(11)\\
5.195 & 0.1 &0.5167(5)& 13.52(34)\\
5.200 & 0.1 &0.5203(4)& 15.85(10)\\
5.205 & 0.1 &0.5222(4)& 15.87(35)\\
5.210 & 0.1 &0.5238(5)& 16.32(51)\\
5.220 & 0.1 &0.5269(5)& 17.28(53)\\
\hline
\end{tabular}
\end{minipage}
\end{center}
\end{table}

\newpage
\bibliography{references}{}
\bibliographystyle{apsrev4-1}

\end{document}